\documentclass[aps,pra,superscriptaddress,showpacs,twocolumn]{revtex4-1}
\usepackage{bm}
\usepackage{amsmath}
\usepackage{slashed}
\usepackage{float}
\allowdisplaybreaks
\usepackage{graphicx}

\newcommand*{\centt}[1]{\multicolumn{1}{c}{#1}}
\usepackage{dcolumn}
\newcolumntype{L}{>{$}l<{$}}

\begin{document}

\title{Nonradiative $\bm{\alpha^7m}$ QED effects in Lamb shift of helium triplet states}

\author{Vojt\v{e}ch Patk\'o\v{s}}
\affiliation{Faculty of Mathematics and Physics, Charles University,  Ke Karlovu 3, 121 16 Prague
2, Czech Republic}

\author{Vladimir A. Yerokhin}
\affiliation{Center for Advanced Studies, Peter the Great St.~Petersburg Polytechnic University,
Polytekhnicheskaya 29, 195251 St.~Petersburg, Russia}

\author{Krzysztof Pachucki}
\affiliation{Faculty of Physics, University of Warsaw,
             Pasteura 5, 02-093 Warsaw, Poland}

\begin{abstract}
  Theoretical predictions for the Lamb shift in helium are limited by unknown quantum
  electrodynamic effects of the order $\alpha^7m$, where $\alpha$ is the fine-structure constant
  and $m$ is the electron mass. We make an important step towards the complete calculation of
  these effects by deriving the most challenging part,
  which is induced by the virtual photon exchange between all three helium particles, the two
  electrons, and the nucleus. The complete calculation of the $\alpha^7m$ effect including the
  radiative corrections will allow comparing of the nuclear charge radii determined from the
  electronic and muonic helium atoms and thus provide a stringent test of the Standard Model of
  fundamental interactions.
\end{abstract}

\maketitle

High-precision spectroscopy of the hydrogen atom enables the determination of the Rydberg
constant \cite{mohr:16:codata} and provides a stringent test of the Standard Model of fundamental
interactions, through a comparison of the Lamb shift of ordinary (electronic) hydrogen and
the muonic hydrogen $\mu$H \cite{pohl:10,antognini:13}. This was made possible by the success of
the theory of the hydrogen atom, notably, progress in calculations of the higher-order two-loop quantum
electrodynamic (QED) effects \cite{yerokhin:18:hydr,karshenboim:18}.

The present investigation is a part of a long and challenging project, the aim of which is to extend the
high-precision spectroscopic tests of the Standard Model to two-electron systems; specifically,
the helium atom and light helium-like ions. Helium is better suited for experimental studies than
hydrogen because of the presence of several narrow lines in the spectrum. As a consequence, a
number of recent helium experiments reached a relative precision of a few parts in $10^{-12}$
\cite{zheng:02,rooij:11,luo:13,notermans:14,luo:16,zheng:17,rengelink:18,kato:18}.
This experimental precision is sufficient for an accurate spectroscopic determination of
the nuclear charge radii of $^3$He and $^4$He.

Such determination is of great importance because it would allow effective comparison
of spectroscopic measurements in ordinary helium and in muonic helium $\mu$He
\cite{pohl:16:leap}. A similar comparison in electronic and muonic hydrogen
\cite{pohl:10,antognini:13} revealed a large discrepancy, which remained unexplained for a decade
and became known as the proton-radius puzzle. Recent experiments
\cite{beyer:17,bezginov:19,xiong:19} indicate that this puzzle was most likely caused by
unidentified systematic effects in several previous measurements in electronic hydrogen.
These findings will have a significant impact on the future development of hydrogen
spectroscopy.

Our recent investigations indicate that similar problems might be present also in helium
spectroscopy. Specifically, in Refs.~\cite{pachucki:15:jpcrd,patkos:16:triplet,patkos:17:singlet}
we obtained accurate theoretical predictions for the isotope shifts of various transition
energies in helium. Combining these predictions with the available experimental data
\cite{shiner:95,pastor:04,rooij:11,pastor:12}, results were obtained for the difference $\delta
r^2$ of the mean-square nuclear charge radii of $^3$He and $^4$He. Comparing values of $\delta
r^2$ from different transitions, we found significant inconsistencies
\cite{pachucki:15:jpcrd,patkos:16:triplet}. Since then, an independent measurement of the
$2^3S$-$2^3P$ transition energy \cite{zheng:17} found a $20\,\sigma$ shift from the previous
experimental result \cite{pastor:12}, which led to a better, albeit not perfect, agreement among
different $\delta r^2$ values. This means that further work is needed for a reliable
determination of $\delta r^2$.

A more stringent test of the helium spectroscopic results could be accomplished if one extracts
the {\em absolute} nuclear radius and compares it with the corresponding value derived from the
spectroscopy of the muonic helium. In order to realize this project, a significant advance in
theory of the helium Lamb shift is needed; specifically, a complete calculation of the QED
effects of order $\alpha^7\,m$. The best existing calculations of the helium energy levels
\cite{pachucki:06:he,pachucki:17:heSummary} are complete through order $\alpha^6m$. The
higher-order $\alpha^7m$ were calculated for simpler cases; namely, for the fine structure
\cite{pachucki:06:prl:he,pachucki:09:hefs,pachucki:10:hefs}, the hydrogen molecular ions, and
the antiprotonic helium \cite{korobov:13,korobov:14}.

The first step on the path towards the complete calculation of the $\alpha^7m$ QED effects was
made in our previous work \cite{yerokhin:18:betherel}, in which we calculated the relativistic
correction to the so-called Bethe logarithm. 
In the present investigation we take the next step, which is the calculation of the $\alpha^7m$
correction induced by an exchange of two and three virtual photons between the two electrons and
between the electrons and the nucleus. The key part is the derivation of the corresponding
effective operator $H^{(7)}_{\rm exch}$. This challenging problem was successfully solved in the
present work, and the final result is compact and very simple. We leave for the future the third
and the last step, which is the calculation of the radiative $\alpha^7m$ corrections. After that,
the calculation of the $\alpha^7m$ QED effects for the triplet states of two-electron atoms will
be completed, enabling the spectroscopic determinations of the nuclear charge radii of the helium
atom and light helium-like ions.

\section{NRQED Approach}

The derivation in the present work is performed within the Nonrelativistic QED (NRQED) method,
originally introduced by Caswell and Lepage \cite{caswell:86}. This method is based on the NRQED
Lagrangian, which we obtain by the Foldy-Wouthuysen transformation of the Dirac Hamiltonian in
the presence of an electromagnetic field as described in Appendix \ref{app:A}.

Once the NRQED Lagrangian is obtained, the Feynman path-integral approach is used to derive
various corrections to the nonrelativistic multi-electron propagator $G(t-t')$, where $t$ and $t'$
are the common time of the {\em out} and the {\em in} electrons, correspondingly. The Fourier
transform over the time variable yields the propagator in the energy-coordinate representation,
\[
G(E) = \frac{1}{E-H_0 -\Sigma(E)},
\]
where $H_0$ is the Schr\"odinger-Coulomb Hamiltonian of an $N$-electron atom. $H_0$ may also
include the nucleus as a dynamic particle, but this is not needed in the present work, so we
assume the nucleus to be a static source of the Coulomb field. The operator $\Sigma(E)$
incorporates various corrections due to the photon exchange, the electron and photon self-energy,
etc.

The energy of a bound state $|\phi\rangle$ is obtained as a position of the pole of the matrix
element of $G(E)$ between the nonrelativistic wave functions of the reference state,
\begin{eqnarray}
\langle\phi|G(E)|\phi\rangle &=& \frac{1}{E-E_0} + \frac{1}{(E-E_0)^2}\,\langle\phi|\Sigma(E)|\phi\rangle
 \nonumber \\&&
+
 \frac{1}{(E-E_0)^2}\,\langle\phi|\Sigma(E)\,\frac{1}{E-H_0}\,\Sigma(E)|\phi\rangle + \ldots
 \nonumber \\
&=& \frac{1}{E - E_0 - \sigma(E)}\,,
\end{eqnarray}
where
\begin{equation}
\sigma(E) =  \langle\phi|\Sigma(E)|\phi\rangle + \langle\phi|\Sigma(E)\,\frac{1}{(E-H_0)'}\,\Sigma(E)|\phi\rangle +\ldots\,.
\end{equation}
The resulting bound-state energy $E$ (i.e. the position of the pole) is
\begin{equation}\label{eq:3}
E = E_0 + \sigma(E_0) + \sigma(E_0)\,\frac{\partial \sigma(E_0)}{\partial E_0} + \ldots\,.
\end{equation}
We assume that $E$ can be expanded in a power series of the
fine-structure constant $\alpha$,
\begin{align}
E(\alpha) =&\ m\alpha^2\,E^{(2)}
+ m\alpha^4\,E^{(4)}
\nonumber \\ &\
+ m\alpha^5\,E^{(5)}
+ m\alpha^6\,E^{(6)}
+ m\alpha^7\,E^{(7)} +\ldots \,,\label{04}
\end{align}
where the expansion coefficients $E^{(n)}$ may contain finite powers of $\ln\alpha$. These
coefficients can be expressed as expectation values of some effective Hamiltonians with the
nonrelativistic wave function. The derivation of these effective Hamiltonians is the central
problem of the NRQED method. While the leading-order expansion terms are simple, formulas become
increasingly complicated for higher orders in $\alpha$.

The present status of the theory of helium energy levels up to the order $\alpha^6 m$ is summarized in
our recent review \cite{pachucki:17:heSummary}. In the present investigation we are interested in
the next-order $\alpha^7m$ contribution $E^{(7)}$.

\section{$\bm{\alpha^7\,m}$ effects}
The contribution of order $\alpha^7m$ can be represented as a sum of four parts,
\begin{equation}\label{eq:1}
  E^{(7)} =
  \big< H^{(7)}_{\rm exch}\big> + \big< H^{(7)}_{\rm rad}\big>
  + 2\,\big< H^{(4)} \, \frac{1}{(E_0-H_0)'} \, H^{(5)}\big>
  +   E_L \,.
\end{equation}
The first two terms here are the expectation values of the effective Hamiltonians of order
$\alpha^7m$ induced by the photon exchange and the radiative effects, respectively. The third
term is the second-order perturbative correction induced by the Breit Hamiltonian $H^{(4)}$ and
the effective Hamiltonian $H^{(5)}$ of order $\alpha^5\,m$, whereas the last term is the
relativistic correction to the Bethe logarithm.

The goal of the present investigation is the derivation of the effective Hamiltonian
$H^{(7)}_{\rm exch}$ , which originates exclusively from the photon-exchange diagrams. It is
convenient to split it into three parts,
\begin{equation} \label{eq:2}
\big< H^{(7)}_{\rm exch}\big> = E_L^{\Lambda}+E_M+E_H\,,
\end{equation}
which are induced by momenta $k$ of the exchanged virtual photons of the order $m\alpha^2$,
$m\alpha$, and $m$, respectively. These three terms will be referred to as the low-energy,
middle-energy, and high-energy parts, correspondingly. The derivation will be performed in the
Coulomb gauge unless explicitly stated otherwise.

The middle-energy and the high-energy parts  $E_M$ and $E_H$ contain singular operators that need
a systematic regularization. In the present work we will use the dimensional regularization, with
the dimension $d = 3 - 2\epsilon$. Singular contributions of order $1/\epsilon$ will be canceled
algebraically in momentum space. After that, the result will be transformed into the coordinate
representation, where it can be calculated numerically.

$E_M$ consists of the two- and three-photon exchange contributions. The two-photon exchange part
is the most complicated one. In order to eliminate possible errors, its derivation is performed
by two independent methods, namely, by the NRQED approach and by the scattering amplitude method.
The derivation of the three-photon part from the scattering amplitude is too complicated to be
feasible, so we perform it only within the NRQED approach. There are only a few three-photon
diagrams and the corresponding derivation within the NRQED approach is sufficiently tractable to
be properly checked. The high-energy part $E_H$ is derived from the two-photon exchange
scattering amplitude, in an analogous way as it was done for the fine structure in
Refs.~\cite{pachucki:06:prl:he,pachucki:09:hefs,pachucki:10:hefs}.

The low-energy contribution $E_L^{\Lambda}$ in Eq.~(\ref{eq:2}) is a complementary part to the
relativistic correction to the Bethe logarithm  ($E_L$ in Eq.~(\ref{eq:1})) calculated in
Ref.~\cite{yerokhin:18:betherel}. The definition of $E_L$ involved the cut-off photon-momentum
parameter $\Lambda$. The low-energy contribution $E_L^{\Lambda}$ converts the $\Lambda$
regularization to the dimensional regularization used in $E_M$ and $E_H$, so that all
singularities can be  consistently canceled.

In our derivation we will need the $d$-dimensional generalization of the Breit Hamiltonian. For a
two-electron atom, it can be written as
\begin{align}
 H^{(4)}&\  = H'^{(4)} + H''^{(4)}\,, \\
 H'^{(4)}&\ = - \frac{\pi\,\alpha}{m^2}\,\delta^d(r)
+ \sum_{a=1,2}\biggl\{ -\frac{p_a^4}{8\,m^3} +\frac{\pi\,Z\alpha}{2\,m^2}\,\delta^d(r_a)
\nonumber \\ &\
-\frac{1}{4\,m^2}\,\sigma_a^{ij}\,\nabla^i \left[\frac{Z\,\alpha}{r_a}\right]_\epsilon p_a^j
+\frac{1}{4\,m^2} \sigma_a^{ij}\,\nabla_a^i\left[\frac{\alpha}{r}\right]_\epsilon p_a^j
\biggr\}\,,
 \nonumber \\
 H''^{(4)} &\ =
-\frac{\alpha}{2\,m^2}\,p_1^i\,
\biggl[\frac{\delta^{ij}}{r}+\frac{r^i\,r^j}{r^3}
\biggr]_\epsilon\,p_2^j
- \frac{\pi\,\alpha}{d\,m^2}\,\sigma_1^{ij}\,\sigma_2^{ij}\,\delta^d(r) \nonumber \\ &\
+\frac{1}{4\,m^2}\,\sigma_1^{ik}\,\sigma_2^{jk}\,
\left(\nabla^i\,\nabla^j-\frac{\delta^{ij}}{d}\,\nabla^2\right)\,
\left[\frac{\alpha}{r}\right]_\epsilon \nonumber \\ &\
-\frac{1}{2\,m^2}\,\biggl(\sigma_1^{ij}\,\nabla^i\left[\frac{\alpha}{r}\right]_\epsilon\,p^j_2
-\sigma_2^{ij}\,\nabla^i\left[\frac{\alpha}{r}\right]_\epsilon\,p^j_1 \biggr)\,,
\label{breit}
\end{align}
where $\vec{r} = \vec{r}_{1}-\vec{r}_{2}$, $\sigma^{ij}$ is defined in Appendix~\ref{app:B},
$\delta^{d}(r)$ is the Dirac $\delta$-function in $d$ dimensions, and $\big[ \ldots
\big]_{\epsilon}$ denotes the $d$-dimensional generalization of the operator $\big[\ldots\big]$
written in $d=3$. The separation of $H^{(4)}$ into two parts is based on the fact that $H'^{(4)}$
comes from the exchange of a Coulomb photon, whereas $H''^{(4)}$ originates from the exchange of
a transverse photon.

The second-order contribution in Eq. (\ref{eq:1}) involves the Breit Hamiltonian $H^{(4)}$ and
the effective $\alpha^5\,m$ operator $H^{(5)}$, which is the sum of the photon-exchange and
radiative parts,
\begin{align}
  H^{(5)} = H^{(5)}_{\rm exch} + H^{(5)}_{\rm rad}\,.
\end{align}
This contribution is relatively simple. The derivation of $H^{(5)}_{\rm exch}$  is presented
in  Appendix~\ref{app:D} following the method described in Ref.~\cite{simple},
and the result in $d=3$ for triplet states is
\begin{align}
  H^{(5)}_{\rm exch}  = &\, -\frac{7}{6\,\pi}\,\frac{\alpha^2}{m^2}\,\frac{1}{r^3}
\end{align}
The radiative $H^{(5)}_{\rm rad}$ part is left for future investigation, alongside $H^{(7)}_{\rm rad}$.

\section{Low-energy part}

The low-energy part comes from the virtual photon momenta of the order $k\propto m\alpha^2$. In
order to derive formulas for the low-energy part, we start with the one-loop nonrelativistic
dipole exchange contribution of the order $\alpha^5m$, which is
\begin{eqnarray}\label{eq:L:0}
E_{L0} &=& \frac{e^2}{m^2}\int_0^\infty \frac{d^dk}{(2\pi)^d2k}\,
 \delta_\perp^{ij}(k)
 \nonumber \\ && \times
\,\bigg\langle\phi\bigg|p_1^i\, \frac{1}{E_0-H_0-k}\,p_2^j\bigg|\phi\bigg\rangle
+(1\leftrightarrow2)\,,
\end{eqnarray}
where $\delta_\perp^{ij}(k) = \delta^{ij} - k^ik^j/k^2$. We are interested in relativistic
corrections to $E_{L0}$ of order $\alpha^7m$. Such corrections arise through: ({\em i})
perturbations of the reference-state wave function $\phi$, the zeroth-order energy $E_0$ and the
zeroth-order Hamiltonian $H_0$ by the Breit Hamiltonian $H^{(4)}$, ({\em ii}) the perturbation of
the current $\vec{p} \to \delta \vec{j}$, and ({\em iii}) the retardation (quadrupole)
correction. The corresponding corrections will be denoted as $E_{L1}$, $E_{L2}$, and $E_{L3}$,
respectively, so
\begin{align}\label{eq:L:00}
E_L^\Lambda = E_{L1}^{\Lambda} + E_{L2}^{\Lambda} + E_{L3}^{\Lambda}.
\end{align}

It is convenient to separate the $k$ integral in Eq.~(\ref{eq:L:0}) into two regions,
\begin{eqnarray}
\int_0^\infty dk = \int_0^\Lambda dk+\int_\Lambda^\infty dk\,,\label{lambda}
\end{eqnarray}
where $\Lambda \equiv m\alpha^2 \lambda$ and $\lambda$ is the dimensionless cutoff parameter.
In the small-$k$ region, $k < \Lambda$, the binding effects should be accounted for to all
orders. Such contributions give rise to relativistic corrections to the Bethe logarithm, already
computed in Ref.~\cite{yerokhin:18:betherel}. In the present investigation we will be concerned
with the large-$k$ region, $k > \Lambda$. In this region, $k$ is much larger than the
characteristic energy of the intermediate electron states and we can use the large-$k$ expansion
of the resolvent $1/(E_0-H_0-k)$.

The perturbation of Eq.~(\ref{eq:L:0}) by the Breit Hamiltonian can be written as
\begin{eqnarray}\label{eq:L:1}
E_{L1}^{\Lambda} &=& \frac{e^2}{m^2}\int_{\Lambda}^\infty \frac{d^dk}{(2\pi)^d2k}\,
 \delta_\perp^{ij}(k)
 \nonumber \\ && \times
\,\delta\bigg\langle\phi\bigg|p_1^i\frac{1}{E_0-H_0-k}p_2^j\bigg|\phi\bigg\rangle
+(1\leftrightarrow2)\,,
\end{eqnarray}
where the symbol $\delta\langle\cdots\rangle$ stands for the first-order perturbation of the
matrix element $\langle\cdots\rangle$ by the Breit-Hamiltonian $H^{(4)}$, which implies
perturbations of the reference-state wave function $\phi$, the energy $E_0$ and the zeroth-order
Hamiltonian $H_0$. Since $k$ is much bigger than $H_0-E_0$, we can expand the integrand of Eq.~(\ref{eq:L:1}) in
large $k$, keeping only the $1/k^2$ term, while $1/k$ contributes at the lower order of $\alpha^6\,m$.
The result is
\begin{widetext}
\begin{eqnarray}\label{eq:L:2}
\delta\bigg\langle\phi\bigg|p_1^i\frac{1}{E_0-H_0-k}p_2^j\bigg|\phi\bigg\rangle + (1\leftrightarrow2)
&=&\frac{1}{k^2}\,\delta\big\langle\phi\big|\big[p_1^i,\big[H_0-E_0,p_2^j\big]\big]\big|\phi\rangle\nonumber\\
&=& \frac{2}{k^2}\,\bigg\langle\phi\bigg|[p_1^i,[V,p_2^j]]\frac{1}{(E_0-H_0)'}H^{(4)}\bigg|\phi\bigg\rangle
+ \frac{1}{k^2}\,\big\langle\phi|[p_1^i,[H^{(4)},p_2^j]]\big|\phi\big\rangle\,,
\end{eqnarray}
where $V$ is the Coulomb potential.  The first term is the second-order perturbation correction
included in the third term of Eq.~(\ref{eq:1}), so we omit it here. We thus have
\begin{eqnarray}
E_{L1}^{\Lambda} = \frac{e^2}{m^2}\frac{d-1}{d}\int_\Lambda^\infty\frac{d^dk}{(2\pi)^d\,2\,k^3}
 \big\langle\phi\big|\,[p_1^i,[\delta H^{(4)},p_2^i]]\,\big|\phi\big\rangle\,,
\end{eqnarray}
where $\delta H^{(4)}$ is the spin-independent part of the Breit Hamiltonian that survives the double
commutator, given by (in momentum representation and in $d$ dimensions)
\begin{equation}
\delta H^{(4)} = -\frac{\alpha\,\pi}{d\,m^2}\sigma_1\cdot\sigma_2
-\frac{\alpha}{m^2}\bigg[\pi + p_1^i\,\frac{4\pi}{q^2}\bigg(\delta^{ij}-\frac{q^iq^j}{q^2}\bigg)\,p'^j_2\bigg]
\,.
\end{equation}
Here, $\vec{q} = \vec p_1{}' -\vec p_1$ is the momentum exchanged between electrons.
All integrations are performed in the momentum space using formulas presented in Appendix C.
After expanding in $\epsilon = (3-d)/2$ and then in $\alpha$, we obtain the result for the effective
operator $H_{L1}$, defined as $E_{L1}^\Lambda = \langle H_{L1}\rangle$,
\begin{eqnarray} H_{L1} &=&
\frac{\alpha^2}{m^4}\bigg\{\sigma_1\cdot\sigma_2\bigg(-\frac{7}{27}-\frac19\ln{\Lambda_\epsilon}
\bigg)\,q^2
+\bigg(-\frac{5}{9}-\frac13\ln{\Lambda_\epsilon}\bigg)\,
\bigg[q^2+4\vec P_1\cdot\vec P_2 - 4\frac{\big(\vec P_1\cdot \vec q\big)\big(\vec P_2\cdot\vec q\big)}{q^2}\bigg]\bigg\}\,,
\end{eqnarray}
where
\begin{eqnarray}
\ln{\Lambda_\epsilon} &=& \frac{1}{\epsilon}+2\ln[\alpha^{-2}]-2\ln(2\lambda)\,
\end{eqnarray}
and
\begin{eqnarray}
\vec P_1 = \frac12(\vec p_1+\vec p_1{}'),\,\,\vec P_2 = \frac12(\vec p_2+\vec p_2{}')
\end{eqnarray}
are sums of the {\em in} and {\em out} momenta of the corresponding electron.

The second term in Eq.~(\ref{eq:L:00}) comes from the correction to current. Specifically,
$\vec{p}_1$ gets a correction $\delta \vec j_1$, which is
\begin{eqnarray}
\delta j^i_1 = i\big[H^{(4)},r_1^i\big]
=-\frac{1}{2m^3} p_1^i p_1^2 - \frac{1}{m^2}\frac{4\pi\alpha}{q'^2}\bigg(\delta^{ij}-\frac{q'^i q'^j}{q'^2}\bigg)\,p_2^j\,,
\end{eqnarray}
where $q'$ is the exchanged momentum, and the same for $\vec{p}_2$. The correction $E_{L2}$ is then
\begin{eqnarray}
E_{L2}^{\Lambda} &=& 2\frac{e^2}{m}\int_\Lambda^\infty \frac{d^dk}{(2\pi)^d2k}\,\delta_\perp^{ij}(k)
\,\bigg\langle\phi\bigg|\delta j_1^i\frac{1}{E_0-H_0-k}p_2^j\bigg|\phi\bigg\rangle
+(1\leftrightarrow2)\,.
\end{eqnarray}
Expanding for large $k$ and performing the angular average, we arrive at
\begin{eqnarray}
E_{L2}^{\Lambda} = \frac{e^2}{m}\frac{d-1}{d}\int_\Lambda^\infty\frac{d^dk}{(2\pi)^d2k^3}
 \big\langle\phi\big|\,[\delta j_1^i,[V,p_2^i]]\,\big|\phi\big\rangle
+(1\leftrightarrow2)\,.
\end{eqnarray}
This expression contains three-photon $\sim\alpha^3$ and two-photon $\sim\alpha^2$ terms.
The result for the effective operator
$H_{L2}$, $E_{L2}^\Lambda=\langle H_{L2}\rangle$, is
\begin{eqnarray}
H_{L2}
&=&\frac{\alpha^3}{m^3}\bigg\{\bigg[p_2^i,\bigg[-\bigg[\frac{Z}{r_2}\bigg]_\epsilon,p_2^j\bigg]\bigg]
\bigg(-\frac{20}{9}-\frac43\ln{\Lambda_\epsilon}
\bigg)\bigg(\delta^{ij}-\frac{q^i q^j}{q^2}\bigg)\frac{1}{q^2}
+(1\leftrightarrow2)\bigg\}\nonumber\\
&&+ \frac{\alpha^2}{m^4}\bigg(-\frac{20}{9}-\frac43\ln{\Lambda_\epsilon}\bigg)
\bigg(\frac12\big(\vec P_1-\vec P_2\big)^2+\vec P_1\cdot\vec P_2+\frac14 q^2
+\frac{\big[\big(\vec P_1-\vec P_2\big)\cdot\vec q\big]^2}{q^2}
+2\frac{\big(\vec P_1\cdot\vec q\big)\big(\vec P_2\cdot\vec q\big)}{q^2}\bigg)\nonumber\\
&&+\pi \frac{\alpha^3}{m^3}\bigg(\frac{4}{9}+\frac23\ln{\Lambda_\epsilon}+\frac43\ln2-\frac43\ln q\bigg)q
\,.
\end{eqnarray}

$E^\Lambda_{L3}$ is the correction due to the retardation. We write it as
\begin{eqnarray}
E_{L3}^{\Lambda} &=& \frac{e^2}{m^2}\int_\Lambda^\infty \frac{d^dk}{(2\pi)^d2k}
\, \delta_\perp^{ij}(k)
\,\delta_{k^2} \bigg\langle\phi\bigg|p_1^i\,e^{i\vec k\cdot\vec r_1}\frac{1}{E_0-H_0-k}p_2^j\,e^{-i\vec k\cdot\vec r_2}\bigg|\phi\bigg\rangle
+(1\leftrightarrow2)\,.
\end{eqnarray}
Here the symbol $\delta_{k^2}\big<\ldots\big>$ means that the exponent functions $e^{i\vec
k\cdot\vec r_1}$ and $e^{-i\vec k\cdot\vec r_2}$ in the matrix element $\big<\ldots\big>$ are
expanded in small $k$ and only the $k^2$ term is left. We now perform the large-$k$ expansion of the resolvent,
\begin{eqnarray}
\frac{1}{E_0-H_0-k} &=& -\frac{1}{k}+\frac{H_0-E_0}{k^2}-\frac{(H_0-E_0)^2}{k^3}+\frac{(H_0-E_0)^3}{k^4}+\cdots\,.
\end{eqnarray}
We have to extend the expansion up to order $k^{-4}$ because of the additional $k^2$ from the
expansion of the exponent functions. The result of the expansion is
\begin{eqnarray}
E_{L3}^{\Lambda} &=& \frac{e^2}{m^2}\int_\Lambda^\infty \frac{d^dk}{(2\pi)^d2k^5}\,\delta_\perp^{ij}(k)
\,\delta_{k^2} \bigg\langle\phi\bigg|p_1^i\,e^{i\vec k\cdot\vec r_1}(H_0-E_0)^3\,p_2^j\,e^{-i\vec k\cdot\vec r_2}\bigg|\phi\bigg\rangle
+(1\leftrightarrow2)\,.
\end{eqnarray}
This expression is evaluated in Appendix~\ref{app:E}, with the result for $E_{L3}^\Lambda =
\langle H_{L3}\rangle$,
\begin{eqnarray}
H_{L3} &=&
\frac{\alpha^3}{m^3}\bigg\{
\bigg[p_2^i,\bigg[-\bigg[\frac{Z}{r_2}\bigg]_\epsilon,p_2^j\bigg]\bigg]
\bigg[\delta^{ij}\bigg(\frac{184}{225}+\frac{4}{15}\ln{\Lambda_\epsilon}\bigg)
+\frac{q^i q^j}{q^2}\bigg(-\frac{436}{225}-\frac{16}{15}\ln{\Lambda_\epsilon}\bigg)\bigg]\frac{1}{q^2}\nonumber\\
&&+\bigg(-\frac{254}{225}-\frac{14}{15}\ln{\Lambda_\epsilon}
\bigg)\bigg[p_2^j,-\bigg[\frac{Z}{r_2}\bigg]_\epsilon\bigg]\frac{q^j}{q^2}
+(1\leftrightarrow2)\bigg\}
+\frac{\alpha^2}{m^4}\bigg\{\bigg(\frac{124}{225}+\frac{4}{15}\ln{\Lambda_\epsilon}\bigg)
\bigg((\vec P_1-\vec P_2)^2+\frac12 q^2\bigg)\nonumber\\
&&+\bigg(\frac{136}{45}+\frac43 \ln{\Lambda_\epsilon}\bigg)\,\vec P_1\cdot\vec P_2
+\bigg(\frac{10}{9}+\frac23\ln{\Lambda_\epsilon}\bigg)\frac{[(\vec P_1-\vec P_2)\cdot\vec q]^2}{q^2}
+\bigg(\frac{112}{45}+\frac43\ln{\Lambda_\epsilon}\bigg)\frac{(\vec P_1\cdot\vec q)(\vec P_2\cdot\vec q)}{q^2}\bigg\}\nonumber\\
&&+\pi\frac{\alpha^3}{m^3}\bigg(\frac{29}{225}-\frac{1}{15}\ln{\Lambda_\epsilon}-\frac{2}{15}\ln2+\frac{2}{15}\ln q\bigg)\,q\,.
\end{eqnarray}

The total result for the low-energy contribution is a sum of three terms in Eq.~(\ref{eq:L:00}),
\begin{eqnarray}
E_L^\Lambda &=& \langle H_{L}\rangle = \langle H_{L1}+H_{L2}+H_{L3}\rangle\,, \nonumber
\end{eqnarray}
\begin{eqnarray}
H_L&=&
\frac{\alpha^3}{m^3}\bigg\{
\bigg[p_2^i,\bigg[-\bigg[\frac{Z}{r_2}\bigg]_\epsilon,p_2^j\bigg]\bigg]
\bigg[\delta^{ij}\bigg(-\frac{316}{225}-\frac{16}{15}\ln{\Lambda_\epsilon}\bigg)
+\frac{q^i q^j}{q^2}\bigg(\frac{64}{225}+\frac{4}{15}\ln{\Lambda_\epsilon}\bigg)\bigg]\frac{1}{q^2}\nonumber\\
&&+\bigg(-\frac{254}{225}-\frac{14}{15}\ln{\Lambda_\epsilon}\bigg)
\bigg[p_2^j,-\bigg[\frac{Z}{r_2}\bigg]_\epsilon\bigg]\frac{q^j}{q^2}
+(1\leftrightarrow2)\bigg\}
+\frac{\alpha^2}{m^4}\bigg[
\bigg(-\frac{14}{25}-\frac25\ln{\Lambda_\epsilon}\bigg)(\vec P_1-\vec P_2)^2\nonumber\\
&&+\bigg(-\frac{188}{225}-\frac{8}{15}\ln{\Lambda_\epsilon}\bigg)q^2
+\bigg(-\frac{64}{45}-\frac43\ln{\Lambda_\epsilon}\bigg)\vec P_1\cdot\vec P_2
+\frac{4}{15}\frac{(\vec P_1\cdot\vec q)(\vec P_2\cdot\vec q)}{q^2}\nonumber\\
&&+\bigg(-\frac{10}{9}-\frac23\ln{\Lambda_\epsilon}\bigg)
\frac{[(\vec P_1-\vec P_2)\cdot\vec q]^2}{q^2}
+\sigma_1\cdot\sigma_2\bigg(-\frac{7}{27}-\frac19\ln{\Lambda_\epsilon}\bigg)\,q^2\bigg]\nonumber\\
&&+\pi\frac{\alpha^3}{m^3}\bigg(\frac{43}{75}+\frac35\ln{\Lambda_\epsilon}+\frac{6}{5}\ln2-\frac{6}{5}\ln q\bigg)q\,.
\nonumber\\
\end{eqnarray}
This concludes the derivation of the low-energy part $E_{L}^{\Lambda}$.

\section{Middle-energy photon-exchange part $\bm{E_M}$}
The derivation of $E_M$ is by far the most complicated. We start with deriving the NRQED Hamiltonian
by the Foldy-Wouthuysen transformation of the Dirac Hamiltonian in the external electromagnetic
field, as described in Appendix \ref{app:A}. The result is
\begin{eqnarray}
H_{\rm FW} &=&
e A^0 + \frac{\pi^2}{2m}  - \frac{e}{4m}\sigma\cdot B - \frac{\pi^4}{8m^3} +
\frac{e}{16m^3}\bigl(\sigma\cdot B\,\vec{\pi}^2 + \vec{\pi}^2 \,\sigma\cdot B\bigr)
-\,\frac{e}{8m^2}\bigl(\vec{\nabla}\,\vec{E_\parallel}
+ \sigma^{ij}\{E_\parallel^i,\pi^j\}\bigr)\nonumber\\
&& + \frac{e^2}{8m^3} \vec{E}^2 - \frac{e^2}{16m^3}B^{ij} B^{ij}
-\frac{e^2}{4m^2}\sigma^{ij} A^i E^j
+\frac{ie}{16m^3}[\sigma^{ij}\{A^i,\pi^j\},\pi^2] -\frac{e^2}{8m^3}A^i\,\nabla^j B^{ij}\,,
\end{eqnarray}
where $\vec E_\parallel = -\vec\nabla V$ and $B^{ij}=\partial^i A^j-\partial^j A^i$.
From this Hamiltonian we obtain the single-transverse
photon vertices,
\begin{eqnarray}\label{eq:ST}
  \delta_{\rm ST}H_{\rm FW} &=& -\frac{e}{m}\Bigl(\vec{p}\cdot\vec{A} + \frac{1}{4}\sigma\cdot B\Bigr)
+\frac{e}{4m^3}\bigl\{\vec{p}\,{}^2,\vec{p}\cdot\vec{A}+\frac14\,\sigma\cdot B\bigr\}
+\frac{e^2}{2m^2}\sigma^{ij} E_\parallel^i A^j
+\frac{e^2}{4\,m^3}\,\vec E_\parallel\,\vec E_\perp
+\frac{ie}{16m^3}[\sigma^{ij}\{A^i,p^j\},p^2]
\,,\nonumber\\
\end{eqnarray}
and the double-transverse photon vertices,
\begin{align}\label{eq:DT}
\delta_{\rm DT}H_{\rm FW} =&
\frac{e^2}{2m}\vec{A}\,{}^2
-\frac{1}{8m^3}\{\,\vec{p}\,{}^2,e^2 \vec{A}\,{}^2\}
- \frac{e^2}{2m^3} (\vec{A}\cdot\vec{p}\,)^2
+\frac{e^2}{4m^2}\,\sigma^{ij} {E}_\perp^i A^j
+ \frac{e^2}{8m^3} \vec{E}^2_\perp - \frac{e^2}{16m^3}B^{ij}B^{ij}
 -\frac{e^2}{8m^3}A^i\,\nabla^j B^{ij}
 \,,
\end{align}
where we omitted terms contributing to the fine structure only. From $\delta_{\rm ST}H_{\rm FW}$
and $\delta_{\rm DT}H_{\rm FW}$ in Eqs.~(\ref{eq:ST}) and (\ref{eq:DT}) we obtain the set of all
possible vertices that enter time ordered diagrams describing the corrections to the energy
of the order $\alpha^7\,m$. There are altogether 17 classes of diagrams contributing to the middle-energy
contribution, listed in Fig.~\ref{fig:Ei} and in Table.~\ref{tab:Ei}. So, the middle-energy
contribution $E_M$ is split into 17 parts, each of which is represented as an expectation value
of the corresponding Hamiltonian,
\begin{align}
  E_M = \sum_{i=1}^{17} E_i \equiv \sum_{i=1}^{17} \langle\phi|H_i|\phi\rangle\,.
\end{align}
In the remaining part of this section we evaluate all diagrams one by one. The calculation is
performed in the momentum representation. Most of the terms are
calculated from these time-ordered diagrams, for which one can use the following formulas  for
$\vec A$ and $\vec E$ in the vertices (\ref{eq:ST}) and (\ref{eq:DT})
\begin{align}
\vec A =&\ \int\frac{d^dk}{\sqrt{(2\,\pi)^d}}\frac{1}{\sqrt{2\,k}}\,
 \vec\epsilon_\lambda\,\bigl(a_{\vec k,\lambda}\,e^{i\,\vec k\,\vec r} +a^+_{\vec k,\lambda}\,e^{-i\,\vec k\,\vec r}\bigr)\,,\\
\vec E =&\ \int\frac{d^dk}{\sqrt{(2\,\pi)^d}}\frac{k}{\sqrt{2}}\,
\vec\epsilon_\lambda\,i\,\bigl(a_{\vec k,\lambda}\,e^{i\,\vec k\,\vec r} -a^+_{\vec k,\lambda}\,e^{-i\,\vec k\,\vec r}\bigr)\,,
\end{align}
with $\vec\epsilon_\lambda$ being the polarization vector and $a_{\vec k,\lambda}$ and $a^+_{\vec
k,\lambda}$ being the annihilation and creation operators, respectively. However, $E_{10}$ and $E_{15}$,
due to their specific structure, are calculated from Feynman diagrams. In the following
derivation we will extensively use the currents
\begin{align}
  j^i(\vec k) =&\ \frac{p^i}{m} + \frac{i}{2\,m}\,k^k\,\sigma^{ki}\,.
\end{align}
The commutator of two currents with $e^{i\,\vec k\,\vec r}$ factor is evaluated as
\begin{align}
  [j^i(\vec k_1)\,e^{i\,\vec k_1\,\vec r}\,,\,j^j(\vec k_2)\,e^{i\,\vec k_2\,\vec r}] =&\
  \Big(j_n^{ij}(\vec p,\vec k_1,\vec k_2) + j_s^{ij}(\vec k_1,\vec k_2)\Big)\,e^{i\,(\vec k_1+\vec k_2)\vec r} \,,
\label{37}
\end{align}
where
\begin{align}
j_n^{ij}(\vec p,\vec k_1,\vec k_2) =&\ \frac{1}{m^2}\big(-p^i\,k_1^j+p^j\, k_2^i\big) \,,\\
j_s^{ij}(\vec k_1,\vec k_2) =&\
\frac{i\,\sigma^{ab}}{2\,m^2}\,j_s^{ij,ab}(\vec k_1,\vec k_2) \,, \\
j_s^{ij,ab}(\vec k_1,\vec k_2)
=&\  -\vec k_1\cdot\vec k_2\,\delta^{ia}\,\delta^{jb} -\delta^{ij}k_1^a k_2^b
+k_1^j k_2^b\,\delta^{ia}
+k_1^a k_2^i\,\delta^{jb}
-k_1^j k_1^a\,\delta^{ib}
-k_2^i k_2^b \,\delta^{ja}\,.
\end{align}
Moreover, we will often suppress writing arguments of currents $j_n^{ij}(\vec p,\vec
k_1,\vec k_2)$ and $j_s^{ij}(\vec k_1,\vec k_2)$.

\begin{table}
\caption{
  List of 17 diagrams contributing to the middle-energy part $E_M$. The exponent in $(H-E)^n$
  corresponds to the retardation order, {\em i.e.}, the order in the large-$k$ expansion of
  $1/(E-H-k)$. The term $E_5$ and a part of $E_6$ are contained in  $E_2$ (the transverse photon
  exchange combined with the Breit interaction) and thus are excluded in order to avoid
  the double counting. Terms $E_1$, a part of $E_2$, and terms $E_7-E_{11}$ are due to the single
  transverse photon exchange; the remaining terms correspond to the double transverse photon exchange.
 \label{tab:Ei}}
\begin{ruledtabular}
  \begin{tabular}{LLLLL}
\centt{vertex} &\centt{vertex} & \centt{vertex/comment} &\centt{retardation order} &\centt{diagram}\\[2pt] \hline
  -\frac{e}{m}\Bigl(\vec{p}\cdot\vec{A}+ \frac{1}{4}\sigma\cdot B\Bigr)&
  -\frac{e}{m}\Bigl(\vec{p}\cdot\vec{A}+ \frac{1}{4}\sigma\cdot B\Bigr)&&
  (H-E)^3&E_1\\[8pt]
  -\frac{e}{m}\Bigl(\vec{p}\cdot\vec{A}+ \frac{1}{4}\sigma\cdot B\Bigr)&
  -\frac{e}{m}\Bigl(\vec{p}\cdot\vec{A}+ \frac{1}{4}\sigma\cdot B\Bigr)&\mbox{\rm perturbed by}\, H^{(4)}&
  (H-E)&E_2\\[8pt]
  \frac{e^2}{2m}\vec{A}\,{}^2&
  \frac{e^2}{2m}\vec{A}\,{}^2&&
(H-E)^2&E_3\\[8pt]
  \frac{e^2}{2m}\vec{A}\,{}^2&
  -\frac{e}{m}\Bigl(\vec{p}\cdot\vec{A}+ \frac{1}{4}\sigma\cdot B\Bigr)&
  -\frac{e}{m}\Bigl(\vec{p}\cdot\vec{A}+ \frac{1}{4}\sigma\cdot B\Bigr)&
  (H-E)&E_4\\[8pt]
  -\frac{e}{m}\Bigl(\vec{p}\cdot\vec{A}+ \frac{1}{4}\sigma\cdot B\Bigr)&
  -\frac{e}{m}\Bigl(\vec{p}\cdot\vec{A}+ \frac{1}{4}\sigma\cdot B\Bigr)&\times2:\,\mbox{\rm reducible part}&
  (H-E)^0&E_5\\[8pt]
    -\frac{e}{m}\Bigl(\vec{p}\cdot\vec{A}+ \frac{1}{4}\sigma\cdot B\Bigr)&
  -\frac{e}{m}\Bigl(\vec{p}\cdot\vec{A}+ \frac{1}{4}\sigma\cdot B\Bigr)&\times2:\,\mbox{\rm irreducible part}&
  (H-E)^0&E_6\\[8pt]
  -\frac{e}{m}\Bigl(\vec{p}\cdot\vec{A}+ \frac{1}{4}\sigma\cdot B\Bigr)&
   \frac{e}{4m^3}\bigl\{\vec{p}\,{}^2,\vec{p}\cdot\vec{A}+\frac14\,\sigma\cdot B\bigr\}&&
    (H-E)&E_7\\[8pt]
  -\frac{e}{m}\Bigl(\vec{p}\cdot\vec{A}+ \frac{1}{4}\sigma\cdot B\Bigr)&
  \frac{e^2}{2m^2}\sigma^{ij}{E}_\parallel^i A^j &&
  (H-E)&E_8\\[8pt]
	-\frac{e}{m}\Bigl(\vec{p}\cdot\vec{A}+ \frac{1}{4}\sigma\cdot B\Bigr)&
	\frac{ie}{16m^3}[\sigma^{ij}\{A^i,p^j\},p^2] && (H-E)& E_9\\[8pt]
\frac{e}{2m^2}\sigma^{ij} {E}_\parallel^i A^j &
\frac{e}{2m^2}\sigma^{ij} {E}_\parallel^i A^j &&
(H-E)^0& E_{10}\\[8pt]
    -\frac{e}{m}\Bigl(\vec{p}\cdot\vec{A}+ \frac{1}{4}\sigma\cdot B\Bigr)&
  \frac{e^2}{4\,m^3}\,\vec E_\parallel\,\vec E_\perp&&
  (H-E)^0&E_{11}\\[8pt]
  \frac{e^2}{2m}\vec{A}\,{}^2&
-\frac{1}{8m^3}\{\,\vec{p}\,{}^2,e^2 \vec{A}\,{}^2\}  &&
  (H-E)^0&E_{12}\\[8pt]
  \frac{e^2}{2m}\vec{A}\,{}^2&
  \frac{e^2}{2m^3} (\vec{A}\cdot\vec{p}\,)^2  &&
  (H-E)^0&E_{12}\\[8pt]
  \frac{e^2}{2m}\vec{A}\,{}^2&
 \frac{e^2}{8m^3} \vec{E}^2_\perp &&
    (H-E)^0&E_{13}\\[8pt]
   \frac{e^2}{2m}\vec{A}\,{}^2&
  - \frac{e^2}{16m^3}B^{ij}B^{ij} & &
   (H-E)^0&E_{14}\\[8pt]
\frac{1}{4m^2}e^2 \,\sigma^{ij}{E}_\perp^i A^j&
\frac{1}{4m^2}e^2 \,\sigma^{ij}{E}_\perp^i A^j &&
   (H-E)^0&E_{15}\\[8pt]
  \frac{1}{4m^2}e^2 \,\sigma^{ij}E_\perp^i A^j &
  -\frac{e}{m}\Bigl(\vec{p}\cdot\vec{A}+ \frac{1}{4}\sigma\cdot B\Bigr)&
  -\frac{e}{m}\Bigl(\vec{p}\cdot\vec{A}+ \frac{1}{4}\sigma\cdot B\Bigr)&
  (H-E)^0&E_{16}\\[8pt]
	 \frac{e^2}{2m}\vec{A}\,{}^2&
-\frac{e^2}{8m^3}A^i \,\nabla^j B^{ij}  &&
  (H-E)^0&E_{17}\\[8pt]
\end{tabular}
\end{ruledtabular}
\end{table}

\begin{figure}[H]
\centering
\includegraphics[height=\textheight]{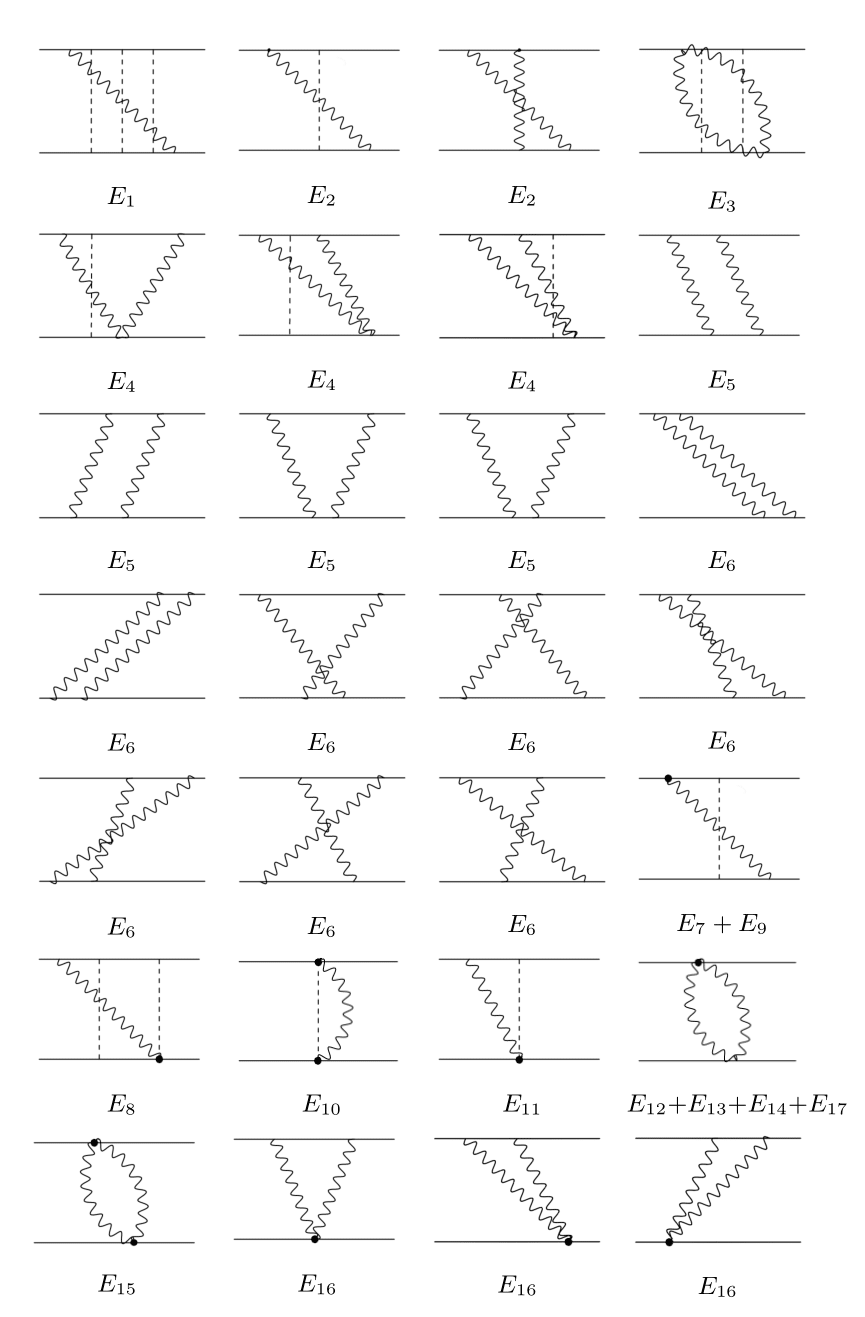}
\caption{
  Individual time-ordered diagrams contributing to the middle-energy contribution.
  We note that $E_{10}$ and $E_{15}$ were calculated from Feynman diagrams.
  The large dot denotes a correction to the vertex according to $\delta_{\rm ST}H_{\rm FW}$ and
  $\delta_{\rm DT}H_{\rm FW}$, as explicitly shown in Table I.
\label{fig:Ei}}
\end{figure}

\subsection{Evaluation of individual diagrams}

\subsubsection{Triple retardation in nonrelativistic single transverse photon exchange $E_1$}
The correction due to the single transverse photon exchange reads
\begin{eqnarray}
\delta E = e^2 \int \frac{d^dk}{(2\pi)^d\,2k}\delta_\perp^{ij}(k)\langle\phi|
j_1^i (k)e^{i\vec{k}\cdot\vec{r}_1}\, \frac{1}{E_0-H_0-k}\,j_2^j(-k) e^{-i\vec{k}\cdot\vec{r}_2}|\phi\rangle + (1\leftrightarrow2)\,.
\end{eqnarray}
Expanding the denominator of the above expression in $k$ and taking only the term contributing to
the order $m\alpha^7$, we obtain
\begin{align}
E_1 =&\  e^2 \int \frac{d^dk}{(2\pi)^d\,2k^5}\delta_\perp^{ij}(k)
\langle\phi|\biggl(\frac{p_1^i}{m}+\frac{1}{2m}\sigma_1^{ki}\nabla_1^k\biggr)e^{i\vec{k}\cdot\vec{r}_1} (H_0-E_0)^3
\biggl(\frac{p_2^j}{m}+\frac{1}{2m}\sigma_2^{lj}\nabla_2^l\biggr)e^{-i\vec{k}\cdot\vec{r}_2}|\phi\rangle + (1\leftrightarrow2)
\nonumber\\ =&\ E_1^{ss} + E_1^{nn}\,,
\end{align}
where
\begin{eqnarray}
  E_1^{ss} &=&  \frac{e^2}{2\,m^2} \int \frac{d^dk}{(2\pi)^d\,k^5}\delta_\perp^{ij}(k)
  \langle\phi|\biggl[\biggl[\frac{i}{2m}\sigma_1^{ki}k^k\,e^{i\vec{k}\cdot\vec{r}_1},H_0-E_0\biggr],\biggl[
H_0-E_0,
\biggl[H_0-E_0,\frac{(-i)}{2m}\sigma_2^{lj}k^l\,e^{-i\vec{k}\cdot\vec{r}_2}\biggr]\biggr]\biggr]|\phi\rangle \nonumber\\
&=&  \frac{e^2}{8\,m^4}\,\frac{\sigma_1\cdot\sigma_2}{d}
\int \frac{d^dk}{(2\pi)^d\,k^3}\,k^m k^n
\langle\phi|\,e^{i\vec{k}\cdot\vec{r}}\,\biggl[p_1^n,\biggl[\biggl[\frac{\alpha}{r}\biggr]_\epsilon,p_2^m\biggr]\biggr]|\phi\rangle\,,\\
E_1^{nn} &=&  \frac{e^2}{2\,m^2} \int \frac{d^dk}{(2\pi)^d k^5}\delta_\perp^{ij}(k)
\langle\phi|\bigl[\bigl[\,p_1^i\,e^{i\vec{k}\cdot\vec{r}_1},H_0-E_0\bigr],\bigl[
H_0-E_0,\bigl[H_0-E_0,
  p_2^j \,e^{-i\vec{k}\cdot\vec{r}_2}\bigr]\bigr]\bigr]|\phi\rangle\nonumber \\ &=&
\frac{e^2}{2\,m^2} \int \frac{d^dk}{(2\pi)^d k^5}\delta_\perp^{ij}(k)
\Bigl\langle\phi\Bigl|\Bigl[\Bigl[\,p_1^i\,e^{i\vec{k}\cdot\vec{r}_1},\frac{p_1^2}{2\,m}+V\Bigr],\Bigl[
\frac{p_1^2}{2\,m}+\frac{p_2^2}{2\,m}+V,\Bigl[\frac{p_2^2}{2\,m}+V,p_2^j \,e^{-i\vec{k}\cdot\vec{r}_2}\Bigr]\Bigr]\Bigr]\Bigr|\phi\Bigr\rangle\,.
\end{eqnarray}
The spin part is easy to evaluate and for this we use integrals from Appendix C with the result
\begin{align}
  H_1^{ss} = \frac{\alpha^2}{m^4}\,\sigma_1\cdot\sigma_2\,\Big(-\frac13-\frac{1}{9\epsilon}+\frac29\ln2+\frac29\ln q\Big)\,q^2.
\end{align}
For the spin-independent part $H_1^{nn}$ the evaluation is quite lengthy and thus is moved to
Appendix~\ref{app:F}. The result for the sum of both parts is
\begin{eqnarray}
H_1 &=&
\frac{\alpha^2}{m^4}\bigg\{\sigma_1\cdot\sigma_2\,\Big(-\frac13-\frac{1}{9\epsilon}+\frac29\ln2+\frac29\ln q\Big)\,q^2
+\frac{4}{45}\big(\vec P_1-\vec P_2\big)^2 - \frac{26}{15}\frac{\big[\big(\vec P_1-\vec P_2\big)\cdot\vec q\big]^2}{q^2}
+\frac{28}{45}q^2 + \frac49\vec P_1\cdot\vec P_2\nonumber\\
&&-\frac{16}{3}\frac{\big(\vec P_1\cdot\vec q\big)\big(\vec P_2\cdot\vec q\big)}{q^2}
+\bigg(\frac{8}{15}\big(\vec P_1-\vec P_2\big)^2+\frac43\frac{\big[\big(\vec P_1-\vec P_2\big)\cdot\vec q\big]^2}{q^2}
-\frac{16}{15}q^2+\frac83\frac{\big(\vec P_1\cdot\vec q\big)\big(\vec P_2\cdot\vec q\big)}{q^2}\bigg) \bigg(-\frac{1}{2\epsilon}+\ln2+\ln q\bigg)\bigg\}\nonumber\\
&&+
\frac{\pi \alpha^3}{m^3}\bigg(-\frac{2}{5}+\frac{1}{15\epsilon}-\frac{4}{15}\ln q\bigg)\,q
+
\frac{\alpha^3}{m^3}\bigg\{\bigg[p_2^j,-\bigg[\frac{Z}{r_2}\bigg]_\epsilon\bigg]
\,\bigg(-\frac{2}{15}+\frac{14}{15\epsilon}-\frac{28}{15}\ln2-\frac{28}{15}\ln q\bigg)\frac{q^j}{q^2}\nonumber\\
&&+\bigg[\bigg[p_2^i,-\bigg[\frac{Z}{r_2}\bigg]_\epsilon\bigg],p_2^j\bigg]\bigg[\delta^{ij}\bigg(-\frac{4}{5}-\frac{4}{15\epsilon}+\frac{8}{15}\ln2+\frac{8}{15}\ln q\bigg)
+\frac{q^i q^j}{q^2}\bigg(\frac{16}{15}+\frac{16}{15\epsilon}-\frac{32}{15}\ln2-\frac{32}{15}\ln q\bigg)\bigg]\frac{1}{q^2}\nonumber\\
&&+(1\leftrightarrow2)
\bigg\}\,.
\end{eqnarray}

\subsubsection{Single transverse photon exchange with Breit correction $E_2$}
This contribution comes from the perturbation of the nonrelativistic Hamiltonian $H_0$, energy
$E_0$ and wave function $\phi$ by the Breit Hamiltonian $H^{(4)}$ in the single transverse photon
exchange. We thus have
\begin{align}
E_2 =&\  e^2 \int \frac{d^dk}{(2\pi)^d\,2k^3}\delta_\perp^{ij}(k)\,
\delta\langle\phi|\biggl(\frac{p_1^i}{m}+\frac{1}{2m}\sigma_1^{ki}\nabla_1^k\biggr)e^{i\vec{k}\cdot\vec{r}_1} (H-E)
\biggl(\frac{p_2^j}{m}+\frac{1}{2m}\sigma_2^{lj}\nabla_2^l\biggr)e^{-i\vec{k}\cdot\vec{r}_2}|\phi\rangle + (1\leftrightarrow2)\nonumber\\
=&\  e^2 \int \frac{d^dk}{(2\pi)^d\,2k^3}\delta_\perp^{ij}(k)\,
\delta\langle\phi|\biggl[\biggl(\frac{p_1^i}{m}+\frac{1}{2m}\sigma_1^{ki}\nabla_1^k\biggr)e^{i\vec{k}\cdot\vec{r}_1},
\biggl[H-E,
\biggl(\frac{p_2^j}{m}+\frac{1}{2m}\sigma_2^{lj}\nabla_2^l\biggr)e^{-i\vec{k}\cdot\vec{r}_2}\biggr]\biggr]|\phi\rangle\nonumber\\
=&\  e^2 \int \frac{d^dk}{(2\pi)^d\,2k^3}\delta_\perp^{ij}(k) \biggl\{
\langle\phi|\biggl[\biggl(\frac{p_1^i}{m}+\frac{1}{2m}\sigma_1^{ki}\nabla_1^k\biggr)e^{i\vec{k}\cdot\vec{r}_1},
\biggl[H^{(4)}-E^{(4)},
\biggl(\frac{p_2^j}{m}+\frac{1}{2m}\sigma_2^{lj}\nabla_2^l\biggr)e^{-i\vec{k}\cdot\vec{r}_2}\biggr]\biggr]|\phi\rangle\nonumber\\
&+\,
\langle\phi|H^{(4)}\,\frac{1}{(E_0-H_0)'}\biggl[\biggl(\frac{p_1^i}{m}+\frac{1}{2m}\sigma_1^{ki}\nabla_1^k\biggr)e^{i\vec{k}\cdot\vec{r}_1},
\biggl[H_0-E_0,
\biggl(\frac{p_2^j}{m}+\frac{1}{2m}\sigma_2^{lj}\nabla_2^l\biggr)e^{-i\vec{k}\cdot\vec{r}_2}\biggr]\biggr]|\phi\rangle\nonumber\\
&+\,
\langle\phi|\biggl[\biggl(\frac{p_1^i}{m}+\frac{1}{2m}\sigma_1^{ki}\nabla_1^k\biggr)e^{i\vec{k}\cdot\vec{r}_1},
\biggl[H_0-E_0,
\biggl(\frac{p_2^j}{m}+\frac{1}{2m}\sigma_2^{lj}\nabla_2^l\biggr)e^{-i\vec{k}\cdot\vec{r}_2}\biggr]\biggr]\,\frac{1}{(E_0-H_0)'}\,H^{(4)}|\phi\rangle\biggr\}\,.\label{Eq3}
\end{align}
The last two lines are a part of the second-order contribution, the third term of Eq. (\ref{eq:1})
and thus will be omitted here, so
\begin{eqnarray}
E_2 &=&
e^2\int\frac{d^dk}{(2\pi)^d\,2k^3} \delta_\perp^{ij}(k)
\langle\phi|\biggl[\biggl(\frac{p_1^i}{m}+\frac{1}{2m}\sigma_1^{ki}\nabla_1^k\biggr)\,e^{i\vec{k}\cdot\vec{r}_1},\biggl[H^{(4)}-E^{(4)},
    \biggl(\frac{p_2^j}{m}+\frac{1}{2m}\sigma_2^{lj}\nabla_2^l\biggr)\,e^{-i\vec{k}\cdot\vec{r}_2}\biggr]\biggr]|\phi\rangle \label{33} \\
 &=& E'_2 + E''_2\,,
\end{eqnarray}
where $E'_2$ is a correction due to $H'^{(4)}$ and $E''_2$ is due to $H''^{(4)}$. $E''_2$ is the
double transverse photon exchange, because $H''^{(4)}$ comes from the transverse photon exchange.
Therefore, this double transverse photon exchange will later be excluded from $E_5+E''_6$ in
Eq.~(\ref{excl}), to avoid double counting. We transform $E'_2$ and $E''_2$ as
\begin{align}
E'_2 =&\
e^2\int\frac{d^dk}{(2\pi)^d\,2k^3} \delta_\perp^{ij}(k)\frac{1}{m^4}
\biggl\langle\phi\biggl|-\frac{\sigma_1^{ki}\,\sigma_1^{mn}}{8}\,k^k\,k^n\,e^{i\vec{k}\cdot\vec{r}}
\,\nabla^j\,\nabla^m\biggl[\frac{\alpha}{r}\biggr]_\epsilon
-\frac{\sigma_2^{mn}\,\sigma_2^{lj}}{8}\,k^l\,k^n\,e^{i\vec{k}\cdot\vec{r}}
\,\nabla^i\,\nabla^m\biggl[\frac{\alpha}{r}\biggr]_\epsilon\nonumber\\
&\ +\bigl[p_1^i\,e^{i\vec{k}\cdot\vec{r}_1},\bigl[ -\pi\,\alpha\,\delta^d(r)\,,\,
p_2^j\,e^{-i\vec{k}\cdot\vec{r}_2}\bigr]\bigr]	
\biggr|\phi\biggr\rangle\,,\\
E''_2 =&\
  -e^4 \int \frac{d^dk_1}{(2\pi)^d\,2k_1^3}\int\frac{d^dk_2}{(2\pi)^d\,k_2^2}\delta_\perp^{ij}(k_1)
\delta_\perp^{mn}(k_2)
\langle\phi|\bigl[j_1^i\,e^{i\vec{k_1}\cdot\vec{r}_1},j_1^{m}\,e^{i\vec{k_2}\cdot\vec{r}_{1}}\bigr]
\bigl[j_2^n\,e^{-i\vec{k_2}\cdot\vec{r}_2},j_2^j\,e^{-i\vec{k_1}\cdot\vec{r}_2}\bigr]|\phi\rangle
\,.
\end{align}
We now evaluate the $E_2'$ part. The term with the delta function vanishes because, due to the delta
function, the exponent function disappears, $e^{i\vec{k}\cdot\vec{r}}=1$, and in the
dimensional regularization, by definition,
\begin{equation}
\int d^d k \, k^\alpha =0\,.
\end{equation}
We thus have
\begin{align}
E_2' =&\
-\frac{e^2}{m^4}\int\frac{d^dk}{(2\pi)^d\,2k^3} \delta_\perp^{ij}(k)
\biggl\langle\phi\biggl|
\frac{\sigma_1^{ki}\,\sigma_1^{mn}}{8}\,k^k\,k^n\,e^{i\vec{k}\cdot\vec{r}}
\,\nabla^j\,\nabla^m\biggl[\frac{\alpha}{r}\biggr]_\epsilon
+\frac{\sigma_2^{mn}\,\sigma_2^{lj}}{8}\,k^l\,k^n\,e^{i\vec{k}\cdot\vec{r}}
\,\nabla^i\,\nabla^m\biggl[\frac{\alpha}{r}\biggr]_\epsilon
\biggr|\phi\biggr\rangle \nonumber \\ =&\
\frac{e^2}{4m^4}\int\frac{d^dk}{(2\pi)^d\,2k} \delta_\perp^{ij}(k)
\biggl\langle\phi\biggl|
\,e^{i\vec{k}\cdot\vec{r}}\,\nabla^j\,\nabla^i\biggl[\frac{\alpha}{r}\biggr]_\epsilon
\biggr|\phi\biggr\rangle\,.
\end{align}
The corresponding operator in the momentum representation is
\begin{eqnarray}
H_2' = \frac{\alpha^2}{m^4}
\biggl(-\frac79-\frac{1}{3\epsilon}+\frac23\ln2+\frac23\ln q^2\biggr)q^2\,.
\end{eqnarray}
Now we turn to contributions due to $E_2''$ and use Eq. (\ref{37}),
\begin{eqnarray}
E_{2}''
&=&  e^4 \int \frac{d^dk_1}{(2\pi)^d\,2k_1^3}\int\frac{d^dk_2}{(2\pi)^d\,k_2^2}\delta_\perp^{ij}(k_1)
\delta_\perp^{mn}(k_2)
\langle\phi|j_n^{im}j_n^{jn} + j_s^{im,ab}j_s^{jn,kl}\,\sigma_1^{ab}\sigma_2^{kl}|\phi\rangle\,.
\end{eqnarray}
This can be spin averaged with the help of Eq. (\ref{B6}). The corresponding operator in the momentum
representation is
\begin{align}
H_2''
=\frac{\alpha^2}{m^4}\bigg(
-\frac83\vec P_1\cdot\vec P_2+\frac83\frac{\big(\vec P_1\cdot\vec q\big)\big(\vec P_2\cdot\vec q\big)}{q^2}\bigg)
\bigg(\frac12-\frac{1}{2\epsilon}+\ln2+\ln q\biggr)
+ \frac{\alpha^2}{m^4}\sigma_1\cdot\sigma_2
\biggl(-\frac{10}{9}-\frac{5}{18\epsilon}+\frac59\ln2+\frac59\ln q\biggr)\,q\,,
\end{align}
and the result for $H_2=H_2'+H_2''$ is then
\begin{eqnarray}
H_2 &=&\frac{\alpha^2}{m^4}\bigg\{\sigma_1\cdot\sigma_2
\biggl(-\frac{10}{9}-\frac{5}{18\epsilon}+\frac59\ln2+\frac59\ln q\biggr)\,q^2
+\biggl(-\frac79-\frac{1}{3\epsilon}+\frac23\ln2+\frac23\ln q^2\biggr)q^2\bigg\}\nonumber\\
&&+\frac{\alpha^2}{m^4}\bigg(
-\frac83\vec P_1\cdot\vec P_2+\frac83\frac{\big(\vec P_1\cdot\vec q\big)\big(\vec P_2\cdot\vec q\big)}{q^2}\bigg)
\bigg(\frac12-\frac{1}{2\epsilon}+\ln2+\ln q\biggr)
\,.
\end{eqnarray}

\subsubsection{Retardation in the double seagull $E_3$}
In the case of the double seagull diagram we have to take the double retardation to obtain
correction of the order $m\alpha^7$,
\begin{eqnarray}
E_3 &=& \biggl(\frac{e^2}{2m}\biggr)^2\,2 \int\frac{d^dk_1}{(2\pi)^d\,2k_1} \int\frac{d^dk_2}{(2\pi)^d\,2k_2}
 \delta_\perp^{ij}(k_1)\delta_\perp^{ij}(k_2)
\langle\phi|e^{i(\vec{k}_1+\vec{k}_2)\cdot\vec{r}_1}\,\frac{1}{E_0-H_0-k_1-k_2}\,e^{-i(\vec{k}_1+\vec{k}_2)\cdot\vec{r}_2}|\phi\rangle
+(1\leftrightarrow2)\nonumber\\
&\approx& -\biggl(\frac{e^2}{2m}\biggr)^2\,2 \int\frac{d^dk_1}{(2\pi)^d\,2k_1} \int\frac{d^dk_2}{(2\pi)^d\,2k_2}
 \delta_\perp^{ij}(k_1)\delta_\perp^{ij}(k_2)
\langle\phi|e^{i(\vec{k}_1+\vec{k}_2)\cdot\vec{r}_1}\,\frac{(H_0-E_0)^2}{(k_1+k_2)^3}\,e^{-i(\vec{k}_1+\vec{k}_2)\cdot\vec{r}_2}|\phi\rangle
+(1\leftrightarrow2)\nonumber\\
&=&
-\Big(\frac{e^2}{2m}\Big)^2\int\frac{d^d k_1}{(2\pi)^d\,k_1}\int\frac{d^d k_2}{(2\pi)^d\,k_2}
\frac{1}{(k_1+k_2)^3}\delta_\perp^{ij}(k_1)\delta_\perp^{ij}(k_2)
\bigg\langle\phi\bigg|\bigg[\frac{p_2^2}{2m},\bigg[e^{i(\vec{k}_1+\vec{k}_2)\cdot\vec{r}},\frac{p_1^2}{2m}\bigg]\bigg]\bigg|\phi\bigg\rangle\,.
\end{eqnarray}
The result in momentum representation is
\begin{eqnarray}
H_3
&=&\frac{\alpha^2}{m^4}\,\frac{(-9+8\ln2)}{5}\frac{(\vec P_1\cdot\vec q)(\vec P_2\cdot\vec q)}{q^2}\,.
\end{eqnarray}

\subsubsection{Retardation in the single seagull $E_4$}
For the single seagull diagram we have to take a single retardation correction in order to obtain
the contribution of the order $m \alpha^7$. Using the definition of $j_1^i(k)$, we write $E_4$ as
\begin{eqnarray}
E_4 &=&
\frac{e^4}{m}\, \int \frac{d^dk_1}{(2\pi)^d\,2k_1} \int \frac{d^dk_2}{(2\pi)^d\,2k_2}
\delta_\perp^{in}(k_1)\delta_\perp^{im}(k_2)\nonumber\\
&&\times
\biggl\{\langle\phi|j_1^n(k_1)\,e^{i\vec{k}_1\cdot\vec{r}_1}\,\frac{1}{E_0-H_0-k_1}\,e^{-i(\vec{k}_1+\vec{k}_2)\cdot\vec{r}_2}\,
\frac{1}{E_0-H_0-k_2}\,j_1^m(k_2)\,e^{i\vec{k}_2\cdot\vec{r}_1}|\phi\rangle\nonumber\\
&&+\,
\langle\phi|j_1^n(k_1)\,e^{i\vec{k}_1\cdot\vec{r}_1}\,\frac{1}{E_0-H_0-k_1}\,j_1^m(k_2)\,e^{i\vec{k}_2\cdot\vec{r}_1}\,
\frac{1}{E_0-H_0-k_1-k_2}\,e^{-i(\vec{k}_1+\vec{k}_2)\cdot\vec{r}_2}|\phi\rangle\nonumber\\
&&+\,
\langle\phi|e^{-i(\vec{k}_1+\vec{k}_2)\cdot\vec{r}_2}\,\frac{1}{E_0-H_0-k_1-k_2}\,j_1^n(k_1)\,e^{i\vec{k}_1\cdot\vec{r}_1}\,
\frac{1}{E_0-H_0-k_2}\,j_1^m(k_2)\,e^{i\vec{k}_2\cdot\vec{r}_1}|\phi\rangle\biggr\} + (1\leftrightarrow2)\,.
\end{eqnarray}
After expanding in $H_0-E_0$ and further simplifications of the expression, it can be transformed
to
\begin{eqnarray}
E_4
&=&
-\frac{e^4}{m}\, \int \frac{d^dk_1}{(2\pi)^d\,2k_1} \int \frac{d^dk_2}{(2\pi)^d\,2k_2}
\delta_\perp^{in}(k_1)\delta_\perp^{im}(k_2)\nonumber\\
&& \times
\biggl\{
\frac{1}{k_1^2}\,\Bigl(\frac{1}{k_2} + \frac{1}{k_1+k_2}\Bigr)\langle\phi|
\bigl[\bigl[j_1^n(k_1)\,e^{i\vec{k}_1\cdot\vec{r}_1}\,,\, H_0-E_0\bigr]\,,\,
j_1^m(k_2)\,e^{i\vec{k}_2\cdot\vec{r}_1}\bigr]\,e^{-i(\vec{k}_1+\vec{k}_2)\cdot\vec{r}_2}
|\phi\rangle\nonumber\\
&&+\frac{1}{k_1\,(k_1+k_2)^2}\,\langle\phi|\bigl[j_1^n(k_1)\,e^{i\vec{k}_1\cdot\vec{r}_1}\,,\,j_1^m(k_2)\,e^{i\vec{k}_2\cdot\vec{r}_1}\bigr]\,
\bigl[H_0-E_0\,,\,e^{-i(\vec{k}_1+\vec{k}_2)\cdot\vec{r}_2}\bigr]|\phi\rangle \biggr\} + (1\leftrightarrow2)\,.
\end{eqnarray}
Using the  the Jacobi identity
\begin{align}
  [[A_1\,,\,B]\,,\,A_2]  - [[A_2\,,\,B]\,,\,A_1] =   [[A_1\,,\,A_2]\,,\,B]\,,
\end{align}
we obtain
\begin{align}
E_4 =&\ \label{E4}
-\frac{e^4}{2\,m}\, \int \frac{d^dk_1}{(2\pi)^d\,2k_1} \int \frac{d^dk_2}{(2\pi)^d\,2k_2}
\delta_\perp^{in}(k_1)\delta_\perp^{im}(k_2)\nonumber\\
& \times
\biggl\{
\Bigl(\frac{1}{k_1^2\,k_2} + \frac{1}{k_1\,k_2^2} +
\frac{1}{k_1^2}\,\frac{1}{k_1+k_2} + \frac{1}{k_2^2}\,\frac{1}{k_1+k_2}\Bigr)\langle\phi|
\bigl[\bigl[j_1^n(k_1)\,e^{i\vec{k}_1\cdot\vec{r}_1}\,,\, H_0-E_0\bigr]\,,\,
j_1^m(k_2)\,e^{i\vec{k}_2\cdot\vec{r}_1}\bigr]\,e^{-i(\vec{k}_1+\vec{k}_2)\cdot\vec{r}_2}
|\phi\rangle\nonumber\\
&\ +\frac{1}{k_1^2}\,\Bigl(\frac{1}{k_2} + \frac{1}{k_1+k_2}+\frac{2\,k_1}{(k_1+k_2)^2}\Bigr)\,
\langle\phi|\bigl[j_1^n(k_1)\,e^{i\vec{k}_1\cdot\vec{r}_1}\,,\,j_1^m(k_2)\,e^{i\vec{k}_2\cdot\vec{r}_1}\bigr]\,
\bigl[H_0-E_0\,,\,e^{-i(\vec{k}_1+\vec{k}_2)\cdot\vec{r}_2}\bigr]|\phi\rangle
\biggr\}  + (1\leftrightarrow2) \nonumber\\ =&\ E_{4,1}+E_{4,2}+E_{4,3}+E_{4,4}\,.
\end{align}
Here, $E_{4,4}$ corresponds to the second term in the curly brackets, while $E_{4,1}$, $E_{4,2}$,
and $E_{4,3}$ are parts of the first term in the curly brackets. Specifically, $E_{4,1}$
corresponds to the spin-dependent part, and $E_{4,2}$ and $E_{4,3}$ are coming from the two- and
three-photon contributions in the spin-independent part, correspondingly.

We start with the first term in the curly brackets in Eq.~(\ref{E4}),
\begin{eqnarray}
  E_{4,1}+E_{4,2}+E_{4,3} &= &-\frac{e^4}{2\,m^3} \int \frac{d^dk_1}{(2\pi)^d\,2\,k_1} \int \frac{d^dk_2}{(2\pi)^d\,2\,k_2}
\Bigl(\frac{1}{k_1^2\,k_2} + \frac{1}{k_1\,k_2^2} + \frac{1}{k_1^2}\,\frac{1}{k_1+k_2} + \frac{1}{k_2^2}\,\frac{1}{k_1+k_2}\Bigr)
\,\delta^{in}_{\perp}(k_1)\,\delta^{im}_{\perp}(k_2)\nonumber\\ && \hspace{-7ex}
\times
\langle\phi|\,\biggl[\,e^{-i(\vec{k}_1+\vec{k}_2)\cdot\vec{r}_2}\,\biggl(p_1^m+\frac{i}{2}\sigma_1^{rm}k_2^r\biggr)
\,e^{i\vec{k}_2\cdot\vec{r}_1},
\biggl[\,\frac{p_1^2}{2m}+V,\,\biggl(p_1^n+\frac{i}{2}\sigma_1^{sn}k_1^s\biggr)\,e^{i\vec{k}_1\cdot\vec{r}_1}\biggr]\biggr]\,|\phi\rangle+
(1\leftrightarrow2)\,.
\end{eqnarray}
The spin-dependent part $E_{4,1}$ is
\begin{eqnarray}
E_{4,1}
&=&
\frac{e^4}{16m^3} \int \frac{d^dk_1}{(2\pi)^d} \int \frac{d^dk_2}{(2\pi)^d}
\frac{1}{k_1^3k_2}\,\Bigl(\frac{1}{k_2} + \frac{1}{k_1+k_2}\Bigr)
\,\delta^{in}_{\perp}(k_1)\,\delta^{im}_{\perp}(k_2)\,\sigma_1^{rm}\sigma_1^{sn}k_2^rk_1^s\nonumber\\
&& \times
\langle\phi|\,e^{-i(\vec{k}_1+\vec{k}_2)\cdot\vec{r}_2}\biggl[e^{i\vec{k}_2\cdot\vec{r}_1},
\biggl[\,\frac{p_1^2}{2m},e^{i\vec{k}_1\cdot\vec{r}_1}\biggr]\biggr]\,|\phi\rangle+ (1\leftrightarrow2)\nonumber\\
&=&
-\frac{e^4}{16m^4} \int \frac{d^dk_1}{(2\pi)^d} \int \frac{d^dk_2}{(2\pi)^d}
\frac{1}{k_1^3k_2}\,\Bigl(\frac{1}{k_2} + \frac{1}{k_1+k_2}\Bigr)
\,\delta^{in}_{\perp}(k_1)\,\delta^{im}_{\perp}(k_2)
\big(\delta^{mn}\big(\vec{k}_1\cdot\vec{k}_2\big)^2
-k_1^mk_2^n\,\vec{k}_1\cdot\vec{k}_2\big)
\langle\phi|\,e^{i(\vec{k}_1+\vec{k}_2)\cdot\vec{r}}|\phi\rangle\nonumber\\
&&+ (1\leftrightarrow2)\,,
\end{eqnarray}
where we performed spin averaging. The result in momentum representation is
\begin{equation}
H_{4,1} = \frac{\alpha^2}{m^4}\,\Big(-\frac{5}{72}+\frac{23}{24\epsilon}-\frac{23}{12}\ln q\Big)\,q^2\,.
\end{equation}

The spin-independent two-photon exchange part is
\begin{align}
E_{4,2}
=&\
-\frac{e^4}{8\,m^3} \int \frac{d^dk_1}{(2\pi)^d} \int \frac{d^dk_2}{(2\pi)^d}\,\frac{1}{k_1\,k_2}
\Bigl(\frac{1}{k_1^2\,k_2} + \frac{1}{k_1\,k_2^2} + \frac{1}{k_1^2}\,\frac{1}{k_1+k_2} + \frac{1}{k_2^2}\,\frac{1}{k_1+k_2}\Bigr)
\,\delta^{in}_{\perp}(k_1)\,\delta^{im}_{\perp}(k_2)\nonumber\\
& \times
\langle\phi|\,\biggl[\,e^{-i(\vec{k}_1+\vec{k}_2)\cdot\vec{r}_2}\,p_1^m
\,e^{i\vec{k}_2\cdot\vec{r}_1},
\biggl[\,\frac{p_1^2}{2m},\,p_1^n\,e^{i\vec{k}_1\cdot\vec{r}_1}\biggr]\biggr]\,|\phi\rangle+ (1\leftrightarrow2)\nonumber\\
=&\
-\frac{e^4}{8\,m^4} \int \frac{d^dk_1}{(2\pi)^d} \int \frac{d^dk_2}{(2\pi)^d}
\frac{1}{k_1^3k_2}\,\Bigl(\frac{1}{k_2} + \frac{1}{k_1+k_2}\Bigr)
\langle\phi|\,p_1^m\biggl(
-2\,\vec k_1\cdot\vec k_2\,\delta_{\perp}^{in}(k_1)\,\delta_{\perp}^{im}(k_2)
\nonumber\\ &
-(k_1^n-k_2^n)\, k_2^k\,\delta_{\perp}^{ik}(k_1) \delta_{\perp}^{im}(k_2)
+(k_1^m-k_2^m)\, k_1^k\,\delta_{\perp}^{in}(k_1)\, \delta_{\perp}^{ik}(k_2)
\biggr) e^{i(\vec{k}_1+\vec{k}_2)\cdot\vec{r}}\,p_1^n|\phi\rangle
+(1\leftrightarrow2)\,.
\end{align}
The result for this term is
\begin{eqnarray}
H_{4,2} &=&
\frac{\alpha^2}{m^4}\bigg[
\frac{32}{45}\big(\vec P_1-\vec P_2\big)^2-\frac{83}{90}q^2+\frac{17}{15}\frac{\big[\big(\vec P_1-\vec P_2\big)\cdot\vec q\big]^2}{q^2}
+\frac{64}{45}\vec P_1\cdot\vec P_2+\frac{34}{15}\frac{\big(\vec P_1\cdot\vec q\big)\big(\vec P_2\cdot\vec q\big)}{q^2}\nonumber\\
&&+\biggl(
-\frac{8}{15}\big(\vec P_1-\vec P_2\big)^2+\frac{16}{15}q^2
-\frac{8}{5}\frac{\big[\big(\vec P_1-\vec P_2\big)\cdot\vec q\big]^2}{q^2}
-\frac{16}{15}\vec P_1\cdot\vec P_2-\frac{16}{5}\frac{\big(\vec P_1\cdot\vec q\big)\big(\vec P_2\cdot\vec q\big)}{q^2}\bigg)\ln2\nonumber\\
&&+\big(2\big(\vec P_1-\vec P_2\big)^2-q^2+4\vec P_1\cdot\vec P_2\big)\bigg(-\frac{1}{2\epsilon}+\ln q\bigg)
\bigg]\,.
\end{eqnarray}

The spin-dependent three-photon part $E_{4,3}$ is
\begin{eqnarray}
E_{4,3}
&=&
-\frac{e^4}{4m^3} \int \frac{d^dk_1}{(2\pi)^d} \int \frac{d^dk_2}{(2\pi)^d}
\frac{1}{k_1^3k_2}\,\Bigl(\frac{1}{k_2} + \frac{1}{k_1+k_2}\Bigr)
\delta_{\perp}^{in}(k_1)\delta_{\perp}^{im}(k_2)
\langle\phi|\,\bigl[\,p_1^m,
\bigl[V,\,p_1^n\bigr]\bigr]\,\,e^{i(\vec{k}_1+\vec{k}_2)\cdot\vec{r}}|\phi\rangle+ (1\leftrightarrow2)\,.\nonumber\\
\end{eqnarray}
This expression contains two-body and three-body terms. Evaluating it we obtain
\begin{eqnarray}
H_{4,3} &=&
\frac{\alpha^3}{m^3}
\bigg\{\biggl[\,p_1^m,
\biggl[-\bigg[\frac{Z}{r_1}\bigg]_\epsilon,\,p_1^n\biggr]\biggr]
\bigg[\delta^{mn}\bigg(\frac{4}{15}+\frac{4}{3\epsilon}-\frac{8}{15}\ln2-\frac{8}{3}\ln q\bigg)
+\frac{q^m q^n}{q^2}\bigg(-\frac{12}{5}-\frac{4}{3\epsilon}+\frac{32}{15}\ln2+\frac{8}{3}\ln q\bigg)\bigg]\frac{1}{q^2}	
\nonumber\\ &&+ (1\leftrightarrow2)\bigg\}+
\pi\frac{\alpha^3}{m^3}\bigg(-\frac{8}{15}-\frac{2}{3\epsilon}-\frac{4}{15}\ln2+\frac83\ln q\bigg)\,q\,.
\end{eqnarray}

The last part $E_{4,4}$ is evaluated as
\begin{align}
E_{4,4} =&\
-\frac{e^4}{2\,m}\, \int \frac{d^dk_1}{(2\pi)^d\,2k_1} \int \frac{d^dk_2}{(2\pi)^d\,2k_2}\,
\frac{1}{k_1^2}\,\Bigl(\frac{1}{k_2} + \frac{1}{k_1+k_2}+\frac{2\,k_1}{(k_1+k_2)^2}\Bigr)
\delta_\perp^{ia}(k_1)\delta_\perp^{ib}(k_2)\nonumber\\
&\ \times
\langle\phi|\bigl[j_1^a(k_1)\,e^{i\vec{k}_1\cdot\vec{r}_1}\,,\,j_1^b(k_2)\,e^{i\vec{k}_2\cdot\vec{r}_1}\bigr]\,
\bigl[H_0-E_0\,,\,e^{-i(\vec{k}_1+\vec{k}_2)\cdot\vec{r}_2}\bigr]|\phi\rangle
+ (1\leftrightarrow2)\\ =&\
-\frac{e^4}{4\,m}\, \int \frac{d^dk_1}{(2\pi)^d\,2k_1} \int \frac{d^dk_2}{(2\pi)^d\,2k_2}\,
\frac{1}{k_1^2}\,\Bigl(\frac{1}{k_2} + \frac{1}{k_1+k_2}+\frac{2\,k_1}{(k_1+k_2)^2}\Bigr)
\nonumber\\ &\times
\langle\phi| \bigl[p_2^2\,,\, \delta_{\perp}^{ia}(k_1)\,\delta_{\perp}^{ib}(k_2)\,
\big(-p_1^a\,k_1^b+p_1^b\, k_2^a\big) \,e^{i(\vec{k}_1+\vec{k}_2)\cdot\vec{r}}\bigr]|\phi\rangle
+ (1\leftrightarrow2)
\\ =&\
-\frac{e^4}{4\,m}\, \int \frac{d^dk_1}{(2\pi)^d\,2k_1} \int \frac{d^dk_2}{(2\pi)^d\,2k_2}\,
\frac{1}{k_1^2}\,\Bigl(\frac{1}{k_2} + \frac{1}{k_1+k_2}+\frac{2\,k_1}{(k_1+k_2)^2}\Bigr)
\nonumber\\ &\times
\langle\phi| \bigl[p_2^2\,,\, p_1^a\,\Big(
\delta_{\perp}^{ib}(k_1)\,\delta_{\perp}^{ia}(k_2)\,k_2^b
-\delta_{\perp}^{ia}(k_1)\,\delta_{\perp}^{ib}(k_2)\,k_1^b\Big)
  \,e^{i(\vec{k}_1+\vec{k}_2)\cdot\vec{r}}\bigr]|\phi\rangle
+ (1\leftrightarrow2)
\nonumber \\ =&\ 0\,.
\end{align}
This expression vanishes because the two terms in curly brackets in the matrix element cancel each other.
The total result for $H_4=H_{4,1}+H_{4,2}+H_{4,3}$ is
\begin{eqnarray}
H_4 &=&
\frac{\alpha^2}{m^4}\bigg[
\frac{32}{45}\big(\vec P_1-\vec P_2\big)^2+\frac{17}{15}\frac{\big[\big(\vec P_1-\vec P_2\big)\cdot\vec q\big]^2}{q^2}
+\frac{64}{45}\vec P_1\cdot\vec P_2+\frac{34}{15}\frac{\big(\vec P_1\cdot\vec q\big)\big(\vec P_2\cdot\vec q\big)}{q^2}\nonumber\\
&&+\biggl(
-\frac{8}{15}\big(\vec P_1-\vec P_2\big)^2-\frac{8}{5}\frac{\big[\big(\vec P_1-\vec P_2\big)\cdot\vec q\big]^2}{q^2}
-\frac{16}{15}\vec P_1\cdot\vec P_2-\frac{16}{5}\frac{\big(\vec P_1\cdot\vec q\big)\big(\vec P_2\cdot\vec q\big)}{q^2}\bigg)\ln2\nonumber\\
&&+\big(2\big(\vec P_1-\vec P_2\big)^2+4\vec P_1\cdot\vec P_2\big)\bigg(-\frac{1}{2\epsilon}+\ln q\bigg)
+\bigg(-\frac{119}{120}+\frac{35}{24\epsilon}+\frac{16}{15}\ln2-\frac{35}{12}\ln q\bigg)\,q^2\bigg]\nonumber\\
&&+
\frac{\alpha^3}{m^3}
\bigg\{\biggl[\,p_1^m,
\biggl[-\bigg[\frac{Z}{r_1}\bigg]_\epsilon,\,p_1^n\biggr]\biggr]
\bigg[\delta^{mn}\bigg(\frac{4}{15}+\frac{4}{3\epsilon}-\frac{8}{15}\ln2-\frac{8}{3}\ln q\bigg)
+\frac{q^m q^n}{q^2}\bigg(-\frac{12}{5}-\frac{4}{3\epsilon}+\frac{32}{15}\ln2+\frac{8}{3}\ln q\bigg)\bigg]\frac{1}{q^2}	
\nonumber\\ &&+ (1\leftrightarrow2)\bigg\}
+\pi\frac{\alpha^3}{m^3}\bigg(-\frac{8}{15}-\frac{2}{3\epsilon}-\frac{4}{15}\ln2+\frac83\ln q\bigg)\,q\,.
\end{eqnarray}

\subsubsection{Retardation in the non-overlapping double transverse photon exchange $E_5$}
There are altogether 12 double transverse photon exchange diagrams, which we split into two
parts, $E_5$ and $E_6$. $E_5$ corresponds to four diagrams where the photons lines do not overlap
with each other, while the $E_6$ corresponds to the remaining 8 diagrams where the transverse
photons are overlapping. $E_5$ is written as
\begin{eqnarray}
  E_5 &=&
  \int\frac{d^dk_1}{(2\pi)^d\,2k_1} \int\frac{d^dk_2}{(2\pi)^d\,2k_2}\,
	\delta_\perp^{ij}(k_1)\delta_\perp^{kl}(k_2)\nonumber\\
&& \times
\biggl\{
\langle\phi|j_1^i(k_1)\,e^{i\vec{k}_1\cdot\vec{r}_1}\,\frac{1}{E_0-H_0-k_1}\,j_2^j(-k_1)\,e^{-i\vec{k_1}\cdot\vec{r}_2}\frac{1}{(E_0-H_0)'}
j_1^k(k_2)\,e^{i\vec{k}_2\cdot\vec{r}_1}\,\frac{1}{E_0-H_0-k_2}\,j_2^l(-k_2)\,e^{-i\vec{k}_2\cdot\vec{r}_2}|\phi\rangle\nonumber\\
&&+\,
\langle\phi|j_2^i(k_1)\,e^{i\vec{k}_1\cdot\vec{r}_2}\,\frac{1}{E_0-H_0-k_1}\,j_1^j(-k_1)\,e^{-i\vec{k_1}\cdot\vec{r}_1}\frac{1}{(E_0-H_0)'}
j_2^k(k_2)\,e^{i\vec{k}_2\cdot\vec{r}_2}\,\frac{1}{E_0-H_0-k_2}\,j_1^l(-k_2)\,e^{-i\vec{k}_2\cdot\vec{r}_1}|\phi\rangle\nonumber\\
&&+\,
\langle\phi|j_1^i(k_1)\,e^{i\vec{k}_1\cdot\vec{r}_1}\,\frac{1}{E_0-H_0-k_1}\,j_2^j(-k_1)\,e^{-i\vec{k_1}\cdot\vec{r}_2}\frac{1}{(E_0-H_0)'}
j_2^k(k_2)\,e^{i\vec{k}_2\cdot\vec{r}_2}\,\frac{1}{E_0-H_0-k_2}\,j_1^l(-k_2)\,e^{-i\vec{k}_2\cdot\vec{r}_1}|\phi\rangle\nonumber\\
&&+\,
\langle\phi|j_2^i(k_1)\,e^{i\vec{k}_1\cdot\vec{r}_2}\,\frac{1}{E_0-H_0-k_1}\,j_1^j(-k_1)\,e^{-i\vec{k_1}\cdot\vec{r}_1}\frac{1}{(E_0-H_0)'}
j_1^k(k_2)\,e^{i\vec{k}_2\cdot\vec{r}_1}\,\frac{1}{E_0-H_0-k_2}\,j_2^l(-k_2)\,e^{-i\vec{k}_2\cdot\vec{r}_2}|\phi\rangle\biggr\}
\,.\nonumber\\
\end{eqnarray}
Expanding in $H_0-E_0$ and simplifying, we obtain
\begin{eqnarray}
  E_5
&=&-2\int\frac{d^dk_1}{(2\pi)^d\,2k_1} \int\frac{d^dk_2}{(2\pi)^d\,2k_2}\,
\delta_\perp^{ij}(k_1)\delta_\perp^{kl}(k_2)\nonumber\\
&&\times
\biggl\{\langle\phi|[\,j_1^i\,e^{i\vec{k}_1\cdot\vec{r}_1},[H_0-E_0,j_2^j\,e^{-i\vec{k}_1\cdot\vec{r}_2}]]\,\frac{1}{(E_0-H_0)'}
\,j_1^k\,e^{i\vec{k}_2\cdot\vec{r}_1}\,j_2^l\,e^{-i\vec{k}_2\cdot\vec{r}_2}|\phi\rangle\frac{1}{k_1^2 k_2}\nonumber\\
&&+\,
\langle\phi|j_1^i\,e^{i\vec{k}_1\cdot\vec{r}_1}\,j_2^j\,e^{-i\vec{k}_1\cdot\vec{r}_2}\,\frac{1}{(E_0-H_0)'}
\,[\,j_1^k\,e^{i\vec{k}_2\cdot\vec{r}_1},[H_0-E_0,j_2^l\,e^{-i\vec{k}_2\cdot\vec{r}_2}]]|\phi\rangle\frac{1}{k_1 k_2^2}\nonumber\\
&&+\,\langle\phi|j_1^i\,e^{i\vec{k}_1\cdot\vec{r}_1}\,j_2^j\,e^{-i\vec{k}_1\cdot\vec{r}_2}\,\frac{H_0-E_0}{(E_0-H_0)'}
\,j_1^k\,e^{i\vec{k}_2\cdot\vec{r}_1}\,j_2^l\,e^{-i\vec{k}_2\cdot\vec{r}_2}|\phi\rangle\biggl(\frac{1}{k_1^2 k_2}+\frac{1}{k_1 k_2^2}\biggr)\biggr\}\label{42}\,.
\end{eqnarray}
The first two terms in curly brackets in Eq. (\ref{42}) correspond to a part of the second-order term
$\langle H^{(4)}\frac{1}{(E-H)'}H^{(5)}\rangle$, which is considered separately,
and thus is omitted here. The result is
\begin{eqnarray}
E_5&=& 2\int\frac{d^dk_1}{(2\pi)^d\,2k_1} \int\frac{d^dk_2}{(2\pi)^d\,2k_2}\,
\delta_\perp^{ij}(k_1)\delta_\perp^{kl}(k_2)\nonumber\\
&&\times
\langle\phi|j_1^i\,e^{i\vec{k}_1\cdot\vec{r}_1}\,j_2^j\,e^{-i\vec{k}_1\cdot\vec{r}_2}\,(I-|\phi\rangle\langle\phi|\,)
\,j_1^k\,e^{i\vec{k}_2\cdot\vec{r}_1}\,j_2^l\,e^{-i\vec{k}_2\cdot\vec{r}_2}|\phi\rangle\biggl(\frac{1}{k_1^2 k_2}+\frac{1}{k_1 k_2^2}\biggr)\,.
\label{e5}
\end{eqnarray}
The part with $|\phi\rangle\langle\phi|$ is cancelled by $\sigma(E_0)\sigma'(E_0)$, see
Eq.~(\ref{eq:3}), so $E_5$ becomes
\begin{align}
E_5 =&\ 2\int\frac{d^dk_1}{(2\pi)^d\,2k_1} \int\frac{d^dk_2}{(2\pi)^d\,2k_2}\,
\delta_\perp^{ij}(k_1)\delta_\perp^{kl}(k_2) \langle\phi|j_1^i\,e^{i\vec{k}_1\cdot\vec{r}_1}\,j_2^j\,e^{-i\vec{k}_1\cdot\vec{r}_2}\,
\,j_1^k\,e^{i\vec{k}_2\cdot\vec{r}_1}\,j_2^l\,e^{-i\vec{k}_2\cdot\vec{r}_2}|\phi\rangle\,\frac{k_1+k_2}{k_1^2 k_2^2}\,.
\label{ee5}
\end{align}
$E_5$  will later be combined  with $E''_6$ to give  $E''_2$, so we leave it in this form for
now.

\subsubsection{Retardation in the overlapping double transverse photon exchange $E_6$}
The contribution due to the overlapping two transverse photon exchange is
\begin{eqnarray}
E_6 &=&
-2\,\int\frac{d^dk_1}{(2\pi)^d\,2k_1} \int\frac{d^dk_2}{(2\pi)^d\,2k_2}\,
\delta_\perp^{ij}(k_1)\delta_\perp^{kl}(k_2)\nonumber\\
&&\times
\biggl\{\langle\phi|j_1^i\,e^{i\vec{k}_1\cdot\vec{r}_1}\,j_1^k\,e^{i\vec{k}_2\cdot\vec{r}_1}
j_2^j\,e^{-i\vec{k}_1\cdot\vec{r}_2}\,j_2^l\,e^{-i\vec{k}_2\cdot\vec{r}_2}|\phi\rangle \frac{1}{k_1}\frac{1}{k_1+k_2}\frac{1}{k_2}\nonumber\\
&&+\,\langle\phi|j_1^i\,e^{i\vec{k}_1\cdot\vec{r}_1}\,j_1^k\,e^{i\vec{k}_2\cdot\vec{r}_1}
j_2^l\,e^{-i\vec{k}_2\cdot\vec{r}_2}\,j_2^j\,e^{-i\vec{k}_1\cdot\vec{r}_2}|\phi\rangle
\biggl(\frac{1}{k_1}\frac{1}{k_1+k_2}\frac{1}{k_2} + \frac{1}{k_1^2}\frac{1}{k_1+k_2} + \frac{1}{k_2^2}\frac{1}{k_1+k_2}\biggr)\biggr\}\nonumber\\
&=&
-2\int\frac{d^dk_1}{(2\pi)^d\,2k_1} \int\frac{d^dk_2}{(2\pi)^d\,2k_2}\,
\delta_\perp^{ij}(k_1)\delta_\perp^{kl}(k_2)\nonumber\\
&&
\times \biggl\{\langle\phi|j_1^i\,e^{i\vec{k}_1\cdot\vec{r}_1}\,j_1^k\,e^{i\vec{k}_2\cdot\vec{r}_1}
[\,j_2^j\,e^{-i\vec{k}_1\cdot\vec{r}_2},\,j_2^l\,e^{-i\vec{k}_2\cdot\vec{r}_2}]|\phi\rangle \frac{1}{k_1k_2(k_1+k_2)}\nonumber\\
&&+\,\langle\phi|j_1^i\,e^{i\vec{k}_1\cdot\vec{r}_1}\,j_1^k\,e^{i\vec{k}_2\cdot\vec{r}_1}
j_2^l\,e^{-i\vec{k}_2\cdot\vec{r}_2}\,j_2^j\,e^{-i\vec{k}_1\cdot\vec{r}_2}|\phi\rangle
\frac{k_1+k_2}{k_1^2k_2^2}\biggr\}
= E_6' + E_6''\,.
\end{eqnarray}
$E_6''$ is combined with $E_5$ to give  $E''_2$, as follows
\begin{align}
E_5 + E''_6=&\  2\int\frac{d^dk_1}{(2\pi)^d\,2k_1} \int\frac{d^dk_2}{(2\pi)^d\,2k_2}\,
\delta_\perp^{ij}(k_1)\delta_\perp^{kl}(k_2)\,\frac{k_1+k_2}{k_1^2k_2^2}\nonumber\\
\times
&\
\Bigl\langle\phi\Bigl|j_1^i\,e^{i\vec{k}_1\cdot\vec{r}_1}\,j_2^j\,e^{-i\vec{k}_1\cdot\vec{r}_2}\,
\,j_1^k\,e^{i\vec{k}_2\cdot\vec{r}_1}\,j_2^l\,e^{-i\vec{k}_2\cdot\vec{r}_2}
-j_1^i\,e^{i\vec{k}_1\cdot\vec{r}_1}\,j_1^k\,e^{i\vec{k}_2\cdot\vec{r}_1}
j_2^l\,e^{-i\vec{k}_2\cdot\vec{r}_2}\,j_2^j\,e^{-i\vec{k}_1\cdot\vec{r}_2}\Bigr|\phi\Bigr\rangle \label{excl}\\
=&\ \int\frac{d^dk_1}{(2\pi)^d\,2\,k_1^3} \int\frac{d^dk_2}{(2\pi)^d\,k_2^2}\,
\delta_\perp^{ij}(k_1)\delta_\perp^{kl}(k_2)\nonumber\\
&\times
   \Bigl\langle\phi\Bigl|j_1^i\,e^{i\vec{k}_1\cdot\vec{r}_1}\,j_2^j\,e^{-i\vec{k}_1\cdot\vec{r}_2}\,
\,j_1^k\,e^{i\vec{k}_2\cdot\vec{r}_1}\,j_2^l\,e^{-i\vec{k}_2\cdot\vec{r}_2}
-j_1^i\,e^{i\vec{k}_1\cdot\vec{r}_1}\,j_1^k\,e^{i\vec{k}_2\cdot\vec{r}_1}
j_2^l\,e^{-i\vec{k}_2\cdot\vec{r}_2}\,j_2^j\,e^{-i\vec{k}_1\cdot\vec{r}_2}\nonumber \\ &\
+ j_1^k\,e^{i\vec{k}_2\cdot\vec{r}_1}\,j_2^l\,e^{-i\vec{k}_2\cdot\vec{r}_2}\,
\,j_1^i\,e^{i\vec{k}_1\cdot\vec{r}_1}\,j_2^j\,e^{-i\vec{k}_1\cdot\vec{r}_2}
- j_2^j\,e^{-i\vec{k}_1 \cdot\vec{r}_2}\,j_2^l\,e^{-i\vec{k}_2\cdot\vec{r}_2}
j_1^k\,e^{i\vec{k}_2\cdot\vec{r}_1}\,j_1^i\,e^{i\vec{k}_1\cdot\vec{r}_1}
\Bigr|\phi\Bigr\rangle \nonumber \\
=&\ - \int\frac{d^dk_1}{(2\pi)^d\,2\,k_1^3}\,\delta_\perp^{ij}(k_1)
\Bigl\langle\phi\Bigl|j_1^i\,e^{i\vec{k}_1\cdot\vec{r}_1}\,j_2^j\,e^{-i\vec{k}_1\cdot\vec{r}_2}\, H''^{(4)}
-j_1^i\,e^{i\vec{k}_1\cdot\vec{r}_1}\,H''^{(4)}\,j_2^j\,e^{-i\vec{k}_1\cdot\vec{r}_2} \nonumber \\ &\
+ H''^{(4)}\,j_1^i\,e^{i\vec{k}_1\cdot\vec{r}_1}\,j_2^j\,e^{-i\vec{k}_1\cdot\vec{r}_2}
- j_2^j\,e^{-i\vec{k}_1 \cdot\vec{r}_2}\,H''^{(4)}\,j_1^i\,e^{i\vec{k}_1\cdot\vec{r}_1}
\Bigr|\phi\Bigr\rangle \nonumber \\
=&\  \int\frac{d^dk}{(2\pi)^d\,2\,k^3}\,\delta_\perp^{ij}(k)
\bigg\langle\phi\bigg|\bigg[j_1^i\,e^{i\vec{k}\cdot\vec{r}_1}\,,\,\bigg[H''^{(4)}\,,\,j_2^j\,e^{-i\vec{k}\cdot\vec{r}_2}
\bigg]\bigg]\biggr|\phi\biggr\rangle = E_2''\,,
\end{align}
which has already been accounted for as a part of $E_2$. The remainder $E'_6$ then represents the
double transverse photon exchange correction,
\begin{eqnarray}
E_6' &=&
-\int\frac{d^dk_1}{(2\pi)^d\,2k_1^2} \int\frac{d^dk_2}{(2\pi)^d\,2k_2^2}\frac{1}{k_1+k_2}\,
\delta_\perp^{ij}(k_1)\delta_\perp^{kl}(k_2)
\langle\phi|[j_1^i\,e^{i\vec{k}_1\cdot\vec{r}_1},\,j_1^k\,e^{i\vec{k}_2\cdot\vec{r}_1}]
[\,j_2^j\,e^{-i\vec{k}_1\cdot\vec{r}_2},\,j_2^l\,e^{-i\vec{k}_2\cdot\vec{r}_2}]|\phi\rangle\,.
\nonumber \\
\end{eqnarray}
Evaluating commutators of currents, we have
\begin{align}
E_6' =&\ -e^4\int\frac{d^dk_1}{(2\pi)^d\,2k_1^2} \int\frac{d^dk_2}{(2\pi)^d\,2k_2^2}\frac{1}{k_1+k_2}\,
\delta_\perp^{ij}(k_1)\delta_\perp^{kl}(k_2)\nonumber\\ &\
\times
\langle\phi|\bigl(j_{n}^{ik}(\vec p_1,\vec k_1,\vec k_2)+j_{s}^{ik,ab}\sigma_1^{ab}\bigr)
\,e^{i(\vec{k}_1+\vec{k}_2)\cdot\vec{r}}\,\bigl(j_{n}^{jl}(\vec p_2,-\vec k_1,-\vec k_2)+j_{s}^{jl,mn}\sigma_2^{mn}\bigr)|\phi\rangle
\\ =&\
-e^4\int\frac{d^dk_1}{(2\pi)^d\,2k_1^2} \int\frac{d^dk_2}{(2\pi)^d\,2k_2^2}\frac{1}{k_1+k_2}\,
\delta_\perp^{ij}(k_1)\delta_\perp^{kl}(k_2)\nonumber\\ \times &\
\bigg\langle\phi\bigg|\Bigl(-p_1^i\,k_1^k+p_1^k\,k_2^i\Bigr)\,e^{i(\vec{k}_1+\vec{k}_2)\cdot\vec{r}}\,\Bigl(k_1^l\,p_2^j-k_2^j\,p_2^l\Bigr)
+ \frac{\sigma_1\cdot\sigma_2}{d(d-1)} j_{s}^{ik,ab}
\,e^{i(\vec{k}_1+\vec{k}_2)\cdot\vec{r}}\,j_{s}^{jl,mn}\,(\delta^{am}\delta^{bn}-\delta^{an}\delta^{bm})\bigg|\phi\bigg\rangle
\,,
\end{align}
where we omitted terms contributing to the fine structure only. The result for the corresponding
effective operator $H_6$, $E_6' = \langle H_6\rangle$, is
\begin{eqnarray}
H_6 &=&
\frac{\alpha^2}{m^4} \bigg[\bigg(-\frac{8}{15}\vec P_1\cdot\vec P_2+\frac{8}{15}\frac{\big(\vec P_1\cdot\vec q\big)\big(\vec P_2\cdot\vec q\big)}{q^2}\bigg)
\big(1-7\ln2)
+\sigma_1\cdot\sigma_2\,\Big(\frac{179}{216}+\frac{3}{32\epsilon}-\frac79\ln2-\frac{3}{16}\ln q\Big)\,q^2\bigg]\,.
\end{eqnarray}

\subsubsection{$E_7$}
We now turn to the single transverse photon exchange contributions coming from vertices in
$\delta_{\rm ST}H_{\rm FW}$ in Eq. (\ref{eq:ST}). The first contribution,
denoted as $E_7$, comes from the transverse photon exchange with vertices
$-(e/m)\bigl(\vec{p}\cdot\vec{A} + \frac{1}{4}\sigma\cdot B\bigr)$ and
$(e/4m^3)\bigl\{\vec{p}\,{}^2,\vec{p}\cdot\vec{A}+\frac14\,\sigma\cdot B\bigr\}$,
\begin{eqnarray}
E_7 &=& -\frac{e^2}{4m^3}\int\frac{d^dk}{(2\pi)^d \,2k^3}\delta_\perp^{ij}(k)
\langle\phi|\biggl(\frac{p_1^i}{m}+\frac{1}{2m}\sigma_1^{ki}\nabla_1^k\biggr)\,e^{i\vec{k}\cdot\vec{r}_1}\,(H_0-E_0)
\,\biggl\{p_2^2,\biggl(p_2^j+\frac12\sigma_2^{lj}\nabla_2^l\biggr)\,e^{-i\vec{k}\cdot\vec{r}_2}\biggr\}\,|\phi\rangle
+\textrm{h.c.} + (1\leftrightarrow2)\nonumber\\
&=&
-\frac{e^2}{4m^3}\int\frac{d^dk}{(2\pi)^d \,2k^3}\delta_\perp^{ij}(k)
\langle\phi|\biggl[\biggl[\biggl(\frac{p_1^i}{m}+\frac{1}{2m}\sigma_{1}^{ki}\nabla_1^k\biggr)\,e^{i\vec{k}\cdot\vec{r}_1},H_0-E_0\biggr]\,,\,
\biggl\{p_2^2,\biggl(p_2^j+\frac12\sigma_2^{lj}\nabla_2^l\biggr)\,e^{-i\vec{k}\cdot\vec{r}_2}\biggr\}\biggr]\,|\phi\rangle
+ (1\leftrightarrow2)\nonumber\\
&=&
-\frac{e^2}{4\,m^4}\int\frac{d^dk}{(2\pi)^d \,2k^3}\delta_\perp^{ij}(k)\,
\langle\phi|\Bigl[\bigl[p_1^i\,e^{i\vec{k}\cdot\vec{r}_1},V\bigr]\,,\,
\bigl\{p_2^2\,,\,p_2^j\,e^{-i\vec{k}\cdot\vec{r}_2}\bigr\}\Bigr]\,|\phi\rangle
+ (1\leftrightarrow2)\,.
\end{eqnarray}
The operator in the matrix element can be rewritten as
\begin{align}
\Big[[\,p_1^i,V],\{\,p_2^2,e^{i\vec k\cdot\vec r}\}\,p_2^j\Big] &= V^{ij}\, [\,p_2^k,[\,p_2^k,e^{i\vec k\cdot\vec r}]]
+ 2\,p_2^k\,V^{ij}\,e^{i\vec k\cdot\vec r}\,p_2^k
+2\,[\,p_2^k,[\,p_2^j,V^{ik}\,e^{i\vec k\cdot\vec r}]]
+ 4\,p_2^k\,V^{ik}\,e^{i\vec k\cdot\vec r}\,p_2^j\,,
\end{align}
where $V^{ij}=[\,p_1^i,[V,p_2^j]]$. Performing now the momentum integration, we arrive at
\begin{align}
H_7 =&\ \frac{\alpha^2}{m^4}\bigg\{
\frac{2}{9}\big(\vec P_1-\vec P_2\big)^2 + \frac{8}{3}\frac{\big[\big(\vec P_1-\vec P_2\big)\cdot\vec q\big]^2}{q^2}
-\frac{7}{9}q^2 + \frac49\vec P_1\cdot\vec P_2+\frac{16}{3}\frac{\big(\vec P_1\cdot\vec q\big)\big(\vec P_2\cdot\vec q\big)}{q^2}\nonumber\\
&\ +\bigg(-\frac{4}{3}\big(\vec P_1-\vec P_2\big)^2-\frac83\frac{\big[\big(\vec P_1-\vec P_2\big)\cdot\vec q\big]^2}{q^2}
+\frac{2}{3}q^2 - \frac83\vec P_1\cdot\vec P_2-\frac{16}{3}\frac{\big(\vec P_1\cdot\vec q\big)\big(\vec P_2\cdot\vec q\big)}{q^2}\bigg)
\bigg(-\frac{1}{2\epsilon}+\ln2+\ln q\bigg)\bigg\}\,.
\end{align}

\subsubsection{$E_8$}
The second contribution comes from the term $(e^2/2m^2)\,\sigma^{ij} E_\parallel^i A^j$ in
$\delta_{\rm ST}H_{\rm FW}$. This gives a correction to the single seagull diagram with one Coulomb and one
transverse photon with retardation,
\begin{eqnarray}
  E_8 &=&\frac{e^2}{2m^2}\int\frac{d^dk}{(2\pi)^d \,2k^3}\delta_\perp^{ij}(k)
\langle\phi|\biggl(\frac{p_1^i}{m}+\frac{1}{2m}\sigma_1^{ki}\nabla_1^k\biggr)\,e^{i\vec{k}\cdot\vec{r}_1}
\,(H_0-E_0)\,e^{-i\vec{k}\cdot\vec{r}_2}\,\sigma_2^{lj}\,\nabla_2^l\,V|\phi\rangle
+\textrm{h.c.} + (1\leftrightarrow2)\nonumber \\
&=&
\frac{e^2}{8\,m^4}\int\frac{d^dk}{(2\pi)^d \,2k^3}\delta_\perp^{ij}(k)
\langle\phi|\bigl[\sigma_1^{ki}\nabla_1^k\,e^{i\vec{k}\cdot\vec{r}_1}
\,,\,\bigl[p_1^2\,,\,e^{-i\vec{k}\cdot\vec{r}_2}\,\sigma_2^{lj}\,\nabla_2^l\,V\bigr]\bigr]|\phi\rangle
+ (1\leftrightarrow2)\,.
\end{eqnarray}
We evaluate it as
\begin{eqnarray}
E_8
 &=&
-\frac{e^2}{8\,m^4}\int\frac{d^dk}{(2\pi)^d \,k^3}\delta_\perp^{ij}(k) k^k k^m
\sigma_1^{ki}\sigma_2^{lj}
\langle\phi|\,e^{i\vec{k}\cdot\vec{r}}\,\nabla_1^m\nabla_2^l\,V|\phi\rangle
+ (1\leftrightarrow2)\,.
\end{eqnarray}
After spin averaging, we obtain
\begin{eqnarray}
E_8
 &=&
-\frac{e^2}{8\,m^4}\frac{\sigma_1\cdot\sigma_2}{d}
\int\frac{d^dk}{(2\pi)^d \,k^3} k^l k^m
\langle\phi|\,e^{i\vec{k}\cdot\vec{r}}\,\nabla_1^m\nabla_2^l\,V|\phi\rangle
+ (1\leftrightarrow2)\,.
\end{eqnarray}
The result in momentum representation is
\begin{equation}
H_8 = \frac{\alpha^2}{m^4}\,\sigma_1\cdot\sigma_2\Big(\frac23+\frac{2}{9\epsilon}-\frac49\ln2-\frac49\ln q\Big)\,q^2\,.
\end{equation}

\subsubsection{$E_9$}
The next contribution is the single transverse photon exchange with one vertex of the form
$(ie/16m^3)\,[\sigma^{ij}\{A^i,p^j\},p^2]$. It is given by
\begin{eqnarray}
E_9 &=& -\frac{e^2}{16m^4}\int\frac{d^dk}{(2\pi)^d\,2\,k^3}\delta_\perp^{ij}(k)
\langle\phi|\frac{i}{2}\sigma_1^{ki} k^k e^{i\vec k\cdot\vec r_1}\,(H_0-E_0)
\,i[\sigma_2^{jl}\{p_2^l,e^{-i\vec k\cdot\vec r_2}\},p_2^2]|\phi\rangle
+\textrm{h.c.}+(1\leftrightarrow2)\nonumber\\
&=& \frac{e^2}{64m^4}\sigma_1^{ki}\sigma_2^{jl}\int\frac{d^dk}{(2\pi)^d}\frac{k^k}{k^3}\delta_\perp^{ij}(k)
\langle\phi|  \biggl[e^{i\vec k\cdot\vec r_1},\biggl[H_0-E_0,[\{p_2^l,e^{-i\vec k\cdot\vec r_2}\},p_2^2]\biggr]\biggr]|\phi\rangle +
(1\leftrightarrow2)= 0\,.
\end{eqnarray}
Thus $H_9 = 0$.

\subsubsection{$E_{10}$}
The next term is a double seagull contribution with one transverse and one Coulomb photon coming
from the term $(e^2/2m^2)\,\sigma^{ij} E_\parallel^i A^j$ in both vertices. In order to derive
this contribution, we start with a general (Feynman) diagram for the two-photon exchange in the
Coulomb gauge,
\begin{eqnarray}
E_{10} &=&
\frac{e^4}{4m^4}\int\frac{d^D k_1}{(2\pi)^Di}\int\frac{d^D k_2}{(2\pi)^D i}\frac{1}{(\omega_1^2-\vec k_1^2+i\epsilon)}\frac{1}{\vec k_2^2}
\delta_\perp^{ij}(k_1)\nonumber\\
&&\times
\langle\phi|\sigma_1^{ki}\,(-ik_2^k) \,e^{i(\vec k_1+\vec k_2)\cdot\vec r_1}\,\frac{1}{E_0-H_0-\omega_1-\omega_2+i\epsilon}\,\sigma_2^{lj}(ik_2^l)\,e^{-(\vec k_1+\vec k_2)\cdot\vec r_2}|\phi\rangle
+(1\leftrightarrow2)
\,.
\end{eqnarray}
Integration over $\omega_1$ and $\omega_2$ is of the form
\begin{eqnarray}
\int\frac{d\omega_1}{2\pi i}\int\frac{d\omega_2}{2\pi i}
\frac{1}{(\omega_1^2-\vec k_1^2+i\epsilon)}\frac{1}{E_0-H_0-\omega_1-\omega_2+i\epsilon}.
\end{eqnarray}
Shifting the integration variable in the second expression
$\omega_2\rightarrow\omega_2-\omega_1$, we get
\begin{eqnarray}
&&\int\frac{d\omega_1}{2\pi i}
\frac{1}{(\omega_1^2-\vec k_1^2+i\epsilon)}\int\frac{d\omega_2}{2\pi i}\frac{1}{E_0-H_0-\omega_2+i\epsilon}\nonumber\\
&&=\int\frac{d\omega_1}{2\pi i}\frac{1}{(\omega_1^2-\vec k_1^2+i\epsilon)}\frac12
\int\frac{d\omega_2}{2\pi i}\biggl[\frac{1}{E_0-H_0-\omega_2+i\epsilon}+\frac{1}{E_0-H_0+\omega_2+i\epsilon}\biggr]\nonumber\\
&&=\frac{1}{4\,k_1}\,.
\end{eqnarray}
The term $E_{10}$ is then
\begin{eqnarray}
E_{10} &=&
\frac{e^4}{16m^4}\int\frac{d^d k_1}{(2\pi)^dk_1}\int\frac{d^d k_2}{(2\pi)^d k_2^2}
\delta_\perp^{ij}(k_1)\,k_2^k\,k_2^l\,
\langle\phi|\sigma_1^{ki}\,e^{i(\vec k_1+\vec k_2)\cdot\vec r}\sigma_2^{lj}|\phi\rangle
+(1\leftrightarrow2)\nonumber\\
&=&
\frac{e^4}{8m^4}\sigma_1^{ki}\sigma_2^{lj}\int\frac{d^d k_1}{(2\pi)^dk_1}\int\frac{d^d k_2}{(2\pi)^d k_2^2}k_2^k k_2^l
\delta_\perp^{ij}(k_1)
\langle\phi| e^{i(\vec k_1+\vec k_2)\cdot\vec r}|\phi\rangle\,.
\end{eqnarray}
The result in momentum representation is
\begin{equation}
H_{10} = \frac{\alpha^2}{m^4}\sigma_1\cdot\sigma_2\biggl(-\frac{2}{9}-\frac{1}{18\epsilon}+\frac{1}{9}\ln2+\frac{1}{9}\ln q\biggr)\,q^2\,.
\end{equation}

\subsubsection{$E_{11}$}
The next contribution is due to the correction of the form $(e^2/4\,m^3)\,\vec E_\parallel\,\vec
E_\perp$ in one of the vertices,
\begin{eqnarray}
E_{11} &=& -\frac{e^2}{4\,m^3}\int\frac{d^dk}{(2\pi)^d2k}\delta_\perp^{ij}(k)
\langle\phi|\biggl(\frac{p_1^i}{m}+\frac{i}{2m}\sigma_1^{ki}k^k\biggr)e^{i\,\vec{k}\cdot\vec{r}_1}\frac{1}{E_0-H_0-k}(ik)e^{-i\,\vec{k}\cdot\vec{r}_2}\,\nabla_2^jV|\phi\rangle
+\,\textrm{h.c.}+ (1\leftrightarrow2)\nonumber \\
&=& -\frac{e^2}{8m^4}\int \frac{d^dk}{(2\pi)^d\,k}\delta_\perp^{ij}(k)
\,\langle\phi|p_1^i\,e^{i\,\vec{k}\cdot\vec{r}}\,[\,p_2^j,V]|\phi\rangle+\textrm{h.c.}+ (1\leftrightarrow2)\nonumber\\
&=& \frac{e^2}{8m^4}\int \frac{d^dk}{(2\pi)^d\,k}\delta_\perp^{ij}(k)
\,\langle\phi|[\,p_1^i,[V,p_2^j]]\,e^{i\,\vec{k}\cdot\vec{r}}|\phi\rangle+ (1\leftrightarrow2)\,.
\end{eqnarray}
After a straightforward calculation, the result in the momentum representation is
\begin{equation}
H_{11} = \frac{\alpha^2}{m^4}\biggl(\frac{14}{9}+\frac{2}{3\epsilon}-\frac43\ln2-\frac43\ln q\biggr)\,q^2\,.
\end{equation}

\subsubsection{$E_{12}$}

We now turn to the double transverse photon contributions coming from the higher-order terms from
$\delta_{\rm DT}H_{\rm FW}$ given by Eq.~(\ref{eq:DT}). The first contribution comes from the
terms $-1/(8m^3)\{\,\vec{p}\,{}^2,e^2 \vec{A}\,{}^2\}$ and $- e^2/(2m^3)\,
(\vec{A}\cdot\vec{p}\,)^2 $. The corresponding contribution can be evaluated in the
nonretardation approximation,
\begin{eqnarray}
E_{12} &=&
\frac{e^4}{m^4}\, \int\frac{d^dk_1}{(2\pi)^d\,2k_1}\int\frac{d^dk_2}{(2\pi)^d\,2k_2}
\delta_\perp^{ik}(k_1)\delta_\perp^{jk}(k_2)\nonumber\\
&& \times
\langle\phi|e^{i(\vec{k}_1+\vec{k}_2)\cdot\vec{r}_1}\,\frac{(-1)}{k_1+k_2}\biggl[-\frac{\delta^{ij}}{8}\{p_2^2,e^{-i(\vec{k}_1+\vec{k}_2)\cdot\vec{r}_2}\}
-\frac12\,e^{-i\vec{k}_1\cdot\vec{r}_2}\,p_2^i\,e^{-i\vec{k}_2\cdot\vec{r}_2} p_2^j\biggr]|\phi\rangle +\textrm{h.c.} + (1\leftrightarrow2)
\nonumber\\
&=&
-\frac{e^4}{m^4} \int\frac{d^dk_1}{(2\pi)^d\,2k_1}\int\frac{d^dk_2}{(2\pi)^d\,2k_2}
\frac{1}{k_1+k_2}\delta_\perp^{ik}(k_1)\delta_\perp^{jk}(k_2)
\langle\phi|-\frac{\delta^{ij}}{4}\{p_2^2,\,e^{i(\vec{k}_1+\vec{k}_2)\cdot\vec{r}}\}
-p^i_2\,e^{i(\vec{k}_1+\vec{k}_2)\cdot\vec{r}}\,p_2^j
|\phi\rangle + (1\leftrightarrow2)\,.\nonumber\\
\end{eqnarray}
Contracting all indices and integrating in the momentum space, we obtain the
result
\begin{eqnarray}
H_{12} &=& -\frac{7}{18}\big(\vec P_1-\vec P_2\big)^2-\frac{7}{3}\frac{\big[\big(\vec P_1-\vec P_2\big)\cdot\vec q\big]^2}{q^2}
+\frac{19}{36}q^2-\frac{7}{9}\vec P_1\cdot\vec P_2-\frac{14}{3}\frac{\big(\vec P_1\cdot\vec q\big)\big(\vec P_2\cdot\vec q\big)}{q^2}\nonumber\\
&&+\biggl(
\frac{4}{3}\big(\vec P_1-\vec P_2\big)^2+\frac{8}{3}\frac{\big[\big(\vec P_1-\vec P_2\big)\cdot\vec q\big]^2}{q^2}
-\frac23q^2+\frac{8}{3}\vec P_1\cdot\vec P_2+\frac{16}{3}\frac{\big(\vec P_1\cdot\vec q\big)\big(\vec P_2\cdot\vec q\big)}{q^2}\bigg)\ln2\nonumber\\
&&+\bigg(-\frac53\big(\vec P_1-\vec P_2\big)^2-\frac16q^2-\frac{10}{3}\vec P_1\cdot\vec P_2\bigg)\bigg(-\frac{1}{2\epsilon}+\ln q\bigg)\,.
\end{eqnarray}

\subsubsection{$E_{13}$}
The next contribution is due to the correction $e^2/(8m^3)\, \vec{E}^2_\perp$ in one of the
vertices,
\begin{eqnarray}
E_{13} &=&
\frac{e^4}{8\,m^4}\, \int\frac{d^dk_1}{(2\pi)^d\,2k_1}\int\frac{d^dk_2}{(2\pi)^d\,2k_2}
\delta_\perp^{ij}(k_1)\delta_\perp^{ij}(k_2)
\langle\phi|\,e^{i(\vec{k}_1+\vec{k}_2)\cdot\vec{r}_1}\,\frac{(-1)}{k_1+k_2}\,
(-i)^2 k_1 k_2\,e^{-i(\vec{k}_1+\vec{k}_2)\cdot\vec{r}_2}|\phi\rangle
+\textrm{h.c.}+ (1\leftrightarrow2)\nonumber\\
&=&
\frac{e^4}{4\,m^4} \int\frac{d^dk_1}{(2\pi)^d\,2k_1}\int\frac{d^dk_2}{(2\pi)^d\,2k_2}
\delta_\perp^{ij}(k_1)\delta_\perp^{ij}(k_2)\,\frac{k_1 k_2}{k_1+k_2}\,
\langle\phi|\,e^{i(\vec{k}_1+\vec{k}_2)\cdot\vec{r}}|\phi\rangle + (1\leftrightarrow2)\,
\,.
\end{eqnarray}
We contract all indices, use the identity
\begin{equation}
\vec k_1\cdot\vec k_2=\frac12\Bigl[\vec q\,{}^2-(k_1+k_2)^2+2k_1 k_2\Bigr]\,,
\end{equation}
where $\vec q=\vec k_1+\vec k_2$, and perform the momentum integration with help of formulas from
Appendix \ref{app:C} with the result
\begin{equation}
H_{13} = \frac{\alpha^2}{m^4}\,\Big(-\frac{9}{8}-\frac{5}{24\epsilon}+\frac43\ln2+\frac{5}{12}\ln q\Big)\,q^2\,.
\end{equation}

\subsubsection{$E_{14}$}
This correction is due to term $-e^2/(16m^3)\,B^{ij}B^{ij}$. It can be again evaluated in the
nonretardation approximation. Using the identity
\begin{equation}
B^{ij}B^{ij} = 2\big[(\nabla^i A^j)^2 - \nabla^i A^j\,\nabla^j A^i\big]\,,
\end{equation}
we get
\begin{eqnarray}
E_{14} &=&
-\frac{e^4}{8\,m^4}\, \int\frac{d^dk_1}{(2\pi)^d\,2k_1}\int\frac{d^dk_2}{(2\pi)^d\,2k_2}
\delta_\perp^{ik}(k_1)\delta_\perp^{jk}(k_2)\nonumber\\
&& \times
\langle\phi|\,e^{i(\vec{k}_1+\vec{k}_2)\cdot\vec{r}_1}\,\frac{(-1)}{k_1+k_2}\,(-i)^2\bigl(\delta^{ij}
(\vec{k}_1\cdot\vec{k}_2) - k_1^j\,k_2^i\bigr)
\,e^{-i(\vec{k}_1+\vec{k}_2)\cdot\vec{r}_2}|\phi\rangle
+\textrm{h.c.}+ (1\leftrightarrow2)\nonumber\\
&=&
-\frac{e^4}{4\,m^4} \int\frac{d^dk_1}{(2\pi)^d\,2k_1}\int\frac{d^dk_2}{(2\pi)^d\,2k_2}
\delta_\perp^{ik}(k_1)\delta_\perp^{jk}(k_2)
\frac{\bigl(\delta^{ij}\,\vec{k}_1\cdot\vec{k}_2 - k_1^j\,k_2^i\bigr)}{k_1+k_2}\,
\,\langle\phi|\,e^{i(\vec{k}_1+\vec{k}_2)\cdot\vec{r}}|\phi\rangle
+ (1\leftrightarrow2)\,.
\end{eqnarray}
Contracting all the indices and integrating in the momentum space, we obtain the following result
\begin{equation}
H_{14} = \frac{\alpha^2}{m^4}\Big(-\frac{13}{72}-\frac{5}{24\epsilon}+\frac{5}{12}\ln q\Big)\,q^2\,.
\end{equation}

\subsubsection{$E_{15}$}
The next correction comes from the double seagull with the term
$e^2/(4\,m^2)\,\sigma^{ij}\,E^i_\perp\,A^j$ in each vertex.
It cannot be calculated from time-ordered diagrams because it contains
two powers of $\omega$ in the numerator. Such terms, as explained in Ref. \cite{pachucki:06:he},
shall be obtained using Feynman diagrams, and it is of the form
\begin{eqnarray}
E_{15} &=& \biggl(\frac{e^2}{4\,m^2}\biggr)^2\int\frac{d^Dk_1}{(2\pi)^Di}\int\frac{d^Dk_2}{(2\pi)^Di}\frac{1}{(w_1^2-\vec{k}_1^2+i\epsilon)(w_2^2-\vec{k}_2^2+i\epsilon)}
\biggl(\delta^{ik}-\frac{k_1^ik_1^k}{k_1^2}\biggr)\biggl(\delta^{jl}-\frac{k_2^j k_2^l}{k_2^2}\biggr)\,
\frac{(\omega_1-\omega_2)^2}{2}\nonumber\\
&& \times\langle\phi|e^{i(\vec k_1+\vec k_2)\cdot\vec r_1}\,\sigma_1^{ij}
\frac{1}{E_0-H_0-\omega_1-\omega_2+i\epsilon}\sigma_2^{kl}\,e^{-i(\vec k_1+\vec k_2)\cdot\vec r_2}|\phi\rangle
+(1\leftrightarrow2)\,.
\end{eqnarray}
The $\omega_1$ and $\omega_2$ integration leads to
\begin{align}
\int\frac{d\omega_1}{2\pi i}\int\frac{d\omega_2}{2\pi i}
\frac{(\omega_1-\omega_2)^2}{(w_1^2-\vec{k}_1^2+i\epsilon)(w_2^2-\vec{k}_2^2+i\epsilon)}\frac{1}{E_0-H_0-\omega_1-\omega_2+i\epsilon} \approx \frac{1}{k_1+k_2}
\,.
\end{align}
and this correction becomes
\begin{align}
E_{15} =&\  \frac{e^4}{16\,m^4}\, \int\frac{d^dk_1}{(2\pi)^d} \int\frac{d^dk_2}{(2\pi)^d}
 \delta_\perp^{ij}(k_1)\delta_\perp^{kl}(k_2)
 \langle\phi|e^{i(\vec{k}_1+\vec{k}_2)\cdot\vec{r}_1}\,
\frac{1}{k_1+k_2}\,e^{-i(\vec{k}_1+\vec{k}_2)\cdot\vec{r}_2}|\phi\rangle\,
 \sigma_1^{ik}\,\sigma_2^{jl}\,.
\end{align}
In momentum representation, the result is
\begin{equation}
H_{15} = \frac{\alpha^2}{m^4}\sigma_1\cdot\sigma_2\biggl(\frac{23}{216}+\frac{1}{96\epsilon}-\frac{1}{9}\ln2-\frac{1}{48}\ln q\biggr)\,q^2\,.
\end{equation}

\subsubsection{$E_{16}$}
Next, there is one more correction to the single seagull diagram. The double vertex is
$e^2/(4\,m^2)\,\sigma^{ij}\,E^i_\perp\,A^j$, whereas the single interaction vertices are both
$(-e/m)(\vec{p}+\frac14\sigma\cdot B)$. This correction can also be calculated in the
nonretardation approximation. We thus have
\begin{eqnarray}
  E_{16} &=& \frac{ie^4}{4\,m^4}\,\int\frac{d^dk_1}{(2\pi)^d2k_1}\int\frac{d^d k_2}{(2\pi)^d 2k_2}\,
  \delta_\perp^{im}(k_1)\delta_\perp^{jn}(k_2)
	\,\biggl(-\frac{k_1+k_2}{k_1\,k_2} + \frac{k_1-k_2}{k_2(k_1+k_2)} + \frac{k_2-k_1}{k_1\,(k_1+k_2)}\biggr)\nonumber\\
&&\times\langle\phi|\Bigl(p_1^m+\frac{i}{2}\sigma_1^{rm} k_1^r\Bigr) e^{i\vec k_1\cdot\vec r_1}
  \,\Bigl(p_1^n+\frac{i}{2}\sigma_1^{sn} k_2^s\Bigr) e^{i\vec k_2\cdot\vec r_1}\, \sigma_2^{ij}\,e^{-i(\vec k_1+\vec k_2)\cdot\vec r_2}|\phi\rangle\nonumber
+(1\leftrightarrow 2)\nonumber\\
&=& \frac{e^4}{2\,m^4}\int\frac{d^dk_1}{(2\pi)^d\,2\,k_1}\int\frac{d^d k_2}{(2\pi)^d\,2\,k_2}\frac{(-i)}{k_1+k_2}
    \delta_\perp^{im}(k_1)\delta_\perp^{jn}(k_2)\nonumber\\
&&\times\,\langle\phi|\biggl[\Bigl(p_1^m+\frac{i}{2}\sigma_1^{rm} k_1^r\Bigr) e^{i\vec k_1\cdot\vec r}
\,,\,\Bigl(p_1^n+\frac{i}{2}\sigma_1^{sn} k_2^s\Bigr) e^{i\vec k_2\cdot\vec r}\biggr]\,\sigma_2^{ij}|\phi\rangle
+(1\leftrightarrow 2)
\,.
\end{eqnarray}
We calculate it as
\begin{eqnarray}
E_{16}
&=& \frac{e^4}{4\,m^4}\int\frac{d^dk_1}{(2\pi)^d\,2\,k_1}\int\frac{d^d k_2}{(2\pi)^d\,2\,k_2}\frac{1}{k_1+k_2}
    \delta_\perp^{im}(k_1)\delta_\perp^{jn}(k_2)
		\langle\phi|j_s^{mn,ab}\,\sigma_1^{ab}\,\sigma_2^{ij}|\phi\rangle+(1\leftrightarrow 2)\,.
\end{eqnarray}
After spin averaging and performing momentum integration, we obtain the result
\begin{equation}
H_{16} = \frac{\alpha^2}{m^4}\sigma_1\cdot\sigma_2\biggl(-\frac{7}{54}+\frac{11}{144\epsilon}
+\frac49\ln2-\frac{11}{72}\ln q\biggr)\,q^2\,.
\end{equation}

\subsubsection{$E_{17}$}
Finally, we account for the contribution due to the double seagull correction with the term
$-e^2/(8m^3)\, A^i \nabla^j B^{ij}$ in one of the vertices. Omitting part contributing to the
fine structure only, we obtain
\begin{eqnarray}
  E_{17} &=& \frac{e^4}{8\,m^4}\,\int\frac{d^dk_1}{(2\pi)^d 2k_1}\int\frac{d^d k_2}{(2\pi)^d 2k_2}
  \delta_\perp^{ij}(k_1)\delta_\perp^{ij}(k_2)
\langle\phi|e^{i(\vec k_1+\vec k_2)\cdot\vec r}|\phi\rangle\,\frac{(k_1^2 + k_2^2)}{(k_1+k_2)}
+(1\leftrightarrow2)\,.
\end{eqnarray}
We rewrite this as
\begin{align}
E_{17}
=&\ \frac{e^4}{16\,m^4}\,\int\frac{d^dk_1}{(2\pi)^d}\int\frac{d^dk_2}{(2\pi)^d}
\delta_\perp^{ij}(k_1)\delta_\perp^{ij}(k_2)
\,\frac{(k_1 + k_2)^2-2\,k_1\,k_2}{k_1\,k_2\,(k_1+k_2)}
\langle\phi|e^{i(\vec k_1+\vec k_2)\cdot\vec r}|\phi\rangle\,.
\end{align}
The result is
\begin{align}
H_{17}=&\ \frac{\alpha^2}{m^4}\biggl(\frac{25}{72}-\frac{1}{8\epsilon}-\frac23\ln2+\frac14\ln q\biggr)\,q^2\,.
\end{align}

\subsection{Total middle-energy contribution}
The total result for the middle-energy contribution is the sum
\begin{align} H_M =
\sum_{i=1}^{17} H_i = H^{\rm 2ph}_M  + H^{\rm 3ph}_M\,.
\end{align}
The two-photon part is given by
\begin{eqnarray} \label{eq:2phM}
H^{\rm 2ph}_M &=&\frac{\alpha^2}{m^3}\bigg\{ \frac{19}{30}(\vec{P}_1-\vec{P}_2)^2 - \frac{1}{3}\vec{P}_1\cdot\vec{P}_2 - \frac45 q^2
-\frac73 \frac{(\vec{P}_1\cdot\vec{q})(\vec{P}_2\cdot\vec{q})}{q^2}
-\frac{4}{15}\frac{\bigl((\vec{P}_1-\vec{P}_2)\cdot\vec{q}\,\bigr)^2}{q^2}\nonumber\\
&&+\biggl(-\frac{7}{15}(\vec{P}_1-\vec{P}_2)^2 - \frac{14}{3}\vec{P}_1\cdot\vec{P}_2 -\frac{46}{15}q^2
-\frac43 \frac{\bigl((\vec{P}_1-\vec{P}_2)\cdot\vec{q}\,\bigr)^2}{q^2}\biggr)\ln q\nonumber\\
&&+\,\frac{1}{\epsilon}\biggl[
\frac{7}{30}(\vec{P}_1-\vec{P}_2)^2 + \frac73\vec{P}_1\cdot\vec{P}_2 + \frac{23}{15}q^2
+\frac23 \frac{\bigl((\vec{P}_1-\vec{P}_2)\cdot\vec{q}\,\bigr)^2}{q^2}\biggr]\nonumber\\
&&
-\frac{4}{15}\frac{\bigl((\vec{P}_1-\vec{P}_2)\cdot\vec{q}\,\bigr)^2}{q^2}\,\ln2
+ \biggl(-\frac{7q^2}{36}-\frac{q^2}{24\epsilon} + \frac{q^2}{12}\ln q\biggr)\,\sigma_1\cdot\sigma_2\bigg\}\,,
\end{eqnarray}
whereas the three-photon part is
\begin{eqnarray}
H_M^{\rm 3ph} &=&\frac{\alpha^3}{m^3}\bigg\{
\bigg(-\frac{14}{15}-\frac{3}{5\epsilon}-\frac{4}{15}\ln2+\frac{12}{5}\ln q\bigg)\,q\nonumber\\
&&+\bigg[p_2^i,-\frac{Z}{r_2}\bigg]_\epsilon\,\bigg(-\frac{2}{15}+\frac{14}{15\epsilon}-\frac{28}{15}\ln2
-\frac{28}{15}\ln q\bigg)\frac{q^i}{q^2}\nonumber\\
&&+\bigg[p_2^i,\bigg[-\frac{Z}{r_2},p_2^j\bigg]_\epsilon\bigg]\,
\bigg(\delta^{ij}\bigg(-\frac{8}{15}+\frac{16}{15\epsilon}-\frac{32}{15}\ln q\bigg)
+\frac{q^i q^j}{q^2}\bigg(-\frac43-\frac{4}{15\epsilon}+\frac{8}{15}\ln q\bigg)\bigg)\frac{1}{q^2}\bigg\}\,.
\end{eqnarray}
The result for the two-photon part will be verified in the next section by an independent
calculation of the scattering amplitude.

\section{Scattering amplitude approach}

\begin{figure}[H]
\centering
\includegraphics[scale=0.3]{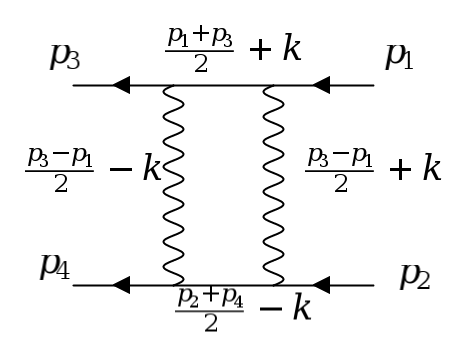}
\includegraphics[scale=0.3]{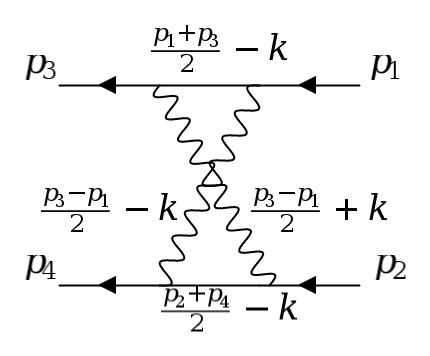}
\caption{Two-photon exchange scattering amplitude. \label{fig:twoph}}
\end{figure}
In this section we apply the scattering amplitude approach to derive the two-photon part of the
middle-energy contribution and the complete high-energy contribution. This part of the derivation
will be performed in the Feynman gauge. In this section we will use the notations that $p_1$ and
$p_3$ are the {\em in} and {\em out} momenta of the first electron, whereas $p_2$ and $p_4$ are
the same for the second electron. We also define $P_1$ and $P_2$ as
\begin{eqnarray}
\vec P_1 =\frac12(\vec p_1+\vec p_3)\ \ \mbox{\rm and}\ \
\vec P_2 =\frac12(\vec p_2+\vec p_4).
\end{eqnarray}
The conservation of the total momentum and the kinetic energy leads to the following conditions,
\begin{eqnarray}
\vec p_1+\vec p_2 &=& \vec p_3+\vec p_4\,,\nonumber\\
\frac{p_1^2}{2}+\frac{p_2^2}{2} &=& \frac{p_3^2}{2}+\frac{p_4^2}{2}\,.
\end{eqnarray}

The two-photon scattering amplitude is graphically represented in Fig.~\ref{fig:twoph} and
defined as
\begin{eqnarray}
\mathcal V &=& -\frac{e^4}{2}\int\frac{d^Dk}{(2\pi)^D\,i}\,\frac{1}{\big(k+\frac{p_3-p_1}{2}\big)^2}\,\frac{1}{\big(k-\frac{p_3-p_1}{2}\big)^2}\nonumber\\
&&
\times
\bigg\{\overline{u}(p_3)\,\gamma^\mu\,\frac{1}{\slashed{k}+\frac{1}{2}\,(\slashed{p}_1+\slashed{p}_3)-m}\,\gamma^\nu\,u(p_1)
+\overline{u}(p_3)\,\gamma^\nu\,\frac{1}{-\slashed{k}+\frac{1}{2}\,(\slashed{p}_1+\slashed{p}_3)-m}\,\gamma^\mu\,u(p_1)
\bigg\}\nonumber\\ && \times
\bigg\{\overline{u}(p_4)\,\gamma^\mu\,\frac{1}{-\slashed{k}+\frac{1}{2}\,(\slashed{p}_2+\slashed{p}_4)-m}\,\gamma^\nu\,u(p_2)
+\overline{u}(p_4)\,\gamma^\nu\,\frac{1}{\slashed{k}+\frac{1}{2}\,(\slashed{p}_2+\slashed{p}_4)-m}\,\gamma^\mu\,u(p_2)\bigg\}\,,
\end{eqnarray}
where $u(p)$ are the free Dirac spinors. We aim to calculate the scattering amplitude in the middle-energy region (where
$\omega\propto m\alpha,\,k\propto m\alpha$) and in the high-energy range (where $\omega\propto
m,\,k\propto m$). External momenta $p_i$ are always of the order $m\alpha$.
To separate out the spin-independent and spin-dependent parts, we make the following
replacements, respectively,
\begin{eqnarray}
u(p_1)\,\overline{u}(p_3)&&\rightarrow\bigg(\frac{\slashed{p}_1+m}{\sqrt{2E_{p_1}(E_{p_1}+m)}}\bigg)
\bigg(\frac{\gamma^0+I}{4}\bigg)\bigg(\frac{\slashed{p}_3+m}{\sqrt{2E_{p_3}(E_{p_3}+m)}}\bigg)\nonumber\\
&&=\bigg(\frac{\slashed{p}_1+m}{2m}\bigg)
\bigg(\frac{\gamma^0+I}{4}\bigg)\bigg(\frac{\slashed{p}_3+m}{2m}\bigg)\bigg(1-\frac38\vec p_1{}^2-\frac38\vec p_3{}^2\bigg)
\end{eqnarray}
and
\begin{eqnarray}
u(p_1)\,\overline{u}(p_3)&&\rightarrow\bigg(\frac{\slashed{p}_1+m}{\sqrt{2E_{p_1}(E_{p_1}+m)}}\bigg)
\bigg(\frac{\gamma^0+I}{2}\bigg)\bigg(\frac{\sigma^{ij}}{4}\bigg)\bigg(\frac{\slashed{p}_3+m}{\sqrt{2E_{p_3}(E_{p_3}+m)}}\bigg)\nonumber\\
&&=\bigg(\frac{\slashed{p}_1+m}{2m}\bigg)
\bigg(\frac{\gamma^0+I}{2}\bigg)\bigg(\frac{\sigma^{ij}}{4}\bigg)\bigg(\frac{\slashed{p}_3+m}{2m}\bigg)\bigg(1-\frac38\vec p_1{}^2-\frac38\vec p_3{}^2\bigg)
\,,
\end{eqnarray}
where $\sigma^{ij} = (i/2)\big[\gamma^i,\gamma^j\big]$ and we used the fact that for
the antisymmetric spin-dependent operator $\hat Q=Q^{ij}\sigma^{ij}$ the following identity holds
\begin{equation}
\hat Q = \frac14\textrm{Tr}[\hat Q\cdot\sigma^{ij}]\,\sigma^{ij}\,.
\end{equation}

For the spin-independent part of the scattering amplitude we get
\begin{eqnarray}
\mathcal V_{nn} &=& -\frac{(4\pi\alpha)^2}{2}\int\frac{d^Dk}{(2\pi)^D\,i}\,\frac{1}{\big(k+\frac{p_3-p_1}{2}\big)^2}\,\frac{1}{\big(k-\frac{p_3-p_1}{2}\big)^2}
\,\bigg(1-\frac38\vec p_1{}^2-\frac38\vec p_2{}^2-\frac38\vec p_3{}^2-\frac38\vec p_4{}^2\bigg)\nonumber\\
&&
\times\,
\textrm{Tr}\bigg[\bigg(\gamma^\mu\,\frac{1}{\slashed{k}+\frac{1}{2}\,(\slashed{p}_1+\slashed{p}_3)-m}\,\gamma^\nu +
\gamma^\nu\,\frac{1}{-\slashed{k}+\frac{1}{2}\,(\slashed{p}_1+\slashed{p}_3)-m}\,\gamma^\mu\bigg)
\bigg(\frac{\slashed{p}_1+m}{2m}\bigg)
\bigg(\frac{\gamma^0+I}{4}\bigg)\bigg(\frac{\slashed{p}_3+m}{2m}\bigg)\bigg]\nonumber\\
&&
\times\,
\textrm{Tr}\bigg[\bigg(\gamma^\mu\,\frac{1}{-\slashed{k}+\frac{1}{2}\,(\slashed{p}_2+\slashed{p}_4)-m}\,\gamma^\nu +
\gamma^\nu\,\frac{1}{\slashed{k}+\frac{1}{2}\,(\slashed{p}_2+\slashed{p}_4)-m}\,\gamma^\mu\bigg)
\bigg(\frac{\slashed{p}_2+m}{2m}\bigg)
\bigg(\frac{\gamma^0+I}{4}\bigg)\bigg(\frac{\slashed{p}_4+m}{2m}\bigg)\bigg]\,.
\end{eqnarray}

The spin-dependent part of the scattering amplitude is
\begin{equation}
\mathcal V_{ss} = \mathcal V_{ss}^{ij,ab}\sigma_1^{ij}\sigma_2^{ab}\,,
\end{equation}
where
\begin{eqnarray}
\mathcal V_{ss}^{ij,ab} &=& \frac{(4\pi\alpha)^2}{2}\int\frac{d^Dk}{(2\pi)^D\,i}\,\frac{1}{\big(k+\frac{p_3-p_1}{2}\big)^2}\,\frac{1}{\big(k-\frac{p_3-p_1}{2}\big)^2}
\,\bigg(1-\frac38\vec p_1{}^2-\frac38\vec p_2{}^2-\frac38\vec p_3{}^2-\frac38\vec p_4{}^2\bigg)\nonumber\\
&&
\times\,
\textrm{Tr}\bigg[\bigg(\gamma^\mu\,\frac{1}{\slashed{k}+\frac{1}{2}\,(\slashed{p}_1+\slashed{p}_3)-m}\,\gamma^\nu +
\gamma^\nu\,\frac{1}{-\slashed{k}+\frac{1}{2}\,(\slashed{p}_1+\slashed{p}_3)-m}\,\gamma^\mu\bigg)
\bigg(\frac{\slashed{p}_1+m}{2m}\bigg)
\bigg(\frac{\gamma^0+I}{2}\bigg)\frac18\big[\gamma^i,\gamma^j\big]\bigg(\frac{\slashed{p}_3+m}{2m}\bigg)\bigg]\nonumber\\
&&
\times\,
\textrm{Tr}\bigg[\bigg(\gamma^\mu\,\frac{1}{-\slashed{k}+\frac{1}{2}\,(\slashed{p}_2+\slashed{p}_4)-m}\,\gamma^\nu +
\gamma^\nu\,\frac{1}{\slashed{k}+\frac{1}{2}\,(\slashed{p}_2+\slashed{p}_4)-m}\,\gamma^\mu\bigg)
\bigg(\frac{\slashed{p}_2+m}{2m}\bigg)
\bigg(\frac{\gamma^0+I}{2}\bigg)\frac18\big[\gamma^a,\gamma^b\big]\bigg(\frac{\slashed{p}_4+m}{2m}\bigg)\bigg]
\,.\nonumber\\
\end{eqnarray}
We now perform traces and then rescale the photon and fermion momenta for either middle-energy or
high-energy contributions. Then we expand the expression in $\alpha$ and pick up the contribution
of the order $\alpha^7$. After performing the trivial $\omega$ integration, the remaining
integrals are handled using formulas from Appendix \ref{app:C}. After the angular
averaging using relations from Appendix \ref{app:B}, we obtain the following result
in the momentum representation for the middle-energy contribution,
\begin{eqnarray}
H_M &=& \frac{\alpha^2}{m^4}\bigg\{\biggl(-\frac{7q^2}{36}-\frac{q^2}{24\epsilon} + \frac{q^2}{12}\ln q\biggr)\,\sigma_1\cdot\sigma_2
+\frac{19}{30}(\vec{P}_1-\vec{P}_2)^2 - \frac{1}{3}\vec{P}_1\cdot\vec{P}_2 - \frac45 q^2
-\frac73 \frac{(\vec{P}_1\cdot\vec{q})(\vec{P}_2\cdot\vec{q})}{q^2}\nonumber\\
&&+\biggl(-\frac{7}{15}(\vec{P}_1-\vec{P}_2)^2 - \frac{14}{3}\vec{P}_1\cdot\vec{P}_2 -\frac{46}{15}q^2
\biggr)\bigg(-\frac{1}{2\epsilon}+\ln q\bigg)\bigg\}
\,.
\end{eqnarray}
This result agrees with the formula (\ref{eq:2phM}) obtained within the NRQED approach in the
previous section, with the exception of terms proportional to
$\bigl((\vec{P}_1-\vec{P}_2)\cdot\vec{q}\,\bigr)^2/q^2$. This is because
$(\vec{P}_1-\vec{P}_2)\cdot\vec{q}=\frac12(p_3^2+p_4^2-p_1^2-p_2^2) = 0$ for the scattering
amplitude due to the conservation of the kinetic energy. Therefore, the corresponding
contribution cannot be derived from the two-photon amplitude. This contribution can be in
principle obtained from the three-photon scattering amplitude, by transforming it into the
three-photon exchange form by using the Schr\"odinger equation, as shown in the next section.
However, the calculation of the middle-energy part from the three-photon exchange scattering
amplitude was too complicated for us to be pursued, in contrast to the NRQED approach, where the
corresponding derivation is relatively simple.

The result the high-energy contribution is
\begin{eqnarray}
H_H &=&\frac{\alpha^2}{m^4}\bigg\{
-\frac{56}{45}(\vec{P}_1-\vec{P}_2)^2-\frac23\vec{P}_1\cdot\vec{P}_2 + \frac{13}{10}q^2
+\frac{1}{\epsilon}\biggl[\frac{1}{6}(\vec{P}_1-\vec{P}_2)^2 - \vec{P}_1\cdot\vec{P}_2 -q^2\biggr]\nonumber\\
&&+ \biggl(\frac{11}{54}(\vec{P}_1-\vec{P}_2)^2 +
\frac12\vec{P}_1\cdot\vec{P}_2 + \frac{7}{36}q^2 + \frac{11}{72\epsilon}q^2\biggr)\,\sigma_1\cdot\sigma_2\bigg\}
\,,
\end{eqnarray}
and there is no three-photon high-energy part.

\section{Total result in momentum space}

The total result for $H_{\rm exch}^{(7)}$ is a sum of the low-energy, middle-energy and
high-energy contributions, see Eq.~(\ref{eq:2}). In order to obtain the total result, we perform
a transformation of the two-photon term $\langle[(\vec P_1-\vec P_2)\cdot\vec q]^2\,f(\vec
q)\rangle$, which is in coordinate representation
\begin{align}
  \biggl\langle \frac{1}{4}\bigl[p_1^2+p_2^2\,,\,\bigl[p_1^2+p_2^2\,,\,f(\vec r)\bigr]\bigr]\biggr\rangle =
  -\biggl\langle \frac{1}{2}\bigl[V\,,\,\bigl[p_1^2+p_2^2\,,\,f(\vec r)\bigr]\bigr]\biggr\rangle =
  -\bigl\langle\bigl[V\,,\,\vec p_1\bigr]\,\bigl[\vec p_1\,,\,f(\vec r)\bigr]\bigr\rangle +
  (1\leftrightarrow2)\,
\end{align}
and thus can be rewritten as a  three-photon contribution. The two-body part of this contribution
is further transformed using the integration formula with $\ln q$ in Eq.  (\ref{C23}) from
Appendix \ref{app:C}. Finally, we obtain
\begin{eqnarray} H_{\rm exch}^{(7)} &=& \frac{\alpha^2}{m^4}\bigg\{\,\sigma_1\cdot\sigma_2
\biggl(-\frac{7}{27}q^2+\frac{11}{54}(\vec{P}_1-\vec{P}_2)^2 + \frac12\vec{P}_1\cdot\vec{P}_2
-\frac29q^2\ln[\alpha^{-2}]+\frac29q^2\ln(2\lambda)+\frac{1}{12}q^2\ln q \biggr)\nonumber\\
&&
+\bigg(-\frac{527}{450}-\frac{4}{5}\ln[\alpha^{-2}]
+\frac{4}{5}\ln(2\lambda)-\frac{7}{15}\ln q\bigg)(\vec P_1-\vec P_2)^2\nonumber\\
&&+\bigg(-\frac{151}{450}-\frac{16}{15}\ln[\alpha^{-2}]
+\frac{16}{15}\ln(2\lambda)-\frac{46}{15}\ln q\bigg)q^2\nonumber\\
&&+\bigg(-\frac{109}{45}-\frac{8}{3}\ln[\alpha^{-2}]
+\frac{8}{3}\ln(2\lambda)-\frac{14}{3}\ln q\bigg)\vec P_1\cdot\vec P_2
-\frac{31}{15}\frac{(\vec P_1\cdot\vec q)(\vec P_2\cdot\vec q)}{q^2}\bigg\}\nonumber\\
&&+ \frac{\pi\alpha^3}{m^3}\bigg(-\frac{161}{225}+\frac{8}{15}\ln[\alpha^{-2}]
-\frac{8}{15}\ln(2\lambda)+\frac{4}{5}\ln2+\frac{8}{15}\ln q\bigg)\,q\nonumber\\
&&+ \frac{\alpha^3}{m^3}\bigg\{\langle\phi|\bigg[p_2^j,-\frac{Z}{r_2}\bigg]\bigg(\frac{26}{225}
-\frac{8}{15}\ln[\alpha^{-2}]
+\frac{8}{15}\ln(2\lambda)-\frac85\ln2-\frac{8}{15}\ln q\bigg)\frac{q^j}{q^2}\nonumber\\
&&+\bigg[\bigg[p_2^i,-\frac{Z}{r_2}\bigg],p_2^j\bigg]\bigg[\delta^{ij}
\bigg(-\frac{436}{225}-\frac{32}{15}\ln[\alpha^{-2}]
+\frac{32}{15}\ln(2\lambda)-\frac{32}{15}\ln q\bigg)\nonumber\\&&
+\frac{q^i q^j}{q^2}\bigg(-\frac{236}{225}+\frac{8}{15}\ln[\alpha^{-2}]
-\frac{8}{15}\ln(2\lambda)+\frac{8}{15}\ln q\bigg)\bigg]\frac{1}{q^2}|\phi\rangle
+(1\leftrightarrow2)\bigg\}
\,.
\end{eqnarray}

In order to make the transformation of the above expression into coordinate space more
accessible, we need to express the momenta $\vec P_1$ and $\vec P_2$ in terms $\vec P$, $\vec p$,
and $\vec q$ defined as
\begin{align}
\vec P =&\ \vec p_1+\vec p_2\,, \ \
\vec p = \frac12\big(\vec p_1-\vec p_2\big)\,, \ \ \mbox{\rm and} \ \ \vec{q} = \vec p_1{}' -\vec p_1\,.
\end{align}
We thus have
\begin{eqnarray}
\big(\vec P_1-\vec P_2\big)^2
&=&q^2+4\, \vec p\cdot\vec p \,{}'\,,
\end{eqnarray}
\begin{eqnarray}
\vec P_1\cdot\vec P_2
&=& \frac14\bigg(P^2-q^2-4\, \vec p\cdot\vec p \,{}'\bigg)\,,
\end{eqnarray}
\begin{eqnarray}
\frac{(\vec P_1\cdot\vec q)(\vec P_2\cdot\vec q)}{q^2}
&=&\bigg(\frac14P^iP^j-p^i p'^j\bigg)\frac{q^i q^j-\frac{\delta^{ij}}{3}q^2}{q^2}
-\frac14q^2+\frac{1}{12}P^2-\frac13\,\vec p\cdot\vec p \,{}'\,.
\end{eqnarray}
We also use $\sigma_1\cdot\sigma_2 = 2\,\vec\sigma_1\cdot\vec\sigma_2$, which is valid in $d = 3$,
and neglect operators $P^2\,\delta^3(r)$ and  $p^2\,\delta^3(r)$, which vanish for triplet
states. Furthermore, for triplet states we can assume $\vec\sigma_1\cdot\vec\sigma_2=1$.
The total result for $H_{\rm exch}^{(7)}$ in momentum space after these transformations becomes
\begin{eqnarray}
H^{(7)}_{\rm exch}
&=& \frac{\alpha^2}{m^4}\bigg\{
\frac{14}{5}\vec p\cdot\vec p \,{}'\,\ln q
-\frac76\,P^2\ln q
-\frac{31}{60}\big(P^iP^j-4p^i p'^j\big)\,\biggl(q^i q^j-\frac{\delta^{ij}}{3}q^2\biggr)\frac{1}{q^2}\nonumber\\
&&+\bigg(-\frac{739}{2700}-\frac{62}{45}\ln[\alpha^{-2}]
+\frac{62}{45}\ln(2\lambda)-\frac{11}{5}\ln q\bigg)q^2\bigg\}
\nonumber\\
&&
+\frac{\pi\alpha^3}{m^3}\bigg(-\frac{161}{225}+\frac{8}{15}\ln[\alpha^{-2}]
-\frac{8}{15}\ln(2\lambda)+\frac{4}{5}\ln2+\frac{8}{15}\ln q\bigg)\,q\nonumber\\
&&+ \frac{\alpha^3}{m^3}\bigg\{\bigg[p_2^j,-\frac{Z}{r_2}\bigg]\bigg(\frac{26}{225}
-\frac{8}{15}\ln[\alpha^{-2}]
+\frac{8}{15}\ln(2\lambda)-\frac85\ln2-\frac{8}{15}\ln q\bigg)\frac{q^j}{q^2}\nonumber\\
&&+\bigg[\bigg[p_2^i,-\frac{Z}{r_2}\bigg],p_2^j\bigg]\bigg[\delta^{ij}
\bigg(-\frac{436}{225}-\frac{32}{15}\ln[\alpha^{-2}]
+\frac{32}{15}\ln(2\lambda)-\frac{32}{15}\ln q\bigg)\nonumber\\&&
+\frac{q^i q^j}{q^2}\bigg(-\frac{236}{225}+\frac{8}{15}\ln[\alpha^{-2}]
-\frac{8}{15}\ln(2\lambda)+\frac{8}{15}\ln q\bigg)\bigg]\frac{1}{q^2}
+(1\leftrightarrow2)\bigg\}\,.
\end{eqnarray}
In spite of quite lengthy calculations, this result is relatively simple. We expect that it will
be further simplified after adding the radiative contribution.

\section{Total result in coordinate space}

We now transform the obtained formulas for $H_{\rm exch}^{(7)}$ into the coordinate
representation in atomic units, which is needed for the numerical evaluation of the matrix
element with the nonrelativistic wave function. All momenta are rescaled $p \rightarrow
\alpha\,p$, so $\ln q\rightarrow\ln\alpha+\ln q$, and the overall factor $\alpha^7$ is pulled out
of $H^{(7)}_{\rm exch}$. In order to perform the Fourier transform, we start with the master
integral formula in $d$ dimensions,
\begin{eqnarray}
\int\frac{d^dq}{(2\pi)^d}\,q^m\,e^{i\vec{q}\cdot\vec{r}} &=&
\frac{\pi ^{-d/2} 2^m \left(r^2\right)^{\frac{1}{2} (-d-m)} \Gamma
   \left(\frac{d+m}{2}\right)}{\Gamma \left(-\frac{m}{2}\right)}\,.
\end{eqnarray}
From the above formula, we derive the following results for the $d = 3$ Fourier transforms of
various operators,
\begin{align}
\int\frac{d^3q}{(2\pi)^3} \,e^{i\vec q\cdot\vec r}\,\frac{4\pi}{q^2}
=&\,\frac{1}{r}\,,\\
\int\frac{d^3q}{(2\pi)^3} \,e^{i\vec q\cdot\vec r}\,\frac{4\pi}{q^2}\,\ln q
=&\,-\frac{\gamma+\ln r}{r}\,,\\
\int\frac{d^3q}{(2\pi)^3} \,e^{i\vec q\cdot\vec r}\,4\pi\frac{q^i}{q^2}
=&\,i\frac{r^i}{r^3}\,,\\
\int\frac{d^3q}{(2\pi)^3} \,e^{i\vec q\cdot\vec r}\,4\pi\frac{q^i}{q^2}\,\ln q
=&\,i\frac{r^i}{r^3}(1-\gamma-\ln r)\,,\\
\int\frac{d^3q}{(2\pi)^3} \,e^{i\vec q\cdot\vec r}\,4\pi \frac{q^i q^j-\frac{\delta^{ij}}{3}q^2}{q^2}
=&\,\frac{\delta^{ij}r^2-3r^i r^j}{r^5}\,,\\
\int\frac{d^3q}{(2\pi)^3} \,e^{i\vec q\cdot\vec r}\,4\pi\,\frac{q^i q^j}{q^4}
=&\,\frac{1}{2r^3}\big(\delta^{ij}r^2-r^ir^j\big)\,,\\
\int\frac{d^3q}{(2\pi)^3} \,e^{i\vec q\cdot\vec r}\,4\pi\,\frac{q^i q^j}{q^4}\,\ln q
=&\,-\nabla^i\nabla^j
\bigg[\frac{r}{4}\big(-3+2\gamma+2\ln r\big)\bigg]\nonumber \\
=&\,\frac14\bigg[\frac{\delta^{ij}r^2-3r^i r^j}{r^3}-\frac{\big(\delta^{ij} r^2-r^i r^j\big)}{r^3}
\big(2\gamma+2\ln r\big)\bigg]\,.
\end{align}
The right-hand sides of the above formulas for the Fourier transforms are well-defined functions.
However, there are some operators in $H^{(7)}_{\rm exch}$ that require more careful treatment.
We now take into account that the matrix element of $H^{(7)}_{\rm exch}$ is calculated with the
antisymmetric wave function, which satisfies the condition $\phi(\vec r_1,\vec r_2) = -\phi(\vec
r_2,\vec r_1)$. Therefore, the wave function behaves as $\phi(\vec r) \sim \vec r\cdot(\vec
r_1+\vec r_2)$ for small $|\vec r|$. Under this condition, matrix elements of all operators in
$H^{(7)}_{\rm exch}$ in momentum representation are finite and well defined. Their transformation
in the coordinate space, however, may require special definitions.

We now define the coordinate-space representation of all singular operators in $H^{(7)}_{\rm
exch}$. We start with the well-known operator $(1/r^3)_\varepsilon$, which is defined by the
integral with an arbitrary smooth function $f(\vec r)$ as follows
\begin{align}
\int d^3r\,f(\vec r)\,\Bigl(\frac{1}{r^3}\Bigr)_\varepsilon = \lim_{\varepsilon\rightarrow0}
\int d^3r\,f(\vec r)\,\bigg[\frac{1}{r^3}\theta(r-\varepsilon)+
4\pi \delta^3(r)(\gamma+\ln\varepsilon)\bigg]\,.
\end{align}
The Dirac delta function in the above definition appears because the $(1/r^3)_\varepsilon$
operator is assumed to be sandwiched between two momenta operators, $\vec
p\,(1/r^3)_\varepsilon\,\vec p$. The Fourier transform of $(1/r^3)_\varepsilon$ is evaluated as
\begin{eqnarray}
\int d^3r\,e^{i\vec q\cdot\vec r}\,\Bigl(\frac{1}{r^3}\Bigr)_\varepsilon  &=& \lim_{\varepsilon\rightarrow0}
\int d^3r\,e^{i\vec q\cdot\vec r}\bigg[\frac{1}{r^3}\theta(r-\varepsilon)+
4\pi \delta^3(r)(\gamma+\ln\varepsilon)\bigg]\nonumber\\
&=& \lim_{\varepsilon\rightarrow0}
2\pi\int_{-1}^1 dx\int_\varepsilon^\infty\frac{dr}{r}\,e^{i q r x} + 4\pi(\gamma+\ln\varepsilon)
= 4\pi(1-\ln q)\,.
\end{eqnarray}
The operator $(1/r^4)_\varepsilon$ is defined by
\begin{align}
\int d^3r\,f(\vec r)\,\Bigl(\frac{1}{r^4}\Bigr)_\varepsilon = \lim_{\varepsilon\rightarrow0}
\int d^3r\,f(\vec r)\,\bigg[\frac{1}{r^4}\theta(r-\varepsilon)-
4\pi \delta^3(r)\,\frac{1}{\varepsilon}\bigg]\,.
\end{align}
Its Fourier transform is
\begin{eqnarray}
\int d^3r\,e^{i\vec q\cdot\vec r}\,\Bigl(\frac{1}{r^4}\Bigr)_\varepsilon  &=& \lim_{\varepsilon\rightarrow0}
\int d^3r\,e^{i\vec q\cdot\vec r}\bigg[\frac{1}{r^4}\theta(r-\varepsilon)-
4\pi \delta^3(r)\,\frac{1}{\varepsilon}\bigg]\nonumber\\
&=& \lim_{\varepsilon\rightarrow0}
2\pi\int_{-1}^1 dx\int_\varepsilon^\infty\frac{dr}{r^2}\,e^{i q r x} - \frac{4\pi}{\varepsilon}
=-\pi^2\,q\,.
\end{eqnarray}
Similarly, $(\ln r/r^4)_\varepsilon$ is defined by
\begin{align}
\int d^3r\,f(\vec r)\,\Bigl(\frac{\ln r}{r^4}\Bigr)_\varepsilon = \lim_{\varepsilon\rightarrow0}
\int d^3r\,f(\vec r)\,\bigg[\frac{\ln r}{r^4}\theta(r-\varepsilon)-
4\pi \delta^3(r)\,\frac{1+\ln\varepsilon}{\varepsilon}\bigg]\,.
\end{align}
Its Fourier transform is
\begin{eqnarray}
\int d^3r\,e^{i\vec q\cdot\vec r}\,\Bigl(\frac{\ln r}{r^4}\Bigr)_\varepsilon &=& \lim_{\varepsilon\rightarrow0}
\int d^3r\,e^{i\vec q\cdot\vec r}\bigg[\frac{\ln r}{r^4}\,\theta(r-\varepsilon)-
4\pi \delta^3(r)\,\frac{1+\ln\varepsilon}{\varepsilon}\bigg]\nonumber\\
&=& \lim_{\varepsilon\rightarrow0}
2\pi\int_{-1}^1 dx\int_\varepsilon^\infty\frac{dr}{r^2}\ln r\,e^{i q r x} - 4\pi\,\frac{1+\ln\varepsilon}{\varepsilon}
= \pi^2\bigg(-\frac32+\gamma+\ln q\bigg)\,q\,.
\end{eqnarray}
The operator $(1/r^5)_\varepsilon$ is defined by
\begin{align}
\int d^3r\,f(\vec r)\,\Bigl(\frac{1}{r^5}\Bigr)_\varepsilon = \lim_{\varepsilon\rightarrow0}
\int d^3r\,f(\vec r)\, \bigg[\frac{1}{r^5}\,\theta(r-\varepsilon)-
2\pi \delta^3(r)\,\frac{1}{\varepsilon^2}
+\frac23\pi \nabla^2\delta^3(r)\,(\gamma+\ln\varepsilon)\bigg]\,,
\end{align}
with the corresponding Fourier transform evaluated as
\begin{eqnarray}
\int d^3r\,e^{i\vec q\cdot\vec r}\,\Bigl(\frac{1}{r^5}\Bigr)_\varepsilon &=& \lim_{\varepsilon\rightarrow0}
\int d^3r\,e^{i\vec q\cdot\vec r}\bigg[\frac{1}{r^5}\,\theta(r-\varepsilon)-
2\pi \delta^3(r)\,\frac{1}{\varepsilon^2}
+\frac23\pi \nabla^2\delta^3(r)\,(\gamma+\ln\varepsilon)\bigg]\nonumber\\
&=& 2\pi\int_{-1}^1 dx\int_\varepsilon^\infty\frac{dr}{r^3}\,e^{i q r x} - 2\pi\,\frac{1}{\varepsilon^2}
-\frac23\pi\big(\gamma+\ln\varepsilon)\,q^2
= \pi\bigg(-\frac{11}{9}+\frac23\ln q\bigg)\,q^2\,.
\end{eqnarray}
Using the above formulas,
the final result for $H^{(7)}_{\rm exch}$ in the coordinate representations
is obtained as
\begin{eqnarray}\label{eq:H7exch}
H^{(7)}_{\rm exch} &=&
\bigg(\frac{15\,409}{1350}
-\frac{124}{45}\ln(2\lambda)+\frac{76}{45}\ln\alpha\bigg)\,\vec p\,\delta^{(3)}(r)\,\vec p
-\frac{33}{10\pi\,r^5}\nonumber\\
&&
-\frac{7}{10\pi}\,\vec p\,\frac{1}{r^3}\,\vec p
+\frac{7}{24\pi}P^2\,\frac{1}{r^3}
+\frac{31}{60\pi}\,p^i\,\frac{\delta^{ij}r^2-3r^ir^j}{r^5}\,p^j
-\frac{31}{240\pi}\,P^iP^j\,\frac{\delta^{ij}r^2-3r^ir^j}{r^5}\nonumber\\
&&+\frac{8}{15\pi\, r^4}\bigg(-\frac{19}{120}
-\frac32\ln2+\ln(2\lambda)+\ln\alpha+\gamma+\ln r\bigg)\nonumber\\
&&+\bigg\{-\frac{2\,Z}{15\pi}\frac{\vec r_1\cdot\vec r}{r_1^3r^3}
\bigg(-\frac{47}{60}
-3\ln2+\ln(2\lambda)+\ln\alpha+\gamma+\ln r\bigg)\nonumber\\
&&+\frac{Z}{15\pi}\frac{\big(\delta^{ij}r_1^2-3r_1^ir_1^j\big)r^i r^j}{ r_1^5r^3}
\bigg(\frac{7}{15}
+\ln(2\lambda)+\ln\alpha+\gamma+\ln r\bigg)\nonumber\\
&&+\frac{88}{45}Z\,\delta^{(3)}(r_1)\frac{1}{r}
\bigg(-\frac{193}{165}
+\ln(2\lambda)+\ln\alpha+\gamma+\ln r\bigg)
+(1\leftrightarrow2)\bigg\}\,,
\end{eqnarray}
\end{widetext}
where $\vec p = (\vec p_1-\vec p_2)/2$, $\vec P = \vec p_1+\vec p_2$. Additional terms with $\ln\alpha$ are
obtained from the transformation into atomic units $\ln q\rightarrow\ln q+\ln\alpha$. We stress
that this form of $H^{(7)}_{\rm exch}$ is valid only for antisymmetric states.
Eq.~(\ref{eq:H7exch}) is the main result of this work. Despite the very complicated derivation,
the final result for $H^{(7)}_{\rm exch}$ has a very simple form consisting of just a few
operators.

\section{Summary}

In this work we calculated the QED effects of the order $\alpha^7m$ to the Lamb shift, which
originate from the virtual photon exchange between the electrons and the nucleus and are
represented by two- and three-photon exchange diagrams. The central problem was the derivation of
the effective Hamiltonian $H^{(7)}_{\rm exch}$ in coordinate space. The corresponding result
valid for the antisymmetric (triplet) states of two-electron atoms is given by
Eq.~(\ref{eq:H7exch}). The obtained expression is free from any singularities (provided that the
expectation value is calculated with wave functions of antisymmetric states). The final
expression was obtained after delicate cancellations of numerous divergences present in
individual operators in momentum space. It is finite but still depends on the
photon-momentum cutoff parameter $\lambda$. This dependence will disappear when $H^{(7)}_{\rm
exch}$ is combined together with the radiative and low-energy contributions in Eq.~(\ref{eq:1}).

The numerical evaluation of the expectation value of $H^{(7)}_{\rm exch}$ would be
straightforward, because many operators have already been encountered in our previous studies
\cite{patkos:16:triplet,patkos:17:singlet} and other operators are of a similar complexity.
However, the numerical value of $\langle H^{(7)}_{\rm exch}\rangle$ would not be useful for a
comparison with experimental results, because of dependence on the cutoff parameter $\lambda$.
For this reason we have postponed the numerical evaluation until the radiative corrections are
calculated, which we plan to accomplish in the forthcoming investigation.

Calculations of the remaining part, namely the radiative $\alpha^7m$ effects,  will be greatly
simplified by the fact that these effects are known for the hydrogenic atoms. Specifically, the
one-loop $\alpha^7\,m$ correction (the so-called $A_{60}$ coefficient) was derived in
Ref.~\cite{pachucki:93}. The corresponding two-loop contribution (the so-called $B_{50}$
coefficient) was calculated in Refs.~\cite{pachucki:94,eides:95:pra}, whereas the three-loop
contribution $C_{40}$ was completed in Ref.~\cite{melnikov:00}. The two- and three-loop
corrections are proportional to the electron charge density on the nucleus, so only the one-loop
radiative contribution needs to be re-derived for the helium atom.

In the present work we performed our derivation for the triplet states of helium only. The
restriction to these states was made because their wave function is antisymmetric with
respect to exchange of spatial electron coordinates and vanishes at $\vec r_1=\vec r_2$, which
greatly simplifies the derivation.  An extension of the present derivation to the singlet states
of helium might be possible but would involve a calculation of four-photon exchange diagrams,
the feasibility of which is not clear at present. Finally, the derivation of this work can be
extended to many other bound systems, particularly to positronium, where a calculation of the
$\alpha^7\,m$ QED effects has been required for a long time but has not yet been accomplished.

\begin{acknowledgments}
K.P. and V.P. acknowledge support from the National Science Center (Poland) Grant No.
2017/27/B/ST2/02459, additionally V.P. acknowledges support from the Czech Science Foundation
- GA\v{C}R (Grant No. P209/18-00918S). Work of V.A.Y. was supported by the Russian Science
Foundation (Grant No. 20-62-46006).
\end{acknowledgments}


\appendix

\begin{widetext}

\section{Foldy-Wouthuysen transformation} \label{app:A}
The discussion in this section is  based on our previous work \cite{fw}. The Foldy-Wouthuysen
(FW) transformation is a widely used method to derive the nonrelativistic expansion of the Dirac
Hamiltonian in an external electromagnetic field,
\begin{equation}
H = \vec \alpha \cdot \vec \pi +\beta\,m + e\,A^0\,, \label{06}
\end{equation}
where $\vec \pi = \vec p-e\,\vec A$. The idea of the FW transformation is to apply a unitary
transformation to the Dirac Hamiltonian that decouples the upper and lower components of the
Dirac wave function up to a specified order in the $1/m$ expansion. The result is the FW
Hamiltonian $H_{\rm FW}$ defined as
\begin{equation}
H_{\rm FW} = e^{i\,S}\,(H-i\,\partial_t)\,e^{-i\,S}\,, \label{07}
\end{equation}
where $S$ is the operator that needs to be determined. In the this work we calculate the FW
Hamiltonian up to terms that contribute to the order $\alpha^7\,m$ to the energy.

The choice of the unitary transformation operator $S$, and therefore the resulting FW
Hamiltonian, is not unique. We use an approach that is somewhat different from the one described
in standard textbooks. Specifically, we use the single operator $S$ defined as
\begin{eqnarray}
S &=&-\frac{i}{2\,m}\,\Big[ \beta\,\vec\alpha\cdot\vec\pi -
\frac{1}{3\,m^2}\,\beta\,(\vec\alpha\cdot\vec\pi)^3
+\frac{i\,e}{2\,m}\,\vec\alpha\cdot\vec E
-\frac{\beta\,e}{4\,m^2}\,\vec\alpha\cdot\dot{\vec E}
+Y\Big]\,,\label{08}
\end{eqnarray}
where $Y$ is some odd operator $\{\beta,Y\}=0$ which satisfies the condition
$[Y,e\,A^0-i\,\partial_t]\approx[Y,(\vec\alpha\cdot\vec\pi)^3]\approx 0$.  We will fix the
explicit form of $Y$ in the very end; this choice will allow us to cancel the unwanted
higher-order odd terms. The FW Hamiltonian is expanded in a power series in $S$
\begin{equation}
H_{\rm FW} = \sum_{j=0}^6 {\cal H}^{(j)}+\ldots\,, \label{09}
\end{equation}
where
\begin{eqnarray}
{\cal H}^{(0)} = H\,, \ \ \
{\cal H}^{(1)} = [i\,S\,,{\cal H}^{(0)}-i\,\partial_t]\,,\ \ \
{\cal H}^{(j)} =  \frac{1}{j}\,[i\,S\,,{\cal H}^{(j-1)}]\; {\mbox{\rm for $j>1$}},
\label{10}
\end{eqnarray}
and terms  with $j > 6$ are neglected. The calculations of subsequent commutators is
straightforward but rather tedious. For the reader's convenience we present separate results for
each $H^{(j)}$,
\begin{eqnarray}
{\cal H}^{(1)} &=& \frac{\beta}{m}\,(\vec\alpha\cdot\vec\pi)^2-
            \frac{\beta}{3\,m^3}\,(\vec\alpha\cdot\vec\pi)^4
            -\frac{i\,e}{4\,m^2}[\vec\alpha\cdot\vec\pi,\vec\alpha\cdot\vec E]
            -\frac{e\,\beta}{8\,m^3}\{\vec\alpha\cdot\vec\pi\,,\,\vec\alpha\cdot\dot{\vec E}\}
            +\frac{1}{2\,m}[Y,\vec\alpha\cdot\vec\pi]\nonumber \\ &&
            -\vec\alpha\cdot\vec\pi +
            \frac{1}{3\,m^2}\,(\vec\alpha\cdot\vec\pi)^3 - \beta\,Y
            -\frac{\beta}{6\,m^3}[(\vec\alpha\cdot\vec\pi)^3,e\,A^0-i\,\partial_t]
            -\frac{i\,\beta\,e}{8\,m^2}\,\vec\alpha\cdot\ddot{\vec E}\,,\label{11}\\
{\cal H}^{(2)} &=& -\frac{\beta}{2\,m}\,(\vec\alpha\cdot\vec\pi)^2+
            \frac{\beta}{3\,m^3}\,(\vec\alpha\cdot\vec\pi)^4-
            \frac{\beta}{18\,m^5}\,(\vec\alpha\cdot\vec\pi)^6
            +\frac{i\,e}{8\,m^2}\,[\vec\alpha\cdot\vec\pi,\vec\alpha\cdot\vec
            E]\nonumber \\ &&
            -\frac{1}{2\,m}\,[Y,\vec\alpha\cdot\vec\pi]
            -\frac{i\,e}{24\,m^4}\,[(\vec\alpha\cdot\vec\pi)^3,\vec\alpha\cdot\vec E]
            +\frac{i\,e}{32\,m^3}\,[\vec\alpha\cdot\vec\pi\,,\,\vec\alpha\cdot\ddot{\vec E}]
            \nonumber \\ && +\frac{1}{24\,m^4}[\vec\alpha\cdot\vec\pi,
            [(\vec\alpha\cdot\vec\pi)^3,e\,A^0-i\,\partial_t]]
            +\frac{e\,\beta}{16\,m^3}\,\bigl\{\vec\alpha\cdot\vec\pi\,,\,
            \vec\alpha\cdot\dot{\vec E}\bigr\}\nonumber \\ &&
            -\frac{1}{2\,m^2}\,(\vec\alpha\cdot\vec\pi)^3+
            \frac{1}{3\,m^4}\,(\vec\alpha\cdot\vec\pi)^5-
            \frac{i\,\beta\,e}{16\,m^3}\,[\vec\alpha\cdot\vec\pi,
            [\vec\alpha\cdot\vec\pi,\vec\alpha\cdot\vec E]]\nonumber \\ &&
            -\frac{i\,\beta\,e}{8\,m^3}\,\bigl((\vec\alpha\cdot\vec\pi)^2\,
            \vec\alpha\cdot\vec E + \vec\alpha\cdot\vec E\,
            (\vec\alpha\cdot\vec\pi)^2\bigr)\label{12}\,,\\
{\cal H}^{(3)} &=& -\frac{\beta}{6\,m^3}\,(\vec\alpha\cdot\vec\pi)^4+
            \frac{\beta}{6\,m^5}\,(\vec\alpha\cdot\vec\pi)^6+
            \frac{i\,e}{96\,m^4}\,[\vec\alpha\cdot\vec\pi,[\vec\alpha\cdot\vec\pi,
            [\vec\alpha\cdot\vec\pi,\vec\alpha\cdot\vec E]]]\nonumber \\ &&
            +\frac{i\,e}{48\,m^4}\,[\vec\alpha\cdot\vec\pi,(\vec\alpha\cdot\vec\pi)^2\,
            \vec\alpha\cdot\vec E + \vec\alpha\cdot\vec
            E\,(\vec\alpha\cdot\vec\pi)^2]
            +\frac{i\,e}{24\,m^4}\,[(\vec\alpha\cdot\vec\pi)^3,
            \vec\alpha\cdot\vec E]\nonumber \\ &&
            +\frac{1}{6\,m^2}\,(\vec\alpha\cdot\vec\pi)^3-\frac{1}{6\,m^4}\,
            (\vec\alpha\cdot\vec\pi)^5+
            \frac{i\,\beta\,e}{48\,m^3}\,[\vec\alpha\cdot\vec\pi,
            [\vec\alpha\cdot\vec\pi,\vec\alpha\cdot\vec E]]\nonumber \\ &&
            +\frac{i\,\beta\,e}{24\,m^3}\,\bigl((\vec\alpha\cdot\vec\pi)^3\,
            \vec\alpha\cdot\vec E + \vec\alpha\cdot\vec
            E\,(\vec\alpha\cdot\vec\pi)^3\bigr)\label{13}\,,\\
{\cal H}^{(4)} &=& \frac{\beta}{24\,m^3}\,(\vec\alpha\cdot\vec\pi)^4-
            \frac{\beta}{18\,m^5}\,(\vec\alpha\cdot\vec\pi)^6-
            \frac{i\,e}{384}\,[\vec\alpha\cdot\vec\pi,[\vec\alpha\cdot\vec\pi,
            [\vec\alpha\cdot\vec\pi,\vec\alpha\cdot\vec E]]]\nonumber \\ &&
            -\frac{i\,e}{192\,m^4}\,[\vec\alpha\cdot\vec\pi,(\vec\alpha\cdot\vec\pi)^2\,
            \vec\alpha\cdot\vec E + \vec\alpha\cdot\vec E\,(\vec\alpha\cdot\vec\pi)^2]-
            \frac{i\,e}{96\,m^4}\,[(\vec\alpha\cdot\vec\pi)^3,\vec\alpha\cdot\vec
            E] \nonumber \\ && +\frac{1}{24\,m^4}\,(\vec\alpha\cdot\vec\pi)^5
            \label{14}\,,\\
{\cal H}^{(5)} &=& -\frac{1}{120\,m^4}\,(\vec\alpha\cdot\vec\pi)^5+
            \frac{\beta}{120\,m^5}\,(\vec\alpha\cdot\vec\pi)^6\label{15}\,,\\
{\cal H}^{(6)} &=& -\frac{\beta}{720\,m^5}\,(\vec\alpha\cdot\vec\pi)^6 \label{16}\,.
\end{eqnarray}

At this stage the operator $H_{\rm FW}$ still depends on $Y$. Following the idea of the FW
transformation, $Y$ is now chosen to cancel all the higher order odd terms from $H_{\rm FW}$,
\begin{eqnarray}
Y &=&  \frac{\beta}{5\,m^4}\,(\vec\alpha\cdot\vec\pi)^5+
     \frac{i\,e}{24\,m^3}\,[\vec\alpha\cdot\vec\pi,
            [\vec\alpha\cdot\vec\pi,\vec\alpha\cdot\vec E]]
     -\frac{i\,e}{3\,m^3}\,\Big((\vec\alpha\cdot\vec\pi)^2\,
            \vec\alpha\cdot\vec E + \vec\alpha\cdot\vec
     E\,(\vec\alpha\cdot\vec\pi)^2\Big). \label{17}
\end{eqnarray}
It can be seen that $Y$ fulfills the condition that commutators $[Y,e\,A^0-i\,\partial_t]$ and
$[Y,(\vec\alpha\cdot\vec\pi)^3]$ are of higher orders and thus can be neglected. The resulting FW
Hamiltonian is
\begin{eqnarray}
H_{\rm FW} &=& e\,A^0 + \frac{(\vec\sigma\cdot\vec\pi)^2}{2\,m} -
\frac{(\vec\sigma\cdot\vec\pi)^4}{8\,m^3}
+\frac{(\vec\sigma\cdot\vec\pi)^6}{16\,m^5}
-\frac{i\,e}{8\,m^2}\,[\vec\sigma\cdot\vec\pi,\vec\sigma\cdot\vec E]
-\frac{e}{16\,m^3}\,\bigl\{\vec\sigma\cdot\vec\pi\,,\,\vec\sigma\cdot\dot{\vec E}\bigr\}
+\frac{i\,e}{32\,m^3}\,[\vec\sigma\cdot\vec\pi\,,\,\vec\sigma\cdot\ddot{\vec E}]
\nonumber \\ &&
-\frac{i\,e}{128\,m^4}\,[\vec\sigma\cdot\vec\pi,[\vec\sigma\cdot\vec\pi,
[\vec\sigma\cdot\vec\pi,\vec\sigma\cdot{\vec E}]]]
+\frac{i\,e}{16\,m^4}\,\Bigl((\vec\sigma\cdot\vec\pi)^2\,
[\vec\sigma\cdot\vec\pi,\vec\sigma\cdot\vec E] +
[\vec\sigma\cdot\vec\pi,\vec\sigma\cdot\vec
  E]\,(\vec\sigma\cdot\vec\pi)^2\Bigr),\label{18}
\end{eqnarray}
where we used the following commutator identity to simplify the expression,
\begin{eqnarray}
[(\vec\sigma\cdot\vec\pi)^3,\vec\sigma\cdot\vec E] &=&
-\frac{1}{2}\,[\vec\sigma\cdot\vec\pi,[\vec\sigma\cdot\vec\pi,
[\vec\sigma\cdot\vec\pi,\vec\sigma\cdot{\vec E}]]]
+\frac{3}{2}\,\Bigl((\vec\sigma\cdot\vec\pi)^2\,
[\vec\sigma\cdot\vec\pi,\vec\sigma\cdot\vec E] +
[\vec\sigma\cdot\vec\pi,\vec\sigma\cdot\vec E]\,(\vec\sigma\cdot\vec\pi)^2\Bigr)\,.\label{19}
\end{eqnarray}

Due to non-uniqueness in the operator $S$, the FW Hamiltonian $H_{\rm FW}$ given by
Eq.~(\ref{18}) differs from the one that can be obtained by the standard textbook approach
(relying on the subsequent use of the FW transformations) by the transformation $S$ with some
additional even operator. However, all variants of the FW Hamiltonian have to be equivalent at
the level of matrix elements between the states that satisfy the Schr\"odinger equation.

For the purpose of calculation of
the $\alpha^7\,m$ contribution we use the following FW Hamiltonian
\begin{eqnarray}
H_{\rm FW} &=& eA^0 +\frac{(\vec{\sigma}\cdot\vec{\pi})^2}{2m}-\frac{(\vec{\sigma}\cdot\vec{\pi})^4}{8m^3}+\frac{(\vec{\sigma}\cdot\vec{\pi})^6}{16m^5}
-\frac{i e}{8m^2} [\vec{\sigma}\cdot\vec{\pi},\vec{\sigma}\cdot\vec{E}] - \frac{e}{16m^3} \{\vec{\sigma}\cdot\vec{\pi},\vec{\sigma}\cdot\dot{\vec{E}}\}
\end{eqnarray}
where we omitted the higher-order terms from (\ref{18}), which do not contribute at the $\alpha^7m$
order. A further transformation
\begin{equation}
S = -\frac{e}{16m^3} \{\vec{\sigma}\cdot\vec{\pi},\vec{\sigma}\cdot\vec{E}\}\,,
\end{equation}
gives the correction to the Hamiltonian of the form
\begin{eqnarray}
\delta H &=& \frac{e}{16m^3} \{\vec{\sigma}\cdot\vec{\pi},\vec{\sigma}\cdot\dot{\vec{E}}\} + \frac{e^2}{8m^3}\,\vec{E}^2
\,.
\end{eqnarray}
The transformed FW Hamiltonian is now
\begin{eqnarray}
H_{\rm FW} &=& eA^0 +\frac{(\vec{\sigma}\cdot\vec{\pi})^2}{2m}-\frac{(\vec{\sigma}\cdot\vec{\pi})^4}{8m^3}
-\frac{i e}{8m^2} [\vec{\sigma}\cdot\vec{\pi},\vec{\sigma}\cdot\vec{E}] + \frac{e^2}{8m^3} \vec{E}^2\,.
\end{eqnarray}
It can be further simplified using the identities
\begin{eqnarray}
i\,[\vec{\sigma}\cdot\vec{\pi},\vec{\sigma}\cdot\vec{E}] &=& \vec{\nabla}\cdot\vec{E} + \sigma^{ij}\,\{E^i\,,\,\pi^j\},\nonumber\\
(\vec{\sigma}\cdot\vec{\pi})^2 &=& \vec\pi^2 - \frac{e}{2}\,\sigma\cdot B,\nonumber\\
(\vec{\sigma}\cdot\vec{\pi})^4 &=& \vec\pi^4 - \frac{1}{2}e\,\sigma\cdot B\,\vec{\pi}^2 - \frac12\vec{\pi}^2 e\,\sigma\cdot B
+ \frac{e^2}{4}\,(\sigma\cdot B)^2\,,
\end{eqnarray}
where $B^{ij} = \partial^i\,A^i-\partial^j\,A^i$. With these reductions, the FW Hamiltonian takes
the form
\begin{eqnarray}
H_{\rm FW} &=&
e A^0 + \frac{\vec \pi^2}{2m}  - \frac{e}{4m}\,\sigma\cdot B - \frac{\vec\pi^4}{8\,m^3} +
\frac{e}{16\,m^3}\bigl\{\sigma\cdot B\,,\,\vec{\pi}^2\bigr\}\nonumber\\
&&
-\frac{e}{8m^2}\bigl(\vec{\nabla}\cdot\vec{E_\parallel}
+ \sigma^{ij}\{E^i_\parallel+E^i_\perp\,,\,\pi^j\}\bigr)
+ \frac{e^2}{8m^3} \vec{E}^2 - \frac{e^2}{32\,m^3}\,(\sigma\cdot B)^2\,,
\end{eqnarray}
where we separated $\vec E$ into the parallel and perpendicular parts, $\vec E = \vec
E_\parallel+\vec E_\perp$. Now we apply the third  transformation, which allows us to get rid of
$\vec E_\perp$, with
\begin{equation}
S = \frac{e}{8m^2}\,\sigma^{ij}\,\{A^i\,,\,\pi^j\}\,.
\end{equation}
The correction to the Hamiltonian is
\begin{eqnarray}
\delta H &=&  \frac{e}{8m^2}\,\sigma^{ij}\,\{E^i_\perp\,,\,\pi^j\}
-\frac{e^2}{4m^2}\,\sigma^{ij}\,A^i\,E^j
+\frac{ie}{8m^2}\Bigl[\sigma^{ij}\,\{A^i\,,\,\pi^j\}\,,\,\frac{\vec\pi^2}{2m}-\frac{e}{4m}\,\sigma\cdot B\Bigr]
\,,
\end{eqnarray}
where
\begin{align}
  \bigl[\sigma^{ij}\,\{A^i\,,\,\pi^j\}\,,\,\sigma^{ab}\,B^{ab}\bigr] =&\
  \frac{1}{2}\,\bigl[\sigma^{ij}\,,\,\sigma^{ab}\bigr]\,\{\{A^i\,,\,\pi^j\}\,,\,B^{ab}\}
  +\frac{1}{2}\,\bigl\{\sigma^{ij}\,,\,\sigma^{ab}\bigr\}\,[\{A^i\,,\,\pi^j\}\,,\,B^{ab}]\,.
\end{align}
The first term contributes only to the fine structure and thus will be neglected here, while the
second term is angular averaged to obtain
 \begin{align}
  \bigl[\sigma^{ij}\,\{A^i\,,\,\pi^j\}\,,\,\sigma^{ab}\,B^{ab}\bigr] \approx &\ 2\,[\{A^i\,,\,\pi^j\}\,,\,B^{ij}]
=-4i\, A^i\nabla^j B^{ij}\,.
\end{align}
Furthermore, omitting terms contributing to the fine structure, we can rewrite
\begin{equation}
(\sigma\cdot B)^2 =  2\,B^{ij}B^{ij}\,.
\end{equation}
The final result for the FW Hamiltonian is
\begin{eqnarray}
H_{\rm FW} &=&
e A^0 + \frac{\pi^2}{2m}  - \frac{e}{4m}\sigma\cdot B - \frac{\pi^4}{8m^3} +
\frac{e}{16m^3}\bigl(\sigma\cdot B\,\vec{\pi}^2 + \vec{\pi}^2 \,\sigma\cdot B\bigr)\nonumber\\
&&
-\,\frac{e}{8m^2}\bigl(\vec{\nabla}\cdot\vec{E_\parallel}
+ \sigma^{ij}\{E_\parallel^i,\pi^j\}\bigr) + \frac{e^2}{8m^3} \vec{E}^2 - \frac{e^2}{16m^3}B^{ij} B^{ij}\nonumber\\
&&-\frac{e^2}{4m^2}\sigma^{ij} A^i E^j
+\frac{ie}{16m^3}[\sigma^{ij}\{A^i,\pi^j\},\pi^2] -\frac{e^2}{8m^3}A^i\,\nabla^j B^{ij}\,.
\end{eqnarray}

\section{Spin algebra in $d$-dimensions} \label{app:B}
The following basic formulas for Pauli matrices in $d$ dimensions are extensively used
throughout the paper:
\begin{align}
  \{\sigma^i\,,\,\sigma^j\} =&\ 2\,\delta^{ij}\,, \ \ \mbox{\rm and thus} \ \  \vec\sigma^2= d\,I\,,\\
  \sigma^{ij} =&\ \frac{1}{2\,i}\,[\sigma^i\,,\,\sigma^j]\,,\\
  \sigma^{ij}\,\sigma^{ij} =&\ \sigma\cdot\sigma = d\,(d-1)\,I\,,\\
        [\sigma^{ij},\sigma^{mn}] =&\ 2i\,(\delta^{im}\sigma^{jn}+\delta^{jn}\sigma^{im}-\delta^{in}\sigma^{jm}-\delta^{jm}\sigma^{jn})\,.
\end{align}
The following two formulas are valid after averaging with respect to all directions:
\begin{align}
\frac12\,\big\langle\{\sigma^{mk},\sigma^{nl}\}\big\rangle =&\ \frac{\sigma\cdot\sigma}{d(d-1)}\,(\delta^{mn}\delta^{kl}-\delta^{ml}\delta^{kn})\,,\\
\big\langle\sigma_1^{ij} \sigma_2^{kl}\big\rangle =&\ \frac{\sigma_1\cdot\sigma_2}{d(d-1)}(\delta^{ik}\delta^{jl}-\delta^{il}\delta^{jk})\label{B6}\,.
\end{align}

\section{Momentum integration} \label{app:C}

We describe below the evaluation of basic integrals in momentum space and $d$ dimensions.
The angular average is performed with help of trivial identities (for odd powers of $\hat{k}^i$ the angular average vanishes),
\begin{align}
\int\frac{d\Omega_d}{\Omega_d} \hat{k}^i \hat{k}^j  =&\ \frac{\delta^{ij}}{d}\,, \label{C1}\\
\int\frac{d\Omega_d}{\Omega_d} \hat{k}^i \hat{k}^j \hat{k}^m \hat{k}^n =&\
\frac{\delta^{ij}\delta^{mn}+\delta^{im}\delta^{jn}+\delta^{in}\delta^{jm}}{d(d+2)}\,.\label{C2}
\end{align}
Once all indices are contracted, one can perform scalar integrals starting from the simplest one, namely
\begin{eqnarray}\label{C4}
f_d(\alpha,\beta) \equiv
q^{-d+\alpha+\beta}\,
\int\frac{d^dk}{(2\pi)^d}\frac{1}{k^{\alpha}}\frac{1}{(k-q)^{\beta}} = \frac{1}{[4\pi]^{\frac{d}{2}}}
\frac{\Gamma(\frac{\alpha + \beta -
d}{2})\,\Gamma(\frac{d-\alpha}{2})\,\Gamma(\frac{d-\beta}{2})}
     {\Gamma(d-\frac{\alpha+\beta}{2})\,\Gamma(\frac{\alpha}{2})\,\Gamma(\frac{\beta}{2})}\,.
\end{eqnarray}
We now consider a more complicated basic integral,
\begin{align}
g_d(i,j,l) \equiv q^{-d+i+j+l}\,\int\frac{d^dk}{(2\pi)^d}\frac{1}{(k + |k-q|)^i}\,\frac{1}{k^j\,|k-q|^l}
\,.
\end{align}
For $i = 0$, this integral reduces to Eq.~(\ref{C4}), so $g_d(0,j,l) =f_d(j,l)$. For negative
$i$, $g_d$ can be expressed as a combination of $f_d$'s, specifically,
\begin{align}
  g_d(-1,j,l) =&\ f_d(j-1,l) + f_d(j,l-1)\,,\\
  g_d(-2,j,l) =&\ f_d(j-2,l) + 2\,f_d(j-1,l-1) + f_d(j,l-2)\,,\\
  g_d(-3,j,l) =&\ f_d(j-3,l) + 3\,f_d(j-2,l-1) + 3\,f_d(j-1,l-2) + f_d(j,l-3)\,.
\end{align}
For positive $i$, we use the following identities
\begin{eqnarray}
  &g_d(i,j,l) = g_d(i,l,j)\,,\\
  &g_d(i,j-1,l) + g_d(i,j,l-1) = g_d(i-1,j,l)\,,
\end{eqnarray}
to reduce the calculation to the single case of $j = l$, $  g_d(i,j) \equiv g_d(i,j,j)$\,. Using
the obvious formula
\begin{align}
  \vec q\,\frac{\partial}{\partial\vec q}\; g_d(i,j) = 0\,,
\end{align}
we obtain the following recurrence relation
\begin{align}
4\,(d - 2\,j)\,g_d(i, j) + (i - 2\,j)\,g_d(i,j+1) = j\,g_d(i-4,j+2) + (i - 4\,j)\,g_d(i-2,j+1) - j\,g_d(i-2,j+2)\,.
\label{rec}
\end{align}
In order to use this equation for calculation of $g_d$, we need initial values. First of all, we
note that
\begin{align}
  f_d(2,2) =&\ \int\frac{d^dk}{(2\pi)^d}\frac{1}{k^2}\frac{1}{(k-q)^2}
 =  \int\frac{d^{d-1}k}{(2\pi)^{d-1}}\int\frac{dx}{2\,\pi}\frac{1}{k^2+x^2}\frac{1}{(k-q)^2+x^2}
  \nonumber \\ =&\
  \frac{1}{2}\,\int\frac{d^{d-1}k}{(2\pi)^{d-1}}\frac{1}{k + |k-q|}\,\frac{1}{k\,|k-q|}
=   \frac{1}{2}\,g_{d-1}(1,1)\,
\end{align}
and, therefore,
\begin{align}
  g_d(1,1) = 2\,f_{d+1}(2,2)\,.
\end{align}
For even values of $i$, we choose $j=i/2$ to get the result for $g(i,i/2)$ directly from the
recursion. Specifically,
\begin{align}
g_d(2, 1) = \frac{1}{4\,(d-2)}\bigl(g_d(-2,3) -2\,g_d(0,2) - g_d(0,3)\bigr).
\end{align}
Next, we obtain $g_d(3,2)$ by transforming the integral representation for $f_d(4,4)$,
\begin{align}
  f_d(4,4) =&\ \int\frac{d^dk}{(2\pi)^d}\frac{1}{k^4}\frac{1}{(k-q)^4}
 = \int\frac{d^{d-1}k}{(2\pi)^{d-1}}\int\frac{dx}{2\,\pi}\frac{1}{(k^2+x^2)^2}\frac{1}{((k-q)^2+x^2)^2}
  \nonumber \\ =&\
  \frac{1}{4}\,\int\frac{d^{d-1}k}{(2\pi)^{d-1}}\biggl(
  \frac{1}{(k + |k-q|)^3}\,\frac{1}{k^2\,(k-q)^2} +
 \frac{1}{k + |k-q|}\,\frac{1}{k^3\,(k-q)^3}\biggr)
 =  \frac{1}{4}\,\bigl(g_{d-1}(3,2) + g_{d-1}(1,3)\bigr)\,.
\end{align}
Therefore,
\begin{align}
  g_d(3,2) = 4\,f_{d+1}(4,4) - g_d(1,3)\,.
\end{align}
We now consider the integral with $\ln k$.
After taking the derivative of Eq.~(\ref{C4}) with respect to $\alpha$ we obtain
\begin{eqnarray}\label{C23}
\int\frac{d^dk}{(2\pi)^d}\frac{1}{k^{\alpha}}\frac{1}{(k-q)^{\beta}}\ln k
&=&q^{d-\alpha-\beta}\,\frac{
\Gamma\big(\frac{\alpha+\beta-d}{2}\big)\Gamma\big(\frac{d-\alpha}{2}\big)
\Gamma\big(\frac{d-\beta}{2}\big)}
{2\,[4\pi]^{\frac{d}{2}}\,\Gamma\big(\frac{\alpha}{2}\big)\Gamma\big(\frac{\beta}{2}\big)
\Gamma\big(d-\frac{\alpha+\beta}{2}\big)}\nonumber\\
&&\times
\bigg[2\ln q+\psi\Big(\frac{\alpha}{2}\Big)
-\psi\Big(\frac{\alpha+\beta-d}{2}\Big)+\psi\Big(\frac{d-\alpha}{2}\Big)
-\psi\Big(d-\frac{\alpha+\beta}{2}\Big)\bigg]\,,
\end{eqnarray}
where $\psi(x)=\Gamma'(x)/\Gamma(x)$ is a digamma function. These are all
integrals in the momentum space, which were  used throughout this work.

\section{Derivation of $H^{(5)}_{\rm exch}$}
\label{app:D}

The Hamiltonian $H^{(5)}_{\rm exch}$ is split into the low-energy, middle-energy, and high-energy contributions,
\begin{equation}
H^{(5)}_{\rm exch} = H^{(5)}_L + H^{(5)}_M + H^{(5)}_S + H^{(5)}_H\,.
\end{equation}
The low-energy term $E_L^\Lambda=\langle H^{(5)}_L\rangle$ can be written as
\begin{eqnarray}
E_L^\Lambda &=&  \frac{e^2}{m^2} \int_\Lambda^\infty \frac{d^dk}{(2\pi)^d\,2k^3}\delta_\perp^{ij}(k)
\langle\phi|\,p_1^i (H_0-E_0)\,p_2^j|\phi\rangle + (1\leftrightarrow2)\nonumber\\
&=&  \frac{e^2}{4m^2} \frac{d-1}{d}\int_\Lambda^\infty \frac{d^dk}{(2\pi)^d k^3}
\langle\phi|\big[p_1^i, \big[V, p_2^i\big]\big]|\phi\rangle + (1\leftrightarrow2)\,.
\end{eqnarray}
Performing the momentum integration we get
\begin{eqnarray}
H_L^{(5)}
&=&\frac{\alpha^2}{m^2}\bigg(\frac{20}{9}+\frac43\ln\Lambda_\epsilon\bigg)\,.
\end{eqnarray}

The middle-energy contribution consists of two parts, the retardation correction to the single
transverse photon exchange $H_{M}^{(5)}$ and the double seagull with no retardation
$H_{S}^{(5)}$. The former is
\begin{eqnarray}
E_{M} &=&  e^2 \int \frac{d^dk}{(2\pi)^d\,2k^3}\delta_\perp^{ij}(k)
\langle\phi|\biggl(\frac{p_1^i}{m}+\frac{1}{2m}\sigma_1^{ki}\nabla_1^k\biggr)e^{i\vec{k}\cdot\vec{r}_1} (H_0-E_0)
\biggl(\frac{p_2^j}{m}+\frac{1}{2m}\sigma_2^{lj}\nabla_2^l\biggr)e^{-i\vec{k}\cdot\vec{r}_2}|\phi\rangle + (1\leftrightarrow2)\nonumber\\
&=&\frac{e^2}{4m^2} \int \frac{d^dk}{(2\pi)^d k^3}\delta_\perp^{ij}(k)
\langle\phi|\,\big[p_1^i, \big[V,
p_2^j\big]\big]\,e^{i\vec{k}\cdot\vec{r}}|\phi\rangle + (1\leftrightarrow2)\,.
\end{eqnarray}
Thus,
\begin{eqnarray}
H_{M}
&=&\frac{\alpha^2}{m^2}\,\bigg(-\frac43-\frac{4}{3\epsilon}+\frac83\ln2+\frac83\ln q\bigg)\,.
\end{eqnarray}
The contribution due to the double seagull diagram is
\begin{eqnarray}
E_{S} &=& -\frac{e^4}{2m^2} \int\frac{d^dk_1}{(2\pi)^d\,2k_1} \int\frac{d^dk_2}{(2\pi)^d\,2k_2}
 \delta_\perp^{ij}(k_1)\delta_\perp^{ij}(k_2)
\langle\phi|\,\frac{1}{(k_1+k_2)}\,e^{i(\vec{k}_1+\vec{k}_2)\cdot\vec{r}}|\phi\rangle
+(1\leftrightarrow2)\,.
\end{eqnarray}
After performing the momentum integrations we obtain
\begin{eqnarray}
H_{S}&=&\frac{\alpha^2}{m^2}\,\bigg(\frac53-\frac{1}{\epsilon}-\frac83\ln2+2\ln q\bigg)\,.
\end{eqnarray}
The high-energy part can be evaluated using the scattering amplitude approach, with the result
\begin{equation}
H_H = \frac{\alpha^2}{m^2}\bigg(\frac23+\frac{1}{\epsilon}-\frac12\sigma_1\cdot\sigma_2\bigg)\,.
\end{equation}
The total result in momentum representation is
\begin{equation}
H^{(5)}_{\rm exch} = \frac{\alpha^2}{m^2}\bigg(\frac{29}{9}+\frac83\ln[\alpha^{-2}]-\frac83\ln(2\lambda)
+\frac{14}{3}\ln q-\vec\sigma_1\cdot\vec\sigma_2\bigg)\,.
\end{equation}
Transforming it into coordinate representation and atomic units, we obtain
\begin{eqnarray}
H^{(5)}_{\rm exch} &=&
-\frac{14}{3}\bigg(\frac{1}{4\pi\,r^3}\bigg) + \bigg(-\vec\sigma_1\cdot\vec\sigma_2+\frac{71}{9}
+\frac83\ln[\alpha^{-2}]-\frac83\ln(2\lambda)+\frac{14}{3}\ln\alpha\bigg)\,\delta^3(r)\,.
\end{eqnarray}

\section{Low-energy retardation correction}
\label{app:E}

In this section we present our derivation for the low-energy retardation correction
$E_{L3}^{\Lambda}$. It is given by
\begin{eqnarray}
E_{L3}^{\Lambda} &=& \frac{e^2}{m^2}\int_\Lambda^\infty \frac{d^dk}{(2\pi)^d2k^5}\delta_\perp^{ij}(k)
\,\delta_{k^2} \bigg\langle\phi\bigg|p_1^i\,e^{i\vec k\cdot\vec r_1}(H_0-E_0)^3\,p_2^j\,e^{-i\vec k\cdot\vec r_2}\bigg|\phi\bigg\rangle
+(1\leftrightarrow2)\nonumber\\
&=& \frac{e^2}{4m^2}\int_\Lambda^\infty \frac{d^dk}{(2\pi)^dk^5}\delta_\perp^{ij}(k)
\,\delta_{k^2} \bigg\langle\phi\bigg|\bigg[p_1^i\,e^{i\vec k\cdot\vec r_1},H_0-E_0\bigg],
\bigg[H_0-E_0,\bigg[H_0-E_0,\,p_2^j\,e^{-i\vec k\cdot\vec r_2}\bigg]\bigg]\bigg]\bigg|\phi\bigg\rangle
+(1\leftrightarrow2)\nonumber\\
&=& \sum_{i=1}^8 \langle T_i\rangle\,,
\end{eqnarray}
where $\delta_{k^2}$ is the $k^2$ term in the Taylor expansion at $\vec k=0$.
The evaluation of $E_{L3}^{\Lambda}$ is in many respects similar to that for $E_1^{nn}$ described in
Appendix~\ref{app:F}. For $T_1$ and $T_2$ we have
\begin{eqnarray}
T_1
&=& \frac{e^2}{4m^2}\int_\Lambda^\infty \frac{d^dk}{(2\pi)^dk^5}\delta_\perp^{ij}(k)
\,\delta_{k^2} \bigg(\bigg[\bigg[p_1^i\,e^{i\vec k\cdot\vec r_1},V\bigg],
\bigg[V,\bigg[V,\,p_2^j\,e^{-i\vec k\cdot\vec r_2}\bigg]\bigg]\bigg]\bigg)
+(1\leftrightarrow2)=0\,,
\end{eqnarray}
\begin{eqnarray}
T_2
&=& \frac{e^2}{4m^2}\int_\Lambda^\infty \frac{d^dk}{(2\pi)^dk^5}\delta_\perp^{ij}(k)
\,\delta_{k^2} \bigg(\bigg[\bigg[p_1^i\,e^{i\vec k\cdot\vec r_1},\frac{p_1^2}{2}\bigg],
\bigg[V,\bigg[V,\,p_2^j\,e^{-i\vec k\cdot\vec r_2}\bigg]\bigg]\bigg]\bigg)
+(1\leftrightarrow2) = 0\,.
\end{eqnarray}
Next,
\begin{eqnarray}
T_3
&=& \frac{e^2}{4m^3}\int_\Lambda^\infty \frac{d^dk}{(2\pi)^dk^5}\delta_\perp^{ij}(k)
\,\delta_{k^2} \bigg(\bigg[\bigg[p_1^i\,e^{i\vec k\cdot\vec r_1},V\bigg],
\bigg[V,\bigg[\frac{p_2^2}{2},\,p_2^j\,e^{-i\vec k\cdot\vec r_2}\bigg]\bigg]\bigg]\bigg)
+(1\leftrightarrow2)\,.
\end{eqnarray}
We perform further transformation
\begin{eqnarray}
\bigg[\bigg[p_1^i\,e^{i\vec k\cdot\vec r_1},V\bigg],
\bigg[V,\bigg[\frac{p_2^2}{2},\,p_2^j\,e^{-i\vec k\cdot\vec r_2}\bigg]\bigg]\bigg]
&=&
\bigg[\bigg[p_1^i\,e^{i\vec k\cdot\vec r_1},V\bigg],
\big[V,\,p_2^j\big]\bigg[\frac{p_2^2}{2},\,e^{-i\vec k\cdot\vec r_2}\bigg]
+p_2^j\,\bigg[V,\bigg[\frac{p_2^2}{2},\,e^{-i\vec k\cdot\vec r_2}\bigg]\bigg]\bigg]\,.
\nonumber\\
\end{eqnarray}
The first $V$ potential in the last
equation contributes only with the electron-electron part, whereas the second may contribute with
either the electron-electron or electron-nucleus parts. We will thus have both two-body and three-body
parts in the $T_3$ term. The $k$-integration is separated from the matric element,
its radial is,
\begin{eqnarray}
4\pi\bigg(\frac{e^2}{4}\bigg)\int_\Lambda^\infty \frac{d^dk}{(2\pi)^dk^3}\delta_\perp^{ij}(k)
&=& \alpha\big(2+\ln{\Lambda_\epsilon}\big)
\int\frac{d\Omega_d}{\Omega_d}\delta_\perp^{ij}(k)\,,
\end{eqnarray}
where $\Omega_d$ is the surface area of a $d$-dimensional unit sphere, while the angular integration
is performed using Eqs. (\ref{C1},\ref{C2}). The results for $T_3$ in the momentum space is
\begin{eqnarray}
T_3 &=&\frac{\alpha^3}{m^3}\bigg\{\bigg(-\frac{2}{225}-\frac{2}{15}\ln{\Lambda_\epsilon}
\bigg)\,\bigg[p_2^j,-\bigg[\frac{Z}{r_2}\bigg]_\epsilon\bigg]\frac{q^j}{q^2}
+(1\leftrightarrow2)\bigg\}
+\pi\frac{\alpha^3}{m^3}\bigg(\frac{1}{225}+\frac{1}{15}\ln{\Lambda_\epsilon}+\frac{2}{15}\ln2-\frac{2}{15}\ln q\bigg)\,q\,.
\end{eqnarray}

The term $T_4$ is
\begin{eqnarray}
T_4
&=& \frac{e^2}{4m^3}\int_\Lambda^\infty \frac{d^dk}{(2\pi)^dk^5}\delta_\perp^{ij}(k)
\,\delta_{k^2} \bigg(\bigg[\bigg[p_1^i\,e^{i\vec k\cdot\vec r_1},V\bigg],
\bigg[\frac{p_1^2}{2}+\frac{p_2^2}{2},\bigg[V,\,p_2^j\,e^{-i\vec k\cdot\vec r_2}\bigg]\bigg]\bigg]\bigg)
+(1\leftrightarrow2)\,.
\end{eqnarray}
These double  commutators can be rewritten to
\begin{align}
\bigg[\bigg[p_1^i\,e^{i\vec k\cdot\vec r_1},V\bigg],
\bigg[\frac{p_1^2}{2}+\frac{p_2^2}{2},\bigg[V,\,p_2^j\,e^{-i\vec k\cdot\vec r_2}\bigg]\bigg]\bigg]
+(1\leftrightarrow2) =
2V^{jk}\,\bigg[\bigg[p_2^i, V\,e^{i\vec k\cdot\vec r}\bigg],p_2^k\bigg] +(1\leftrightarrow2)\,.
\end{align}
where $V^{ij}=[\,p_1^i,[V,p_2^j]]$, and the result in the momentum space is
\begin{eqnarray}
T_4&=&
\frac{\alpha^3}{m^3}\bigg\{\bigg(-\frac{28}{25}-\frac45\ln{\Lambda_\epsilon}\bigg)
\bigg[p_2^j,-\bigg[\frac{Z}{r_2}\bigg]_\epsilon\bigg]\frac{q^j}{q^2}\nonumber\\
&&+\bigg[p_2^i,\bigg[-\bigg[\frac{Z}{r_2}\bigg]_\epsilon,p_2^j\bigg]\bigg]
\bigg[\delta^{ij}\bigg(\frac{184}{225}+\frac{4}{15}\ln{\Lambda_\epsilon}\bigg)
+\frac{q^i q^j}{q^2}\bigg(-\frac{436}{225}-\frac{16}{15}\ln{\Lambda_\epsilon}\bigg)\bigg]\frac{1}{q^2}
+(1\leftrightarrow2)\bigg\}\nonumber\\
&&+ \pi\frac{\alpha^3}{m^3}\bigg(\frac{28}{225}-\frac{2}{15}\ln{\Lambda_\epsilon}-\frac{4}{15}\ln2+\frac{4}{15}\ln q\bigg)\,q
\,.
\end{eqnarray}
This concludes the evaluation of the three-photon terms. Now we move to the two-photon terms, starting
with $T_5$,
\begin{eqnarray}
T_5
&=& \frac{e^2}{4m^4}\int_\Lambda^\infty \frac{d^dk}{(2\pi)^dk^5}\delta_\perp^{ij}(k)
\,\delta_{k^2} \bigg(\bigg[\bigg[p_1^i\,e^{i\vec k\cdot\vec r_1},\frac{p_1^2}{2}\bigg],
\bigg[V,\bigg[\frac{p_2^2}{2},\,p_2^j\,e^{-i\vec k\cdot\vec r_2}\bigg]\bigg]\bigg]\bigg)
+(1\leftrightarrow2)\,.
\end{eqnarray}
The result for this term is
\begin{eqnarray}
T_5 &=& \frac{\alpha^2}{m^4}\bigg\{\vec P_1\cdot\vec P_2\bigg(\frac{48}{25}+\frac45\ln{\Lambda_\epsilon}\bigg)
+\frac{(\vec P_1\cdot\vec q)(\vec P_2\cdot\vec q)}{q^2}\bigg(\frac{64}{225}+\frac{4}{15}\ln{\Lambda_\epsilon}\bigg)\bigg\}
\,.
\end{eqnarray}
Term $T_6$ is given by
\begin{eqnarray}
T_6
&=& \frac{e^2}{4m^4}\int_\Lambda^\infty \frac{d^dk}{(2\pi)^dk^5}\delta_\perp^{ij}(k)
\,\delta_{k^2} \bigg(\bigg[\bigg[p_1^i\,e^{i\vec k\cdot\vec r_1},V\bigg],
\bigg[\frac{p_2^2}{2},\bigg[\frac{p_2^2}{2},\,p_2^j\,e^{-i\vec k\cdot\vec r_2}\bigg]\bigg]\bigg]\bigg)
+(1\leftrightarrow2)
\end{eqnarray}
and after evaluation we get
\begin{eqnarray}
T_6 &=&
\frac{\alpha^2}{m^4}\bigg(\frac{248}{225}+\frac{8}{15}\ln{\Lambda_\epsilon}\bigg)
\bigg[\frac14(\vec P_1-\vec P_2)^2+\frac12\vec P_1\cdot\vec P_2+\frac18 q^2
+\frac12\frac{[(\vec P_1-\vec P_2)\cdot\vec q]^2}{q^2}+\frac{(\vec P_1\cdot\vec q)(\vec P_2\cdot\vec q)}{q^2}\bigg]
\,.
\end{eqnarray}
Finally, the term $T_7$ is
\begin{eqnarray}
T_7
&=& \frac{e^2}{4m^4}\int_\Lambda^\infty \frac{d^dk}{(2\pi)^dk^5}\delta_\perp^{ij}(k)
\,\delta_{k^2} \bigg(\bigg[\bigg[p_1^i\,e^{i\vec k\cdot\vec r_1},\frac{p_1^2}{2}\bigg],
\bigg[\frac{p_1^2}{2}+\frac{p_2^2}{2},\bigg[V,\,p_2^j\,e^{-i\vec k\cdot\vec r_2}\bigg]\bigg]\bigg]\bigg)
+(1\leftrightarrow2)\,.
\end{eqnarray}
The result for this term is
\begin{eqnarray}
T_7 &=&
\frac{\alpha^2}{m^4}\bigg\{\bigg(\frac{248}{225}+\frac{8}{15}\ln{\Lambda_\epsilon}
\bigg)\bigg[\frac14(\vec P_1-\vec P_2)^2+\frac12\vec P_1\cdot\vec P_2
+\frac18 q^2+\frac{(\vec P_1\cdot\vec q)(\vec P_2\cdot\vec q)}{q^2}\bigg]
+\bigg(\frac{14}{25}+\frac25\ln{\Lambda_\epsilon}\bigg)
\,\frac{[(\vec P_1-\vec P_2)\cdot \vec q]^2}{q^2}\bigg\}
\,.
\nonumber \\
\end{eqnarray}
The last term $T_8$ vanishes,
\begin{eqnarray}
T_8
&=& \frac{e^2}{4m^4}\int_\Lambda^\infty \frac{d^dk}{(2\pi)^dk^5}\delta_\perp^{ij}(k)
\,\delta_{k^2} \bigg(\bigg[\bigg[p_1^i\,e^{i\vec k\cdot\vec r_1},\frac{p_1^2}{2}\bigg],
\bigg[\frac{p_1^2}{2}+\frac{p_2^2}{2},\bigg[\frac{p_2^2}{2},\,p_2^j\,e^{-i\vec k\cdot\vec r_2}\bigg]\bigg]\bigg]\bigg)
+(1\leftrightarrow2) = 0\,,
\end{eqnarray}
otherwise it would correspond to the one-photon exchange.
The total result for $H_{L3}^{\Lambda}=\sum_{i=1}^8 T_i$ is
\begin{eqnarray}
H_{L3}^{\Lambda} &=&
\frac{\alpha^3}{m^3}\bigg\{\bigg(-\frac{254}{225}-\frac{14}{15}\ln{\Lambda_\epsilon}\bigg)
\bigg[p_2^j,-\bigg[\frac{Z}{r_2}\bigg]_\epsilon\bigg]\frac{q^j}{q^2}\nonumber\\
&&+\bigg[p_2^i,\bigg[-\bigg[\frac{Z}{r_2}\bigg]_\epsilon,p_2^j\bigg]\bigg]
\bigg[\delta^{ij}\bigg(\frac{184}{225}+\frac{4}{15}\ln{\Lambda_\epsilon}\bigg)
+\frac{q^i q^j}{q^2}\bigg(-\frac{436}{225}-\frac{16}{15}\ln{\Lambda_\epsilon}\bigg)\bigg]\frac{1}{q^2}
+(1\leftrightarrow2)\bigg\}\nonumber\\
&&+\frac{\alpha^2}{m^4}\bigg\{\bigg(\frac{124}{225}+\frac{4}{15}\ln{\Lambda_\epsilon}
\bigg)\bigg((\vec P_1-\vec P_2)^2+\frac12 q^2\bigg)
+\bigg(\frac{136}{45}+\frac43\ln{\Lambda_\epsilon}\bigg)\,\vec P_1\cdot\vec P_2
 \nonumber\\ &&
+\bigg(\frac{10}{9}+\frac23\ln{\Lambda_\epsilon}\bigg)\frac{[(\vec P_1-\vec P_2)\cdot\vec q]^2}{q^2}
+\bigg(\frac{112}{45}+\frac43\ln{\Lambda_\epsilon}\bigg)\frac{(\vec P_1\cdot\vec q)(\vec P_2\cdot\vec q)}{q^2}\bigg\}
\nonumber\\&&
+\pi\frac{\alpha^3}{m^3}\bigg(\frac{29}{225}-\frac{1}{15}\ln{\Lambda_\epsilon}-\frac{2}{15}\ln2+\frac{2}{15}\ln q\bigg)\,q
\,.
\end{eqnarray}

\section{Spin independent triple-retardation $E_1^{nn}$}\label{app:F}
The spin-independent part of $E_1$ is the most complicated term to evaluate. We write it as
\begin{eqnarray}
E_1^{nn}
 &=&
\frac{e^2}{2\,m^2} \int \frac{d^dk}{(2\pi)^d k^5}\delta_\perp^{ij}(k)
\Bigl\langle\phi\Bigl|\Bigl[\Bigl[\,p_1^i\,e^{i\vec{k}\cdot\vec{r}_1},\frac{p_1^2}{2\,m}+V\Bigr],\Bigl[
\frac{p_1^2}{2\,m}+\frac{p_2^2}{2\,m}+V,\Bigl[\frac{p_2^2}{2\,m}+V,p_2^j \,e^{-i\vec{k}\cdot\vec{r}_2}\Bigr]\Bigr]\Bigr]\Bigr|\phi\Bigr\rangle\nonumber\\
&= &  \frac{e^2}{4m^2} \int \frac{d^dk}{(2\pi)^d k^5}\delta_\perp^{ij}(k)
\sum_{i=1}^8 \langle\phi|T_i|\phi\rangle+(1\leftrightarrow2)\nonumber \\ &=& \langle H^{nn}_{1,1} + H^{nn}_{1,2}\rangle\,.
\end{eqnarray}
where $H^{nn}_{1,1}$ involved three-body and $H^{nn}_{1,2}$ two-body terms.

Individual $T_i$ are evaluated as follows.
\begin{eqnarray}
T_1 &=& \bigl[\bigl[\,p_1^i\,e^{i\vec{k}\cdot\vec{r}_1},V\bigr],\bigl[
V,\bigl[V,p_2^j \,e^{-i\vec{k}\cdot\vec{r}_2}\bigr]\bigr]\bigr] = 0\,,\\
T_2 &=& \bigl[\bigl[\,p_1^i\,e^{i\vec{k}\cdot\vec{r}_1},\frac{p_1^2}{2m}\bigr],\bigl[
V,\bigl[V,
p_2^j \,e^{-i\vec{k}\cdot\vec{r}_2}\bigr]\bigr]\bigr]=0\,,
\end{eqnarray}

The term $T_3$ is
\begin{eqnarray}
T_3 &=& \frac{1}{m} \bigg[\bigg[\,p_1^i\,e^{i\vec{k}\cdot\vec{r}_1},V\bigg],\bigg[\frac{p_1^2}{2}+\frac{p_2^2}{2},\bigg[V,p_2^j \,e^{-i\vec{k}\cdot\vec{r}_2}\bigg]\bigg]\bigg]\nonumber\\
&=&\frac1m \bigg[e^{i\vec{k}\cdot\vec{r}_1}\,[p_1^i,V],\,\bigg[\frac{p_2^2}{2},e^{-i\vec{k}\cdot\vec{r}_2}\bigg][V,p_2^j]
  +e^{-i\vec{k}\cdot\vec{r}_2}\,\bigg[\frac{p_1^2}{2}+\frac{p_2^2}{2},[V,p_2^j]\bigg]\bigg]\nonumber \\
&=& \frac{V^{jk}}{m}\bigg(\,\bigg[\bigg[\,p_1^i,V e^{i\vec{k}\cdot\vec{r}}\bigg],p_1^k\bigg] + \bigg[\bigg[\,p_2^i,V e^{i\vec{k}\cdot\vec{r}}\bigg],p_2^k\bigg]\,\bigg)\,,
\end{eqnarray}
where $ V^{ij} \equiv [\,p_1^i,[V,p_2^j]]$.

The term $T_4$ is
\begin{eqnarray}
T_4 &=&
\frac{1}{m} \bigg[\bigg[\,p_1^i\,e^{i\vec{k}\cdot\vec{r}_1},V\bigg],\bigg[V,\bigg[\frac{p_2^2}{2},p_2^j \,e^{-i\vec{k}\cdot\vec{r}_2}\bigg]\bigg]\bigg]\nonumber\\
&=&\frac1m\bigg(\,\big[\,p_2^j,V\big]\,\bigg[e^{i\vec{k}\cdot\vec{r}},p_2^k\bigg]\,V^{ik} + \big[\,p_2^k,V\big]\,\bigg[e^{i\vec{k}\cdot\vec{r}},p_2^k\bigg]\,V^{ij}\bigg)\,.
\end{eqnarray}
The sum of terms $T_3$ and $T_4$ provides the complete three-photon part $H_{1,1}^{nn}$ and is
\begin{eqnarray}
H_{1,1}^{nn} &=& \frac{\alpha}{4m^3}\int\frac{d^dk}{(2\pi)^d\,k^5}\delta_\perp^{ij}(k)\nonumber\\&\times&
\biggl\{V^{jk}\,\bigg(\biggl[\biggl[\,p_1^i,V\,e^{i\vec k\cdot\vec r}\biggr],p_1^k\biggr]
+ \biggl[\biggl[\,p_2^i,V\,e^{i\vec k\cdot\vec r}\biggr],p_2^k\biggr]\bigg)
+\bigl[\,p_2^j,V\big]\,\biggl[e^{i\vec k\cdot\vec r},p_2^k\biggr]\,V^{ik}+\big[\,p_2^k,V\big]\,\biggl[e^{i\vec k\cdot\vec r},p_2^k\biggr]\,V^{ij}\biggr\}
\nonumber\\
&&+(1\leftrightarrow2)\nonumber\\
&=&\frac{\pi \alpha^3}{m^3}\bigg(-\frac{2}{5}+\frac{1}{15\epsilon}-\frac{4}{15}\ln q\bigg)\,q
+
\frac{\alpha^3}{m^3}\bigg\{\bigg[p_2^j,-\frac{Z}{r_2}\bigg]\,\bigg(-\frac{2}{15}+\frac{14}{15\epsilon}-\frac{28}{15}\ln2-\frac{28}{15}\ln q\bigg)\frac{q^j}{q^2}\nonumber\\
&&+\bigg[\bigg[p_2^i,-\frac{Z}{r_2}\bigg],p_2^j\bigg]\bigg[\delta^{ij}\bigg(-\frac{4}{5}-\frac{4}{15\epsilon}+\frac{8}{15}\ln2+\frac{8}{15}\ln q\bigg)
+\frac{q^i q^j}{q^2}\bigg(\frac{16}{15}+\frac{16}{15\epsilon}-\frac{32}{15}\ln2-\frac{32}{15}\ln q\bigg)\bigg]\frac{1}{q^2}\nonumber\\
&&+(1\leftrightarrow2)
\bigg\}\,.
\end{eqnarray}

The terms $T_5$ - $T_7$ involve two-photon contributions. The first term $T_5$ is
\begin{eqnarray}
T_5 &=&
\frac{1}{m^2} \Big[\Big[\,p_1^i\,e^{i\vec{k}\cdot\vec{r}_1},\frac{p_1^2}{2}\Big],\Big[V,\Big[\frac{p_2^2}{2},p_2^j \,e^{-i\vec{k}\cdot\vec{r}_2}\Big]\Big]\Big]\,.
\end{eqnarray}
Defining $\tilde T_5$ as its corresponding contribution in the momentum representation, we get
\begin{eqnarray}
\tilde T_5 &=&
\frac{e^2}{4m^2}\int\frac{d^dk}{(2\pi)^d\,k^5}\delta^{ij}_\perp(k)\,T_5+(1\leftrightarrow2)\nonumber\\
&=&\frac{\alpha^2}{m^4}\bigg\{
\frac{4}{15}\vec P_1\cdot\vec P_2-\frac{16}{15}\frac{(\vec P_1\cdot\vec q)(\vec P_2\cdot\vec q)}{q^2}
+\bigg(-\frac{1}{2\epsilon}+\ln2+\ln q\bigg)
\bigg(-\frac{16}{15}\vec P_1\cdot\vec P_2+\frac{8}{15}\frac{(\vec P_1\cdot\vec q)(\vec P_2\cdot\vec q)}{q^2}\bigg)\bigg\}\,.
\end{eqnarray}
The term $T_6$ is
\begin{eqnarray}
T_6 &=&
\frac{1}{m^2} \bigg[\bigg[\,p_1^i\,e^{i\vec{k}\cdot\vec{r}_1},V\bigg],\bigg[\frac{p_2^2}{2},\bigg[\frac{p_2^2}{2},p_2^j \,e^{-i\vec{k}\cdot\vec{r}_2}\bigg]\bigg]\bigg]\nonumber\\
&=& \frac{1}{m^2} \bigg\{\,p_2^j\,\bigg[\big[\,p_1^i,V\big],\bigg[\frac{p_2^2}{2},\bigg[\frac{p_2^2}{2},e^{i\vec{k}\cdot\vec{r}}\bigg]\bigg]\bigg]
+\big[\,p_1^i,\big[V,p_2^j\big]\big]\,\bigg[\frac{p_2^2}{2},\bigg[\frac{p_2^2}{2},e^{i\vec{k}\cdot\vec{r}}\bigg]\bigg]\bigg\}\,.
\end{eqnarray}
The corresponding result in the momentum representation is
\begin{eqnarray}
\tilde T_6 &=&
\frac{\alpha^2}{m^4}\bigg\{\frac{2}{45}\big(\vec P_1-\vec P_2\big)^2 +\frac{14}{45}q^2 + \frac{4}{45}\vec P_1\cdot\vec P_2
-\frac{32}{15}\frac{(\vec P_1\cdot\vec q)(\vec P_2\cdot\vec q)}{q^2}
-\frac{16}{15}\frac{[(\vec P_1-\vec P_2)\cdot\vec q]^2}{q^2}\nonumber\\
&&+\bigg(-\frac{1}{2\epsilon}+\ln2+\ln q\bigg)
\bigg(
\frac{4}{15}\big(\vec P_1-\vec P_2\big)^2 -\frac{8}{15}q^2 + \frac{8}{15}\vec P_1\cdot\vec P_2
+\frac{16}{15}\frac{(\vec P_1\cdot\vec q)(\vec P_2\cdot\vec q)}{q^2}
+\frac{8}{15}\frac{[(\vec P_1-\vec P_2)\cdot\vec q]^2}{q^2}
\bigg)\bigg\}\,.\nonumber\\
\end{eqnarray}
The term $T_7$ is
\begin{eqnarray}
T_7 &=& \frac{1}{m^2}\bigg[\bigg[\,p_1^i\,e^{i\vec{k}\cdot\vec{r}_1},\frac{p_1^2}{2}\bigg],\bigg[\frac{p_1^2}{2}
+\frac{p_2^2}{2},\bigg[V,p_2^j \,e^{-i\vec{k}\cdot\vec{r}_2}\bigg]\bigg]\bigg]\nonumber\\
&=&\frac{1}{m^2}\bigg[\,p_1^i\bigg[e^{i\vec{k}\cdot\vec{r}_1},\frac{p_1^2}{2}\bigg],\big[V,p_2^j\big]\,\bigg[\frac{p_2^2}{2},e^{-i\vec{k}\cdot\vec{r}_2}\bigg]
+\bigg[\frac{p_1^2}{2}+\frac{p_2^2}{2},\big[V,p_2^j\big]\bigg]\,e^{-i\vec{k}\cdot\vec{r}_2}\bigg]\,.
\end{eqnarray}
Evaluating this term in the momentum representation, we get
\begin{eqnarray}
\tilde T_7 &=&
\frac{\alpha^2}{m^4}\bigg\{\frac{2}{45}\big(\vec P_1-\vec P_2\big)^2 +\frac{14}{45}q^2 + \frac{4}{45}\vec P_1\cdot\vec P_2
-\frac{32}{15}\frac{(\vec P_1\cdot\vec q)(\vec P_2\cdot\vec q)}{q^2}
-\frac{2}{3}\frac{[(\vec P_1-\vec P_2)\cdot\vec q]^2}{q^2}\nonumber\\
&&+\bigg(-\frac{1}{2\epsilon}+\ln2+\ln q\bigg)
\bigg(
\frac{4}{15}\big(\vec P_1-\vec P_2\big)^2 -\frac{8}{15}q^2 + \frac{8}{15}\vec P_1\cdot\vec P_2
+\frac{16}{15}\frac{(\vec P_1\cdot\vec q)(\vec P_2\cdot\vec q)}{q^2}
+\frac{4}{5}\frac{[(\vec P_1-\vec P_2)\cdot\vec q]^2}{q^2}
\bigg)\bigg\}\,.\nonumber\\
\end{eqnarray}

The last term vanishes,
\begin{equation}
T_8 = \frac{1}{m^3}\bigg[\bigg[\,p_1^i\,e^{i\vec{k}\cdot\vec{r}_1},\frac{p_1^2}{2}\bigg],\bigg[\frac{p_1^2}{2}+\frac{p_2^2}{2},\bigg[\frac{p_2^2}{2},p_2^j \,e^{-i\vec{k}\cdot\vec{r}_2}\bigg]\bigg]\bigg]=0\,.
\end{equation}

The total two-photon contribution $H^{nn}_{1,2} = \sum_{i=5}^7 \tilde T^i $ and is given by
\begin{eqnarray}
H_{1,2}^{nn} &=&
\frac{\alpha^2}{m^4}\bigg\{\frac{4}{45}\big(\vec P_1-\vec P_2\big)^2 +\frac{28}{45}q^2 + \frac{4}{9}\vec P_1\cdot\vec P_2
-\frac{16}{3}\frac{(\vec P_1\cdot\vec q)(\vec P_2\cdot\vec q)}{q^2}
-\frac{26}{15}\frac{[(\vec P_1-\vec P_2)\cdot\vec q]^2}{q^2}\nonumber\\
&&+\bigg(-\frac{1}{2\epsilon}+\ln2+\ln q\bigg)
\bigg(
\frac{8}{15}\big(\vec P_1-\vec P_2\big)^2 -\frac{16}{15}q^2
+\frac{8}{3}\frac{(\vec P_1\cdot\vec q)(\vec P_2\cdot\vec q)}{q^2}
+\frac{4}{3}\frac{[(\vec P_1-\vec P_2)\cdot\vec q]^2}{q^2}
\bigg)\bigg\}\,.\nonumber\\
\end{eqnarray}
and the total spin-independent part of $H_1$ is the sum of the two-photon and three-photon
contributions,
\begin{equation}
  H_1^{nn} = H_{1,1}^{nn}+H_{1,2}^{nn}\,.
\end{equation}

\end{widetext}

\end{document}